\documentclass[fleqn,usenatbib]{mnras}

\usepackage{newtxtext,newtxmath}

\usepackage[T1]{fontenc}
\usepackage{ae,aecompl}

\usepackage{graphicx}	
\usepackage{amsmath}	
\usepackage{amssymb}	

\newcommand{\dbreak}{D$_n$(4000)\,}
\newcommand{\prospector}{{\sc Prospector}\,}
\newcommand{\oii}{[O\,{\sc ii}]\,} 
\newcommand{\thh}{$^\mathrm{th}$\,}
\newcommand{\mwa}{$t_\mathrm{ mw}$\,}
\newcommand{\lwa}{$t_\mathrm{ lw}$\,}
\newcommand{\tdelta}{$t_{\Delta}$\,}
\newcommand{\tdelay}{$t_\mathrm{delay}$\,}
\newcommand{\aFe}{[$\alpha$/Fe]\,}
\newcommand{\frejuv}{${f}_{{M}_\ast {<} 1\,\mathrm{Gyr}}$\,}

\title[GOGREEN: Age and environment of quiescent galaxies]{
The GOGREEN survey: Post-infall environmental quenching fails to predict the observed age difference between quiescent field and cluster galaxies at $z{>}1$}

\author[K. Webb, et al.]{
Kristi Webb,$^{1,2}$\thanks{E-mail: kristi.webb@uwaterloo.ca} 
Michael L. Balogh,$^{1,2}$
Joel Leja,$^{3}$
Remco F. J. van der Burg,$^{4}$
\newauthor Gregory Rudnick,$^{5}$
Adam Muzzin,$^{6}$
Kevin Boak,$^{1}$
Pierluigi Cerulo,$^{7}$
David Gilbank,$^{8,9}$
\newauthor Chris Lidman,$^{10,11}$
Lyndsay J. Old,$^{12,13}$
Irene Pintos-Castro,$^{13}$
Sean McGee,$^{14}$
\newauthor Heath Shipley,$^{15}$
Andrea Biviano,$^{16,17}$
Jeffrey C. C. Chan,$^{18}$
Michael Cooper,$^{19}$
\newauthor Gabriella De Lucia,$^{16}$
Ricardo Demarco,$^{20}$
Ben Forrest,$^{18}$
Pascale Jablonka,$^{21,22}$
\newauthor Egidijus Kukstas,$^{23}$
Ian G. McCarthy,$^{23}$
Karen McNab,$^{1,2}$
Julie Nantais,$^{24}$
\newauthor Allison Noble,$^{25,26}$
Bianca Poggianti,$^{27}$
Andrew M. M. Reeves,$^{1,2}$
Benedetta Vulcani,$^{27}$
\newauthor Gillian Wilson,$^{18}$
Howard K. C. Yee,$^{13}$
Dennis Zaritsky$^{28}$\\
\noindent{Affiliations are listed at the end of the paper.}
}

\date{Accepted XXX. Received YYY; in original form ZZZ}

\pubyear{2020}

\begin{document}
\label{firstpage}
\pagerange{\pageref{firstpage}--\pageref{lastpage}}
\maketitle

\begin{abstract}
We study the star formation histories (SFHs) and mass-weighted ages of 331 \textit{UVJ}-selected quiescent galaxies in 11 galaxy clusters and in the field at $1{<}z{<}1.5$ from the Gemini Observations of Galaxies in Rich Early ENvironments (GOGREEN) survey. We determine the SFHs of individual galaxies by simultaneously fitting rest-frame optical spectroscopy and broadband photometry to stellar population models. We confirm that the SFHs are consistent with more massive galaxies having on average earlier formation times. Comparing galaxies found in massive clusters with those in the field, we find galaxies with $M_\ast<10^{11.3}$\,M$_{\sun}$ in the field have more extended SFHs. From the SFHs we calculate the mass-weighted ages, and compare age distributions of galaxies between the two environments, at fixed mass. We constrain the difference in mass-weighted ages between field and cluster galaxies to $0.31_{^{-0.33}}^{_{+0.51}}$\,Gyr, in the sense that cluster galaxies are older. We place this result in the context of two simple quenching models and show that neither environmental quenching based on time since infall (without pre-processing) nor a difference in formation times alone can reproduce both the average age difference and relative quenched fractions. This is distinctly different from local clusters, for which the majority of the quenched population is consistent with having been environmentally quenched upon infall. Our results suggest that quenched population in galaxy clusters at z\,${>}$1 has been driven by different physical processes than those at play at $z{=}0$.
\end{abstract}





\section{Introduction} \label{sec:introduction}

Since $z{\sim}2.5$, the galaxy population demonstrates a marked bimodality in star formation rates \citep[SFRs, e.g.][]{brinchmann2004, brammer2011, muzzin2012}, and the quiescent component, representing galaxies with negligible current SFRs, has increased steadily \citep{faber2007, muzzin2013a, tomczak2014, barro2017}. This indicates that the relatively rapid suppression of star formation (quenching) is a fundamental aspect of galaxy evolution, and one that is largely responsible for the steep decline in cosmic SFR density \citep[e.g.][]{renzini2016}. The rate of quenching, and indeed galaxy evolution in general, is observed to depend strongly on both stellar mass and environment. 
In particular, galaxies that are more massive or exist in denser environments are more likely to be quiescent \citep[e.g.][]{kauffmann2003b, kauffmann2004, brinchmann2004, baldry2006, weinmann2006, kimm2009}. 

There have been many studies focused on identifying the main mechanisms that transform galaxies from star forming to quiescent. Simulations which include feedback from active galactic nuclei (AGN) and feedback from star formation have successfully reproduced the SFR bimodality \citep[e.g.][]{croton2006, bower2006, hirschmann2016}, if not quite replicating the observed quenched fractions \citep[e.g.][]{hirschmann2016, xie2020}. However, explaining the dependence of the quenched fraction on local environment appears to require additional processes related to the larger scale environmental densities of galaxies \citep[e.g.][]{baldry2006, peng2012}. Environmental quenching is commonly thought to take place as a galaxy accretes into the halo of a larger structure, either by the removal of its gas reservoir through tidal/ram pressure stripping or by preventing gas in the galaxy halo from accreting and forming new stars, sometimes called strangulation \citep[e.g.][]{gunn1972, larson1980, balogh2000}. Evidence for the removal of gas can be seen by the lack of line emission from galaxies approaching larger haloes \citep[e.g.][]{odekon2016, jaffe2016, zhang2019}.
Denser environments could also favour tidal interactions, or harassment, between galaxies \citep[e.g.][]{merritt1983, moore1996}, which can lead to increased SFRs and accelerated gas consumption \citep{fujita2004}. Given that the fraction of quiescent galaxies increases with the number density of surrounding galaxies (i.e., rich galaxy clusters \textit{vs} galaxy groups, e.g. \citealt{kauffmann2004, wilman2010, peng2010, darvish2016}, or with distance from cluster cores, e.g. \citealt{loh2008, woo2013, lin2014, muzzin2014, jian2017, guglielmo2019, pintos-castro2019}), the effectiveness of environmental quenching is thought to scale with environmental density. 

A simple empirical model of environmental quenching is that, upon infall, the SFR of a galaxy rapidly declines, on an e-folding timescale called the `fading time'. Motivated in part by the non-zero fraction of star forming galaxies in clusters, this quenching is thought to happen at some time after infall, called the delay time. \citet{wetzel2013} used a sample of local galaxies in the Sloan Digital Sky Survey \citep[SDSS;][]{york2000} together with a cosmological $N$-body simulation in the context of this `delayed-then-rapid' model, and found that typical delay times at $z{=}0$ are 2--4\,Gyr, and fading times ${<}$0.8\,Gyr. Galaxy haloes grow hierarchically, however, and this infall-based quenching might happen upon the first infall of a galaxy into a larger halo, which might not be the final cluster halo. So-called `pre-processing' within galaxy group environments may be an important preceding process \citep[e.g.][]{zabludoff1998, fujita2004, mcgee2009, delucia2012, pallero2019}. Observations at higher redshifts have the potential to remove some of the degeneracies associated with this empirical picture, in part because the evolution in galaxy properties like SFRs and gas fractions is decoupled from the rate of dark matter halo mass growth \citep[e.g.][]{mcgee2014}. 

One direct way to trace galaxy evolution is to measure the stellar mass function (SMF) as a function of redshift and environment for passive and active galaxies \citep[e.g.][]{fontana2004, vulcani2011, vulcani2013, muzzin2013c, tomczak2014, nantais2016, vanderburg2013, vanderburg2020}. Similarly, detailed studies of the stellar populations in galaxies compared across redshift epochs can reveal how the overall population of galaxies has evolved \citep[e.g.][]{poggianti1999, trager2000a, sanchez-blazquez2009}.
This is only indirectly connected to changes in star formation, like quenching, and does not allow one to easily identify what subset of the population is undergoing changes at a given time.
A complementary approach is to measure the star formation histories (SFHs) of individual galaxies and thus reconstruct the growth of populations \citep[e.g.][]{heavens2000, heavens2004, panter2003}. Comparing the SFHs of galaxies in isolated and dense environments has the potential to provide new information on the effect of environment-specific quenching processes. 

Measuring the stellar ages of galaxies as a probe of the SFH is very challenging, however. For all but the nearest galaxies individual stars are not resolved; rather, observations measure the integrated luminosity of the stellar population and thus it is necessary to disentangle the contribution of stars of various masses and ages. The galaxy spectral energy distributions (SEDs) also suffer from degeneracies between galaxy properties (e.g. stellar age, metallicity, and dust) particularly at low resolution \citep[e.g.][]{worthey1994}. Many studies rely on studying select spectral features, observed at high resolution, which are well calibrated against such degeneracies \citep[e.g.][]{vazdekis1999, trager2000b} or more recently with full-spectrum fitting \citep[e.g.][]{macarthur2009, sanchez-blazquez2011}. Photometry alone cannot distinguish between such model parameters, and age estimates can be strongly influenced by priors \citep{carnall2019a, leja2019a}. The integrated luminosity is also dominated by bright young stars, `outshining' evidence of older stellar populations \citep{papovich2001}. Galaxies older than ${\sim}$5\,Gyr have very similar SEDs, making it difficult to precisely estimate the stellar age of quiescent galaxies at low redshifts \citep[e.g.][]{gallazzi2005}. Moreover, empirical models of stellar evolution are biased by systematic uncertainties and can significantly impact age estimates, particularly for galaxies dominated by intermediate age stars \citep[e.g.][]{maraston2005, han2018}. While measuring the properties and SFHs of individual galaxies provides the clearest picture of galaxy evolution, this requires relatively large samples of galaxies with sufficiently high signal-to-noise ratio (SNR) continuum spectroscopy. 

The consensus of observations at low to moderate redshifts, despite these challenges, is that there is a trend between the SFHs and stellar mass for quiescent galaxies: the SFRs of massive galaxies peaked at earlier times than lower mass systems \citep[sometimes called `downsizing'; e.g.][]{cowie1996, brinchmann2004, kodama2004}, and correspondingly massive galaxies form their stellar mass earlier (and are therefore older) on average \citep[`archaeological downsizing'; e.g.][]{nelan2005, thomas2005, thomas2010, treu2005a,treu2005b, cimatti2006, gallazzi2014, pacifici2016, carnall2018, estrada-carpenter2020, saracco2020}. These trends together are commonly referred to as `mass dependent evolution'. 

For massive galaxies, the majority of their stellar mass is formed within only 1--2\,Gyr \citep[][]{gallazzi2004, gallazzi2005, glazebrook2004, bell2005, nelan2005, thomas2005, thomas2010, treu2005a, toft2012, mcdermid2015, citro2016}, and have quenched as early as $z{\sim}3$-$4$ \citep[e.g.][]{straatman2014, glazebrook2017, schreiber2018b, forrest2020}. Low-redshift observations of massive quiescent galaxies (typically early-type galaxies, ETGs) find that galaxies in less-dense environments are on average 1--2\,Gyr younger than galaxies in massive clusters \citep[e.g.][and ref's therein]{vandokkum2003, thomas2005, renzini2006}. Notably, age differences at low redshifts could be enhanced by environmental effects which come into play only at late times, such as `rejuvenation' \citep{thomas2010} or `frosting' \citep{trager2000b} of star formation via galaxy mergers or interactions -- which occur more frequently in lower mass haloes \citep[e.g.][]{cooper2010}. \citet{paulino-afonso2020} show that SF can be enhanced for low-to-moderate mass galaxies even at moderate, `filament-like' overdensities. For these reasons, higher redshift observations can provide better leverage on the differences in galaxy properties related to how they evolved in different environments. 

At $z{\sim}1$, the average ages of ETGs in low-density environments are within 0.5\,Gyr of comparable galaxies in galaxy clusters \citep[e.g.][]{gobat2008, thomas2010, rettura2010, rettura2011, raichoor2011, saracco2017, woodrum2017}. The lack of environmental influence at this epoch is supported by Fundamental Plane (FP) studies of ETGs which show that the mass-to-light (M/L) ratios evolve similarly for galaxies in field and cluster environments \citep[e.g.][]{diseregoalighieri2006a,diseregoalighieri2006b, vandokkum2007, saglia2010, woodrum2017}. Studies of SFRs between star forming galaxies in cluster and field environments show mixed results, either predicting little \citep[e.g.][]{peng2010, muzzin2012} or modest \citep[e.g.][and \citealt{mcgee2011} for groups]{vulcani2010, popesso2011, koyama2013, old2020} trends between the star forming main sequence and environment. 

Importantly, the present comparisons at $z{>}1$ between field and cluster galaxies are typically made for small samples and/or with limited stellar mass coverage, relying on the measurement of few absorption line indices \citep[e.g.][]{vanderwel2004, vandokkum2007, gobat2008, woodrum2017, saracco2020} or photometric SEDs \citep[e.g.][]{rettura2010, rettura2011, raichoor2011, saracco2017}. While recent spectroscopic surveys have collected larger samples of quiescent galaxies at higher redshifts \citep[e.g.][]{pacifici2016, thomas2017, carnall2019b, estrada-carpenter2019}, there has not yet been a systematic study of the SFHs with environment. 
We can significantly improve our understanding of the differences in SFHs of galaxies related to their environment with the Gemini Observations of Rich Early ENvironments survey \citep[GOGREEN\footnote{http://gogreensurvey.ca/};][Balogh et al. 2020; in prep.]{balogh2017}. The GOGREEN survey targeted galaxies in clusters and groups at $1{<}z{<}1.5$, and includes isolated `field' galaxies along the line-of-sight of these structures. With galaxies at lower stellar masses, and at higher redshifts, than preceding surveys (e.g. GCLASS, \citealt{muzzin2012}; GEEC2, \citealt{balogh2014}), GOGREEN is better suited to test the predictions of galaxy evolution models \citep[e.g.][]{bower2012, weinmann2012}.

Taking advantage of the well-sampled, homogeneously selected spectroscopy and broad photometric coverage for hundreds of galaxies observed as part of GOGREEN, we measure the SFHs and mass-weighted ages for quiescent galaxies in both average, `field', environments and in massive galaxy clusters. Comparing the star formation timescales between galaxies in clusters and field environments, we test simple quenching models which have been proposed to explain the difference in ages between the two populations. This work complements the comparison of the stellar mass distributions measured in \citep[][]{vanderburg2020}, and the relation between stellar mass and star formation for star-forming galaxies \citep{old2020}.

The paper is outlined as follows. In Section\,\ref{sec:data}, we provide a brief description of the GOGREEN sample and the selection of quiescent galaxies used in our analysis. In Section\,\ref{sec:fitting}, we describe the SFH fitting procedure. In Section\,\ref{sec:sfhs_mwas}, we show the SFHs and estimated average ages, and test the robustness of the results against our selection criteria for quiescent galaxies. Then in Section\,\ref{sec:discussion}, we discuss the SFHs and average ages as a function of stellar mass and environment in the context of the literature. We also discuss our results in the context of two toy models for environmental galaxy quenching scenarios: either galaxies quench upon being accreted into denser environments, or galaxies in denser environments simply formed earlier. Lastly, in Section\,\ref{sec:conclusions} we summarise the results.

The magnitudes reported are on the AB magnitude system. We use a \citet{chabrier2003} IMF, and adopt a flat $\Lambda$CDM cosmology with $\Omega_\mathrm{m}{=}0.3$ and H$_0{=}70$\,km\,s$^{-1}$\,Mpc$^{-1}$.


    \begin{table}
    \centering
    \caption{Description of the GOGREEN galaxy cluster targets. Notes: Coordinates and redshifts for each galaxy system in the GOGREEN sample. Spectroscopic redshifts are from Balogh et al. (2020; in prep). SpARCS1033 was excluded in this study because of the lack of $K$-band photometry. Notes: $\dag$ indicates clusters also in the GCLASS survey. 
    }
    \label{tab:sample}
    \begin{tabular}{lccc}
    \hline
    Full name	& BCG RA, Dec & Redshift  \\
    & (J2000) &     \\
    \hline
    SPT0205					& 02:05:48.19,  -58:28:49.0	& 1.323		\\
    SPT0546					& 05:46:33.67,  -53:45:40.6	& 1.068		\\
    SPT2016 				& 21:06:04.59,  -58:44:27.9 & 1.132		\\
    $^\dag$SpARCS0035-3412	& 00:35:49.68,  -43:12:23.8	& 1.335		\\
    SpARCS0219-0531		    & 02:19:43.56,  -05:31:29.6	& 1.328			\\
    SpARCS0335-2929			& 03:35:03.56,  -29:28:55.8	& 1.368			\\
    SpARCS1034+5818			& 10:34:49.47,  +58:18:33.1	& 1.388			\\
    $^\dag$SpARCS1051+5818	& 10:51:11.23,  +58:18:02.7	& 1.034		\\
    $^\dag$SpARCS1616+545	& 16:16:41.32,  +55:45:12.4	& 1.157			\\
    $^\dag$SpARCS1634+4021	& 16:34:37.00,  +40:21:49.3	& 1.177			\\
    $^\dag$SpARCS1638+4038	& 16:38:51.64,  +40:38:42.9	& 1.194		\\
    \hline
    \end{tabular}
    \end{table}
    
\section{Data and sample selection} \label{sec:data} 

\subsection{The GOGREEN survey}\label{sec:gogreen_survey} 

The GOGREEN survey includes 21 galaxy systems at $1{<}z{<}1.5$ selected to be representative of progenitors of local clusters and groups, described in detail in \citet{balogh2017} and Balogh et al. (2020; in prep). Groups and clusters with a wide range of halo masses were targeted, and within them galaxies with a wide range of stellar masses were targeted. For the present work we include eleven clusters from the GOGREEN survey which have complete spectroscopy and photometry as of 2020. 

Table\,\ref{tab:sample} lists the clusters with their coordinates and redshifts (\citealt{balogh2017}, \citealt{vanderburg2020}, see Biviano et al. 2020 in prep. for halo masses and velocity dispersions). Three of these systems are from the South Pole Telescope survey \citep[SPT,][]{brodwin2010, foley2011, stalder2013}, nine are from the Spitzer Adaptation of the Red-Sequence Cluster survey \citep[SpARCS,][]{wilson2009, muzzin2009, demarco2010}. Five of the SpARCS clusters were also included in the Gemini cluster Astrophysics Spectroscopic Survey \citep[GCLASS,][]{muzzin2012}. We add to the number of low mass galaxies in the GCLASS sample, and increase the sampling at higher masses particularly at $z{<}1.3$.

GOGREEN provides broadband photometry and Gemini Multi-Object Spectrograph (GMOS) spectroscopy for a selection of galaxies in each system. The survey strategy and magnitude limits ($z^\prime{<}$24.25 and [3.6]${<}$22.5) of the GOGREEN survey enables both a large sampling of bright galaxies and very deep spectroscopy of much fainter galaxies. The full survey is statistically complete for all galaxy types with stellar masses $\log\,M_*/\rm{M}_{\sun} \ga 10.3$ at $1{<}z{<}1.5$ (Balogh et al. 2020, in prep). Including the systematic offset between stellar mass estimates (see Section\,\ref{sec:param_nonparam}) the mass completeness of the sample is $\log\,M_*/\rm{M}_{\sun} \ga 10.5$. Completeness here is characterised as a function of stellar mass and clustercentric distance, where above this limit our sample is representative of an unbiased sampling of the full galaxy population. We note that the lower mass selection used throughout this paper is below this mass completeness threshold, and the conclusions drawn from these galaxies are not necessarily statistically robust.

\subsection{Spectroscopic sample}\label{sec:spectroscopy}

Spectroscopy for the GOGREEN galaxies was taken with the GMOS instruments using the R150 filter and three spectral dither positions (8300\,\AA, 8500\,\AA, and 8700\,\AA). Spectral dithers are done to fill in the gaps between the GMOS CCDs where spectral information is lost. This provides continuous wavelength coverage free of second order contamination over 6400--10200\,\AA. For the redshift range $1{<}z{<}1.5$, this corresponds to about 2500--5250\,\AA\, rest-frame.

The GMOS detector field of view is $5.5^\prime\times5.5^\prime$, which roughly matches the size of our clusters (${\sim}$2.8\,Mpc at $z{=}1.3$). With $1^{\prime\prime}$ slits, the spectra have an observed FWHM resolution of ${\sim}$20\,\AA, ($\mathrm{R} {=} 440\pm60$). We used the nod and shuffle mode to maximise the number of slits per exposure, particularly in the cluster centres, and to perform accurate sky subtraction. Specifics of the spectral data reduction can be found in \citet{balogh2017} and Balogh et al. (2020, in prep). The basic steps follow the Gemini {\sc iraf}\footnote{IRAF is distributed by the National Optical Astronomy Observatories, which are operated by the Association of Universities for Research in Astronomy, Inc., under cooperative agreement with the National Science Foundation.} reduction procedure, with additional corrections for scattered light and telluric absorption. Wavelength calibrations were established using CuAr lamp observations taken concurrent to the GMOS observations, with reference to night sky lines to account for flexure shifts. The lack of features below 6400\,\AA\, results in unreliable calibrations at this end of the spectra. The 1D spectra were extracted and combined. Although a relative sensitivity correction was applied, based on standard star observations, the spectra were not absolute flux calibrated. This requires additional corrections in the fitting procedure discussed in Section\,\ref{sec:fitting}.

In this study we included only galaxies for which we could measure a spectroscopic redshift with confidence (quality flag 3 or 4) -- this includes 970 galaxies. Spectroscopic redshifts were derived using the Manual and Automatic Redshifting Software \citep[MARZ,][]{hinton2016}, as described in Balogh et al. (2020, in prep.).

\subsection{Photometric coverage}\label{sec:photo_data}

GOGREEN has broad photometric coverage for each galaxy system. A full description of the photometry, as well as the calculation of stellar masses and rest-frame colours, is provided in \citet[][]{vanderburg2020}. The photometry includes Gemini GMOS ($z^\prime$), Spitzer IRAC\footnote{Supplemented by archival data primarily from SERVS \citep{mauduit2012}, S-COSMOS \citep{sanders2007}, SPUDS \citep{galametz2013}, and SWIRE \citep{lonsdale2003}}, VLT VIMOS\footnote{Program ID: 097.A-0734.} ($U$, $B$, $V$, $R$, $I$, $z$) and HAWK-I ($Y$, $J$, $K_\mathrm{s}$), Subaru SuprimeCam ($g$, $r$, $i$) and HyperSuprimeCam ($z$, $Y$), Magellan Fourstar ($J_1$, $J$, $K_\mathrm{s}$), CFHT WirCam ($J$, $K_\mathrm{s}$) and MegaCam ($U$), and Blanco DECam ($z$). The one GOGREEN cluster not included in our sample (SpARCS1033) did not have $K$-band data as of the 2019A semester.

\begin{figure}
      \begin{center}
        \includegraphics[width=\linewidth]{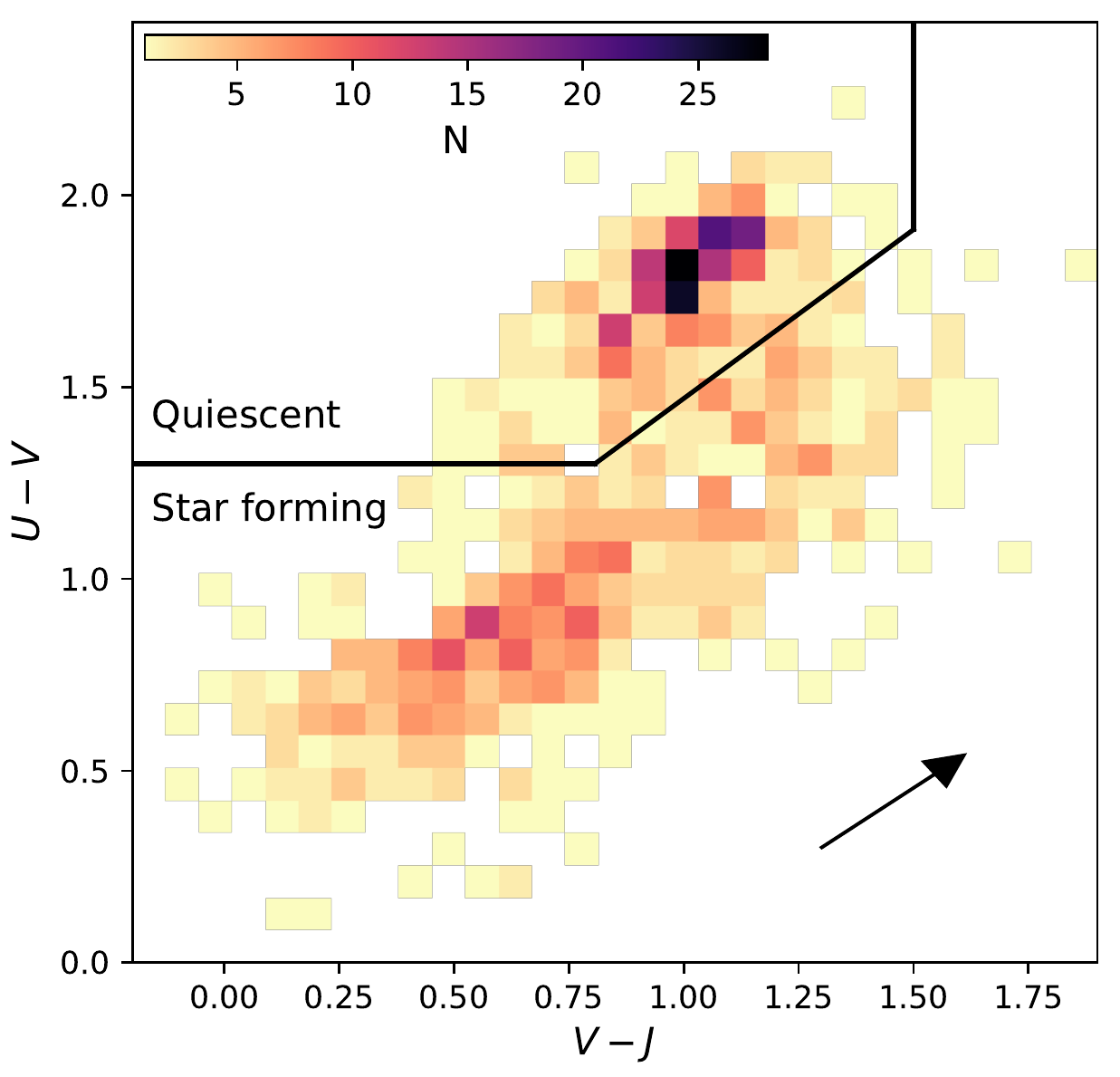}
      \end{center}
      \caption{Rest-frame \textit{UVJ} colours for the GOGREEN spectroscopic sample between $1 {<} z {<} 1.5$, shown as a 2D histogram for both cluster and field galaxies. The black line shows the selection of quiescent galaxies used in this study, as defined by \citet{muzzin2013a}, from star forming galaxies. The arrow indicates the impact of 1\,mag of extinction in the $V$-band, using the \citet{calzetti2000} dust law.  } \label{fig:uvj_diagram}
    \end{figure}

Rest-frame colours were derived from best fit templates to the observed photometry with EAZY \citep{brammer2008}, assuming an exponentially declining SFR, \citet{calzetti2000} dust law, \citet[][BC03]{bruzual2003} stellar library, and solar metallicity. Templates were fixed to the spectroscopically determined redshift, and the redshift-corrected best-fit template was then convolved with $U$,$V$, and $J$ filters (see Figure\,\ref{fig:uvj_diagram}). Galaxies observed in the COSMOS fields have rest-frame colours as provided from the UltraVISTA v4.1 catalogue \citep{muzzin2013b}.

\subsection{Sample selection}\label{sec:sample}

The goal of this paper is to compare the ages and star formation histories of quiescent galaxies in cluster and field environments. Quiescent galaxies were selected based on their position in rest-frame $U$-$V$ and $V$-$J$ colour space, which has been shown to effectively separate star forming and quiescent galaxies \citep{labbe2005, wuyts2007, williams2009, patel2012, whitaker2012b, muzzin2013a} to $z{<}2.5$ \citep{williams2010}. Including the NIR colour allows quiescent galaxies to be more clearly distinguished from dusty star forming galaxies, since dust reddening scatters along the \textit{UVJ}-colour selection vector. We consider alternative selections in Appendix\,\ref{sec:tracers}. 

Figure\,\ref{fig:uvj_diagram} shows the rest-frame $U{-}V$ and $V{-}J$ colours of the GOGREEN spectroscopic sample, with the separation between star forming and quiescent galaxies,
\begin{equation}
 (\mathrm{U}-\mathrm{V}) > 1.3 \,\,\cap\,\, (\mathrm{V}-\mathrm{J})<1.5 \,\,\cap\,\, (\mathrm{U}-\mathrm{V}) > 0.88\,(\mathrm{V}-\mathrm{J}) + 0.59 
 \end{equation}
\noindent as defined in \citet{muzzin2013a} for $1{<}z{<}4$, adapted from \citet{williams2009}. Of the 970 galaxies with spectra and robust redshift measurements, 338 quiescent galaxies were identified. 

Galaxies were identified as cluster members or field based on their spectroscopic redshifts and projected phase space locations. A detailed description will be provided in Biviano et al. (in prep). The field galaxy sample is taken as the galaxies along the line-of-sight of the clusters, not identified as members, and with spectroscopic redshifts within $1{<}z{<}1.5$. We also include galaxies in the five GOGREEN fields within COSMOS \citep{muzzin2013b}. These pointings targeted group-mass systems that are otherwise not considered in this paper. We include galaxies that have a line-of-sight velocity more than 900\,km\,s$^{-1}$ from the targeted group redshift in our field sample. Our sample of quiescent galaxies includes 224 cluster members and 110 field galaxies. 

\begin{figure}
      \begin{center}
        \includegraphics[width=\linewidth]{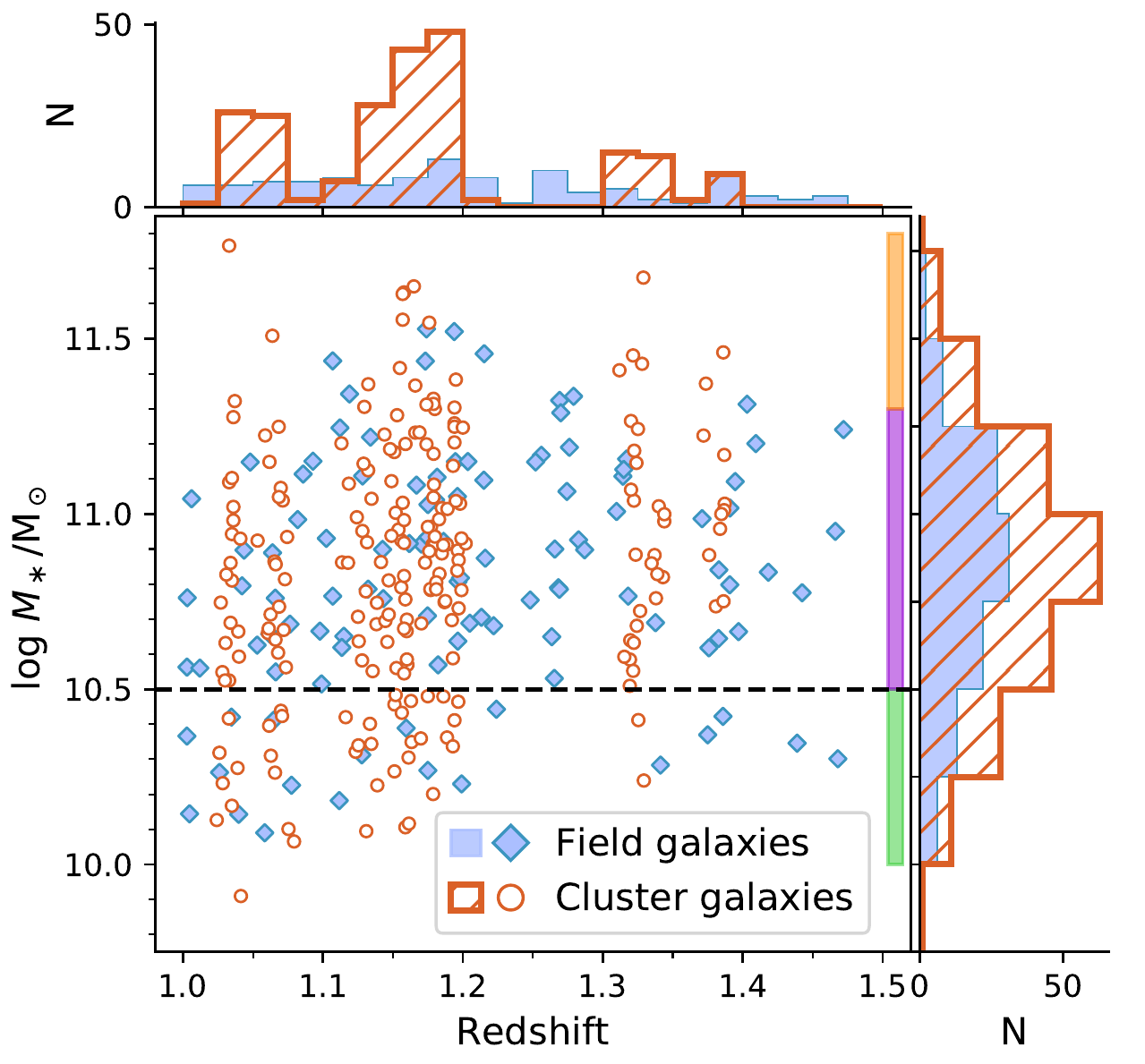}
      \end{center}
      \caption{Stellar masses and redshifts for the \textit{UVJ}-quiescent GOGREEN spectroscopic sample, with corresponding histograms on each axis. Field galaxies are shown as blue diamonds, and cluster member galaxies are red. Coloured blocks indicate the span of mass bins discussed throughout the analysis. A dashed line indicates the mass-completeness of our sample, where the lowest mass bin is below this threshold. Note that the stellar masses shown here are derived with nonparametric SFHs, and are systematically offset from those derived using FAST as reported in other GOGREEN papers, see Section\,\ref{sec:fitting} and Appendix\,\ref{sec:param_nonparam} for details. Our sample ranges between $1{<}z{<}1.5$, and stellar masses between $10^{9.9}\,\mathrm{M}_\odot$ and $10^{11.8}\,\mathrm{M}_\odot$.} \label{fig:z_mass}
    \end{figure}

Figure\,\ref{fig:z_mass} shows the distribution of our quiescent sample as a function of stellar mass and redshift. Cluster galaxies are coloured orange and shown as hatched histograms and field galaxies are blue with solid histograms. Stellar masses were determined from SED fits to the photometry and spectroscopy, discussed further in Section\,\ref{sec:fitting}. The majority of the cluster galaxies are within $1.1{<}z{<}1.2$, while the field galaxies are more evenly spaced in redshift (see Balogh et al. 2020, in prep).

    \begin{table*}
    \centering
    \caption{SFH parameters and priors. Notes: 
    1) Spectroscopic redshift.
    2) Total mass is the sum of total stellar mass and mass lost to outflows. See note 3) for a comment on the prior.
    3) We assume a Milky Way extinction curve \citep{cardelli1989}.
    4) We assume a prior on the stellar mass-metallicity relation (MZR) according to the local trend reported by \citet{gallazzi2005}, where we add the systematic offset between parametric and nonparametric stellar mass estimates (see Appendix\,\ref{sec:param_nonparam}).
    5) Ratio of the SFRs in adjacent bins of the ten-bin nonparametric SFH. The age bins are spaced in lookback time: 0, 30\,Myr, 100\,Myr, 500\,Myr, and 1\,Gyr, five equally spaced bins, and lastly 0.95$\times$ the age of the universe at the observed redshift. For N age bins, there are N-1 free parameters.
    6) The normalization of the spectra is a free parameter to account for systematics in the relative flux calibration. 
    7) The shape of the spectral continuum can be adjusted by a 3$^\mathrm{rd}$ degree Chevyshev polynomial to account for systematics in the relative flux calibration. 
    8) The uncertainty on the spectra can be increased by a given factor, with a likelihood penalty for factors giving reduced $\chi^2{<}1$.
    9) An outlier pixel model can increase the errors for individual pixels by a factor of 50, to accommodate for poor matches between the data and spectral templates.
    }
    \label{tab:params}
    \begin{tabular}{llll}
    \hline
    Note & Parameter	& Description & Prior \\
    \hline
    1 & zred						    & Redshift					& Uniform: $z_\mathrm{spec}\pm0.01$		\\
    2 & $\log\left(M/\mathrm{M}_{\sun}\right)$	& Total mass formed			& MZR: Clipped normal, min = 8, max = 15		\\
    3 & $\hat{\tau}_{\lambda,2}$		& Diffuse dust optical depth 	& Uniform: min = 0, max = 4		\\
    4 & $\log\left(Z/\mathrm{Z}_{\sun}\right)$	& Stellar metallicity			& MZR: Clipped normal, min = -2, max = 0.19		\\
    5 & $\log\left(\frac{\mathrm{SFR}(t)}{\mathrm{SFR}(t+\Delta t)}\right)$
    					             	& Ratio of the SFR ratios in adjacent age bins & Student-t: $\mu$ = 0, $\sigma$ = 0.3, 2 DOF		\\
    6 & spec\_norm				        & Normalization of the spectra 	& Uniform: min=0, max=100		\\
    7 & p$_1$,p$_2$,p$_3$               & Continuum shape correction polynomial coefficients 	& Uniform: min=-0.1/(n+1), max=0.1/(n+1)		\\
    8 & spec\_jitter				    & Spectra white noise model  	& Uniform: min = 1, max = 3		\\
    9 & f$_\mathrm{outlier,\,spec}$	    & Spectra outlier fraction 	        & Uniform: min = 10$^{-5}$, max = 0.5		\\
    \hline
    \end{tabular}
    \end{table*}
    
    \begin{figure*}
      \begin{center}
        \includegraphics[width=0.9\linewidth]{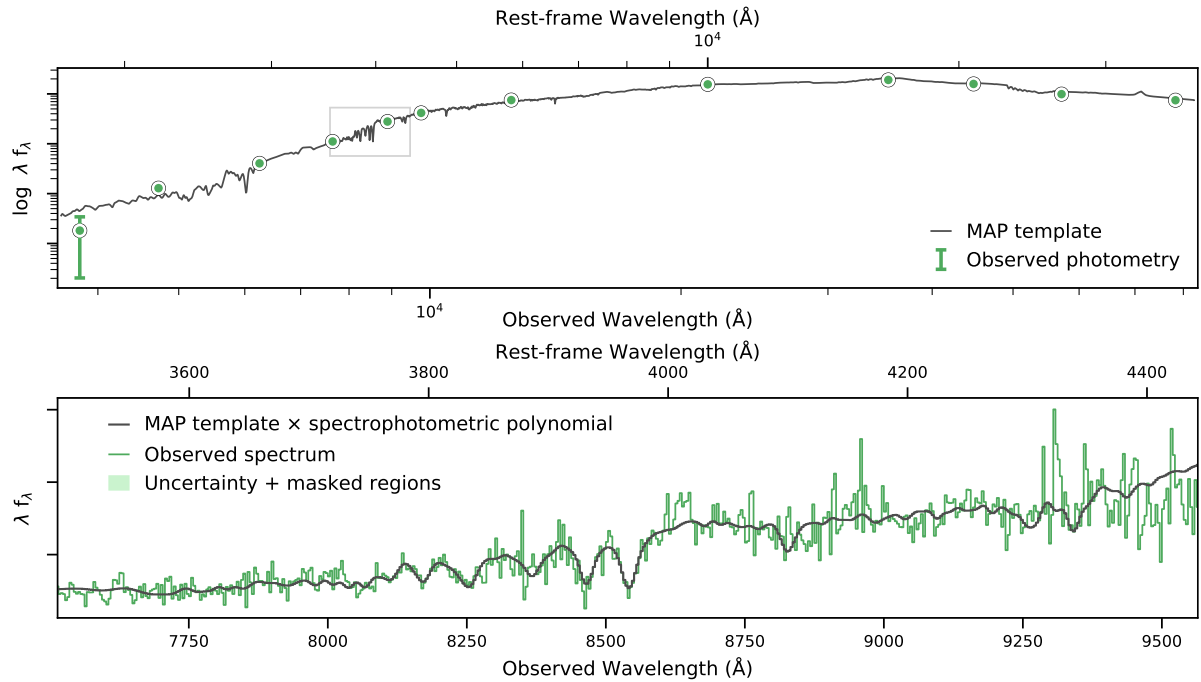}\\
      \end{center}
      \caption{ Example of typical photometric (green circles, top image) and spectroscopic (green line, bottom image) observations shown with the corresponding maximum a posteriori (MAP) template (black line), as a function of observed wavelength. The grey box indicates the wavelength region covered by the spectra relative to the photometry. The MAP template relative to the spectrum is shown with a polynomial `correction' to account for systematics in the relative flux calibration. Green shaded regions indicate the uncertainty and masked regions of the spectrum for the SFH fitting (e.g. the \oii line at 3727\,{\AA} rest-frame). The posteriors for select parameters associated with this fit are shown in Figure\,\ref{fig:fit_params}. } \label{fig:fit}
    \end{figure*}    
    
\section{Fitting star formation histories of quiescent galaxies} \label{sec:fitting} 

SFHs of the quiescent galaxies are constrained by fitting photometric and spectroscopic data with spectral energy templates using the \prospector inference code\footnote{https://github.com/bd-j/prospector} \citep[][v0.3.0]{leja2017, johnson2019}. The physical models are generated from the flexible stellar population synthesis code {\sc FSPS} \citep{conroy2009} with {\sc MIST} stellar evolutionary tracks and isochrones \citep[][based on the {\sc MESA} stellar evolution code \citealt{paxton2011,paxton2013,paxton2015,paxton2018}]{choi2016, dotter2016}, and {\sc MILES}\footnote{http://miles.iac.es/} spectral templates \citep{vazdekis2015}. Biases related to metallicities are discussed further in Appendix\,\ref{sec:mzr}, where we conclude that any such systematics have a negligible impact on our results. 

We assume a nonparametric\footnote{\textit{Nonparametric} here means that the SFH has no specified functional form.} form for the SFHs with a continuity prior \citep[described in][]{leja2019a} and Milky Way extinction curve \citep{cardelli1989}. We mask the only prominent emission line region within our spectral range (\oii) rather than include a nebular line emission model. Table\,\ref{tab:params} lists the free parameters in the fitting procedure: redshift, total mass formed, dust optical depth, stellar metallicity, relative\footnote{Relative with respect to adjacent bins. For N age bins, there are therefore N-1 free parameters. See Table\,\ref{tab:params}.} SFR ratios in ten age bins, spectral normalization, spectral polynomial coefficients, spectral white noise, and spectral outlier fraction. The priors for each parameter are also provided in this table. The age bins are spaced so that the first four bins correspond to 30\,Myr, 100\,Myr, 500\,Myr, and 1\,Gyr in units of lookback time, and the final bin covers the first 5\,per cent of the age of the universe. The remaining age bins are spaced equally in time.\footnote{Convergence tests with some representative galaxies demonstrated that ten bins provide sufficient time resolution, while limiting the number of free parameters in the fitting procedure and the corresponding computational time.} Note that galaxies observed at different redshifts will have different age binning in cosmic time (ie, time since the Big Bang).

Three of the free parameters help to identify systematics in the spectra. The white-noise inflation (spec\_jitter) effectively increases the uncertainties on all spectral points by a multiplicative factor. This is counter-balanced by the standard likelihood penalty term for larger uncertainties. This down-weighting of the spectra is rarely relevant unless the data has high SNR and is more accurate than the models. We also include an outlier pixel model ($\mathrm{f}_\mathrm{outlier,\,spec}$) which modifies the likelihood to be more permissive of large deviations from the model. Such large deviations can come from poor matches to the stellar models (due to differences in, for example, $\alpha$-enhancement) and increases their errors by a factor of 50. The outlier fraction is less than 3\,per cent for the majority (95\,per cent) of our fits. 

In fitting the spectroscopy and photometry together, we need to account for uncertainties in the spectral response calibration, and for the overall flux calibration due to slit losses. Several authors have demonstrated the challenge of simultaneously fitting spectral and photometric data, especially when the spectral continuum is not well characterised \citep{panter2007, vanderwel2016, belli2019, carnall2019b, johnson2019}. As described in Section\,\ref{sec:spectroscopy}, the spectra were not absolute flux calibrated. The flux calibration is uncertain due to slit losses, the lack of atmospheric dispersion correction, and uncertainties in the telluric absorption corrections. To accommodate for these effects, the spectral normalization (spec\_norm) and a spectrophotometric calibration polynomial are calculated from the ratio of the observed and model spectrum, and applied to the template spectrum prior to assessing the goodness of fit. We use a third order Chebyshev polynomial since a higher order polynomial could wash out real spectroscopic features.

The spectral fit was restricted to the wavelength range 3525-4400\,\AA\, rest-frame, covering the majority of useful spectral features (e.g. CaH+K, \dbreak, H$\delta$, G) while minimizing sensitivity to the lowest and highest wavelength ranges where flux calibration is most uncertain due to rapidly changing sensitivity. The lower bound is set by the different resolution of the {\sc MILES} spectral templates at redder wavelengths. Beyond 4400\,\AA, some of the spectra suffer systematic effects due to insufficiently well corrected telluric absorption. Due to the limitations of the template spectra, the metallicity was restricted to $-2<\log\,Z/\mathrm{Z}_{\sun}<0.19$ and the abundance patterns were fixed to solar. Prior to fitting the spectroscopy, the template spectra are smoothed to match the resolution of the observed spectrum. Lastly, we assumed a minimum photometric error of 5\,per cent as a conservative estimate of the calibration uncertainty in the photometry.

\prospector uses {\sc emcee} \citep{foreman-mackey2013} to create an ensemble of walkers which sample the parameter space following an affine invariant algorithm for a given number of steps. We used 64 walkers, iterative `burn-in' in steps of 16, 34, 68, and 124, and a minimum of 1024 iterations thereafter. Each fit was visually confirmed as being converged (i.e., the traces were stable), or the sampling was restarted from the previous maximum probability solution. We take only the last 500 iterations when building the posteriors. The SFHs were sometimes multimodal, particularly where the SNR was poor, which motivated us to use a weighted combination of a differential moves (80\,per cent) and snooker differential moves (20\,per cent) in the MCMC sampling\footnote{As described in \\https://emcee.readthedocs.io/en/stable/user/moves/\#emcee.moves.DEMove}. 

\begin{figure}
      \begin{center}
        \includegraphics[width=\linewidth]{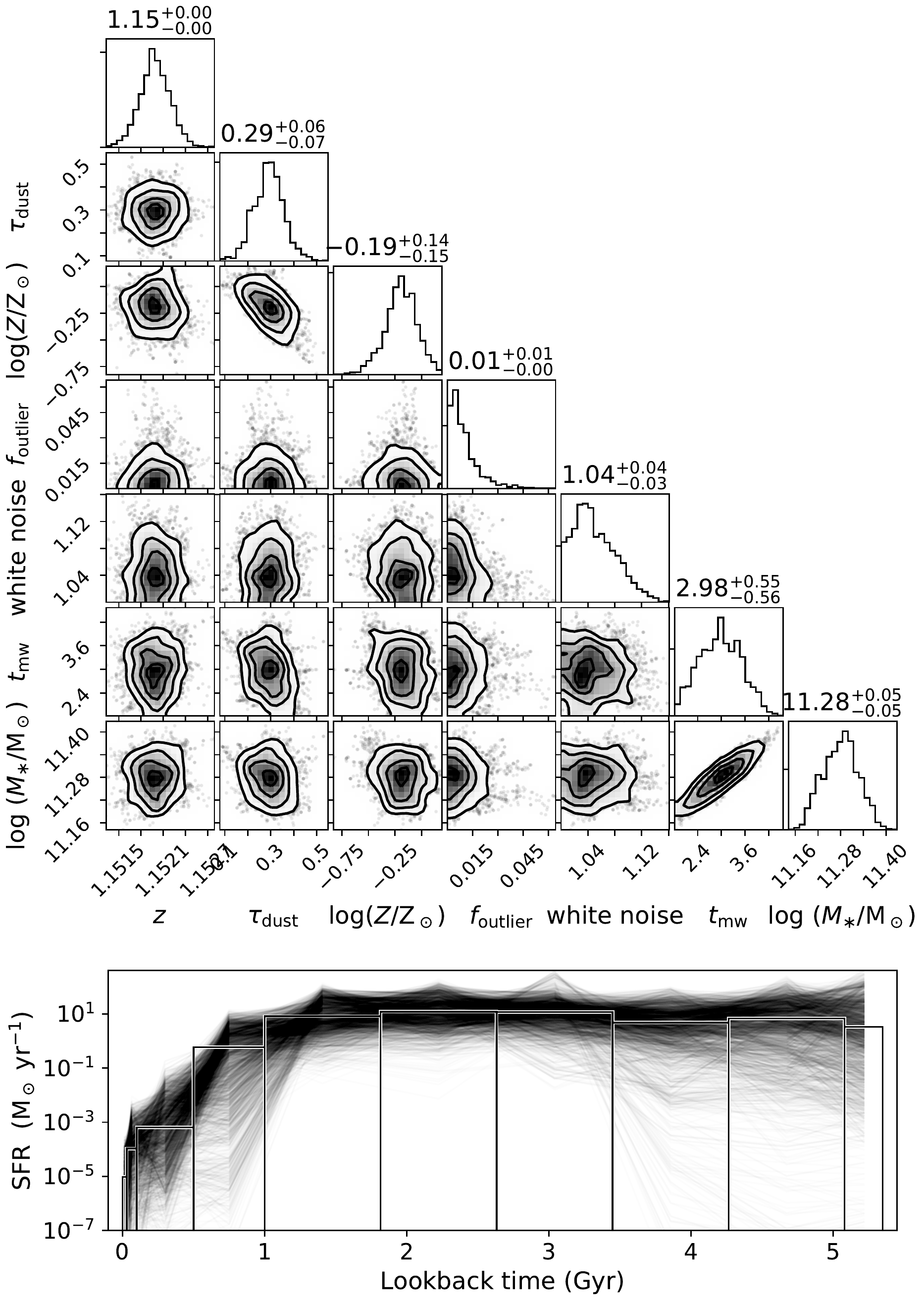}
      \end{center}
      \caption{ Selected posteriors for the fitting result shown in Figure\,\ref{fig:fit}, which is a typical galaxy from our sample of GOGREEN quiescent galaxies. Top: corner plot showing a selection of posterior distributions for SFH parameters: redshift, metallicity, dust opacity, outlier fraction, and white noise, as well as two derived parameters: mass-weighted age and stellar mass (see Section\,\ref{sec:fitting}). Posteriors are shown smoothed with a 1\,$\sigma$ Gaussian, and the 50\thh percentiles are indicated on the top of each histogram with 68 per cent credible regions. Bottom: The posteriors for the SFRs are shown as a function of lookback time, where age bins are drawn with heights equal to the median in each bin.  } \label{fig:fit_params}
    \end{figure}

An example of the output of this fitting procedure is shown in Figures\,\ref{fig:fit} and \ref{fig:fit_params}. In Figure\,\ref{fig:fit}, the observed photometry (top, green circles) and spectrum (bottom, green line) are shown relative to the template with the highest combined likelihood and prior (maximum a posteriori, MAP; black line). The shaded green regions about the spectrum indicate the uncertainty, while masked regions in the fit are shown as faint green lines. The spectrum is shown relative to the MAP after the spectrophotometric polynomial was applied. A selection of the SFH parameters with their posteriors are shown in Figure\,\ref{fig:fit_params} as a corner plot, and the range of SFRs as determined from the relative SFRs. The 50\thh percentile value of each parameter is listed above its posterior, with uncertainties from the 68 per cent confidence regions.

Throughout this work we report the uncertainties as 68 per cent confidence regions (which corresponds to the 16\thh to 84\thh percent range) as the majority of the distributions are non-symmetric. The lower (16\thh--50\thh) and upper (50\thh--84\thh) reported are equivalent to $\pm1\,\sigma$ for a Normal distribution.

From the SFH posteriors we calculate\footnote{{\sc FSPS} calculates \mwa\, when \texttt{compute\_light\_ages}{=}True.} the mass-weighted stellar age (\mwa, discussed in Section\,\ref{sec:sfhs}) and stellar mass. The latter is determined from the posterior of the total stellar mass formed and the corresponding fraction of surviving stellar mass for each sampling. We confirm that the stellar masses derived using nonparametric modelling are systematically offset from than those derived with parametric models such as exponentially declining SFR models (e.g. using {\sc FAST}; \citealt{kriek2009}). This comparison is discussed in Appendix\,\ref{sec:param_nonparam}. We note that the stellar masses reported in other GOGREEN papers \citep[e.g.][]{balogh2017,chan2019,old2020,vanderburg2020} are derived using {\sc FAST}, and therefore differ from the stellar masses in this paper by $+0.2$\,dex. Since the focus of this paper is a differential comparison between galaxies in cluster and field environments, our results are less sensitive to the systematic effects related to model choices.

Only two of the fits clearly failed to reproduce the observations. For both the spectral continuum is dominated by telluric absorption that was not sufficiently corrected. The final sample includes 331 galaxies, 109 of which are field galaxies, and 222 are cluster galaxies.


\section{Results} \label{sec:sfhs_mwas}

In this Section we present the results of the nonparametric SFH fitting applied to the sample of 331 quiescent GOGREEN galaxies. We explore differences related to stellar mass and density of local environment through comparing the SFHs and mass-weighted ages. We then test our result by refining our selection of quiescent galaxies. In Appendix\,\ref{sec:coadds} we compare features in co-added spectra to the results of fitting the individual galaxies. 

%
\subsection{The dependence of star formation histories on mass and environment} \label{sec:sfhs}

\begin{figure*}
      \begin{center}
        \includegraphics[width=\linewidth]{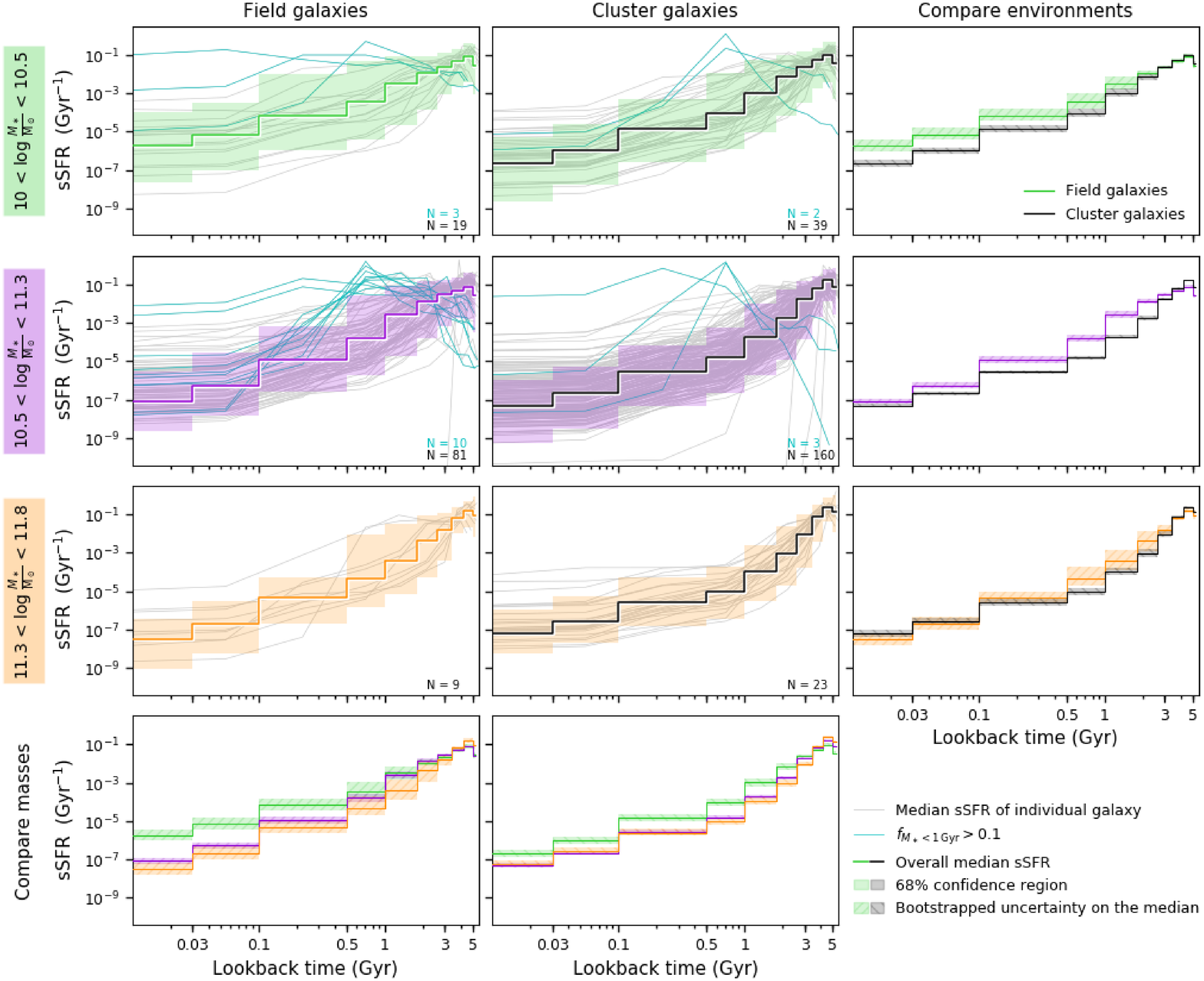}
      \end{center}
      \caption{Specific SFR (SFR($t$)/$M_{\ast, z{=}z_\mathrm{obs}}$) as a function of lookback time for field and cluster galaxies, in three mass bins. Individual sSFRs are shown as grey lines, where galaxies in which more than 10\,per cent of their stellar mass has formed within the last 1\,Gyr (see Section\,\ref{sec:rejuv}) are coloured cyan. The number of galaxies in each mass and environment selection is labelled at the bottom-right of the subplot. The overall median sSFR in each subsample is shown as a bold line, and is also shown in the right-hand column to compare between environments, and in the bottom row to compare between mass selections. The shaded region indicates the 68 per cent confidence region of the combined sSFRs, while the hatched shaded regions show the bootstrapped uncertainty on the overall median. 
      Two trends are apparent from the median SFHs: higher mass galaxies form their mass earlier (i.e., mass-accelerated evolution), and cluster galaxies form their mass earlier.  } \label{fig:sfhs}
    \end{figure*}

Figure\,\ref{fig:sfhs} shows the median sSFRs (star formation rates divided by the final stellar mass) for individual galaxies as a function of lookback time. Subplots separate galaxies according to environment and stellar mass. The overall median sSFRs for each selection of galaxies are shown as a bold lines, and the 68 per cent confidence ranges are shown as shaded regions. The right-hand column compares sSFRs for galaxies between the two environments, at fixed mass. The bottom row compares SFHs for galaxies between mass selections, at fixed environment. The hatched shaded region in the right-hand column and bottom row shows the bootstrapped uncertainties on the medians.

The majority of galaxies follow a steady decline in sSFR, consistent with passive evolution. A few galaxies have more shallow declines or more recent star formation. We indicate galaxies which have more than 10\,per cent of their stellar mass formed within the last 1\,Gyr, with cyan lines in Figure\,\ref{fig:sfhs} (and list the number in each panel), and discuss them in Section\,\ref{sec:rejuv}. This population is not unexpected, as the \textit{UVJ} colour selection can include younger galaxies, or those in transition. Four galaxies have extremely rapid declines in SFR, with negligible star formation within the last 1\,Gyr. 

Comparing galaxies at fixed environment (bottom row of Figure\,\ref{fig:sfhs}), we find that more massive galaxies have overall earlier star formation activity, and form their stars over shorter timescales. Lower mass galaxies, on average, have more extended SFHs. This trend is consistent with the `mass-dependent evolution` scenario \citep[e.g.][]{nelan2005, thomas2005}, sometimes called `archaeological downsizing` \citep{neistein2006}. Interestingly, the galaxies in our moderate mass bin more closely resemble their higher mass counterparts, but have slightly longer star forming timescales. 

Comparing galaxies at fixed mass (right-hand column of Figure\,\ref{fig:sfhs}), galaxies in clusters have overall earlier star formation activity in the sense that the sSFRs decline more quickly. Below masses of $10^{11.3}$\,M$_{\sun}$, the sSFRs of field galaxies are higher within the last ${\sim}2$\,Gyr. In general, field galaxies in our lower mass sample have the flattest (most extended) SFHs. 

    \begin{figure*}
      \begin{center}
        \includegraphics[width=\linewidth]{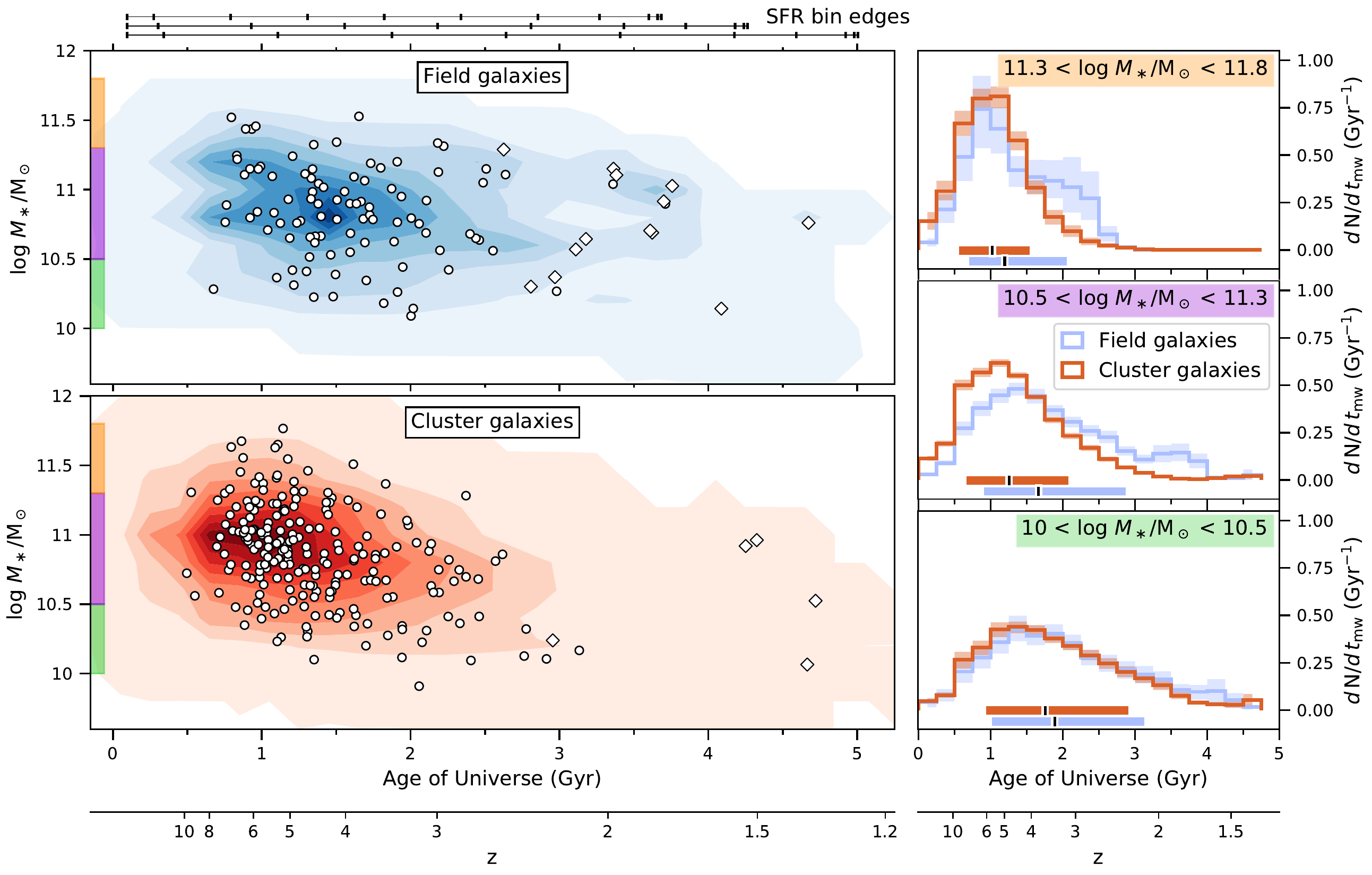}  
      \end{center}
      \caption{ 
      Comparison of stellar masses and mass-weighted ages between field (blue) and cluster (red) galaxies.
        Left: Combined posteriors of stellar masses and \mwa (in units of cosmic time), shown as contours. The medians of the individual posteriors are marked with white circles/diamonds. Diamonds indicate young galaxies, which have formed more than 10\,per cent of their stellar mass within the last 1\,Gyr (discussed in Section\,\ref{sec:rejuv}). Horizontal bars at the top of the figure indicate the edges of the age bins for $z{=}1.5$ (top), $z{=}1.25$ (middle), and $z{=}1$ (bottom). The bins were defined in units of lookback time, and therefore do not match up for galaxies observed at different redshifts. 
      Right: Combined \mwa posteriors for field and cluster galaxies, shown in three mass bins. The medians (black mark) and 68 per cent credible regions (coloured bar) of each distribution are marked at the bottom of each subplot. The shaded regions show the bootstrapped uncertainty of each histogram. 
      Although there are field galaxies that formed as early as the oldest cluster galaxies, and cluster galaxies that formed as late as the youngest field galaxies, \textit{on average} field galaxies have more extended SFHs to reach the same final stellar mass.   } \label{fig:diff_mwa}
    \end{figure*}
    
\citet{rettura2011} estimated the SFHs of massive ETGs in both clusters and the field at $z{\sim}1.3$, based on photometric observations, and concluded that while the formation epochs are similar between environments, field galaxies take longer to assemble than cluster galaxies. Specifically, they found that after 1\,Gyr of star formation, 75\,per cent of cluster galaxies had assembled at least 80\,per cent of their final stellar mass, compared with only 35\,per cent of field ETGs. We find a smaller difference, but also phrase it slightly differently given that we do not use parametric SFHs and do not constrain the onset of star formation: by $z{\sim}5.4$ (${\sim}$1\,Gyr since the Big Bang), 75\,per cent of our higher mass cluster galaxies had formed at least 80\,per cent of their final stellar mass, compared to only 50\,per cent of field galaxies. Although we find a stronger difference between the SFHs of field and cluster galaxies at moderate stellar masses, the difference is smaller than found by \citet{rettura2011} (75\,per cent \textit{vs} 46\,per cent), but consistent within the uncertainties of the SFHs given the systematic differences in modelling \citep[discussed in ][in the context of the \citealt{rettura2011} measurements]{raichoor2011}. We discuss the SFHs in the context of mass-dependent evolution and the literature further in Section\,\ref{sec:mass_discuss}.

\subsection{The dependence of age on mass and environment} \label{sec:mwas}

From the SFHs we calculate the mass-weighted age, or mean stellar age, which broadly describes the average formation time of stars in a given galaxy in units of lookback time,
\begin{equation}\label{eqn:mwa}
 t_\mathrm{mw} = \frac{\int_{t_\mathrm{obs}}^{0} t\,\,\mathrm{SFR}(t)\,\mathrm{d}t}{\int_{t_\mathrm{obs}}^{0} \,\mathrm{SFR}(t)\,\mathrm{d}t}
\end{equation}
\noindent where $t_\mathrm{obs}$ is the age of the Universe at the time of observation. We also express the ages in units of cosmic time, $t_\mathrm{obs} - t_\mathrm{mw}$ (sometimes called the formation time, $t_\mathrm{ form}$), which is convenient when comparing galaxies observed across a range of redshifts. Trends between \mwa and \textit{UVJ} colour are discussed in Appendix\,\ref{sec:mwa_uvj}.

Figure\,\ref{fig:diff_mwa} shows the distribution of the stellar mass and mass-weighted ages, \mwa, in units of cosmic time. Contours show the combined posteriors of the field (blue) and cluster (red) galaxies, where white points indicate the medians of the individual posteriors. The typical uncertainty for the mass-weighted ages is 0.52\,Gyr, and for the stellar masses 0.043\,dex. Diamonds indicate galaxies that have formed more than 10\,per cent of their stellar mass within the last 1\,Gyr ($f_{M_\ast{<}1\,\mathrm{Gyr}}>0.1$), discussed in Section\,\ref{sec:rejuv}. The right-hand column shows combined age histograms for field and cluster galaxies within three mass ranges. The galaxy sample is bootstrap sampled to determine the variances within the age bins. Medians and 68 per cent credible regions of the combined distributions are indicated with horizontal bars near the bottom axis.

The mass-weighted ages of our sample are distributed primarily between $2{<}z{<}8$, where there is a modest mass dependence in that galaxies in our higher mass selection have mass-weighted ages between $3{<}z{<}10$ while the lower mass galaxies fall within $2{<}z{<}6$. The majority (${>}$50\,per cent) of the higher (lower) mass galaxies have formed at least half of their stellar mass by $z{\sim}5.4$ ($z{\sim}$3.3). The shapes of the mass-weighted age distributions are also broader at lower stellar masses, as we saw from the SFHs shown in Figure\,\ref{fig:sfhs} and discussed in the previous section. Specifically, at $z{\sim}3.3$, the at least 90\,per cent of the higher mass galaxies have formed at least half their stellar mass, compared to only 50\,per cent of the lower mass galaxies. 

For the lower and higher mass galaxies in our sample, the differences between the mass-weighted ages of galaxies between environments at fixed mass are smaller than the differences across our stellar mass range at fixed environment. This is apparent in the histograms of the mass-weighted ages shown in the right-hand column of Figure\,\ref{fig:diff_mwa}: the shapes of the distributions at fixed mass are more similar then between the higher and lower mass galaxies. We do note, however, that the age-distributions for field galaxies are shifted towards younger ages, as well as broader. Interestingly, the distribution of mass-weighted ages for the moderate mass cluster galaxies more closely resemble that of their more massive counterparts, while the field galaxies are more similar to their lower mass counterparts. This is to say that the moderate mass galaxies in clusters are largely older, while the field galaxies are both younger overall and have an extended tail towards younger ages.

Next we attempt to compare the intrinsic distribution of ages between the field and cluster environments, accounting for the uncertainties on individual measurements. Comparing the rms uncertainties of individual posteriors to that of combined posteriors of similar mass (i.e., $(\sigma_{i}^2 - \sigma_{\mathrm{comb.}}^2)^{-1/2}$, although neither are necessarily Normal), we find that there are significant intrinsic distributions of ages in both the cluster and field sample, with rms's of 0.74\,Gyr and 0.73\,Gyr, respectively. The distributions are consistent between environments, however. 

\begin{figure}
      \begin{center}
        \includegraphics[width=\linewidth]{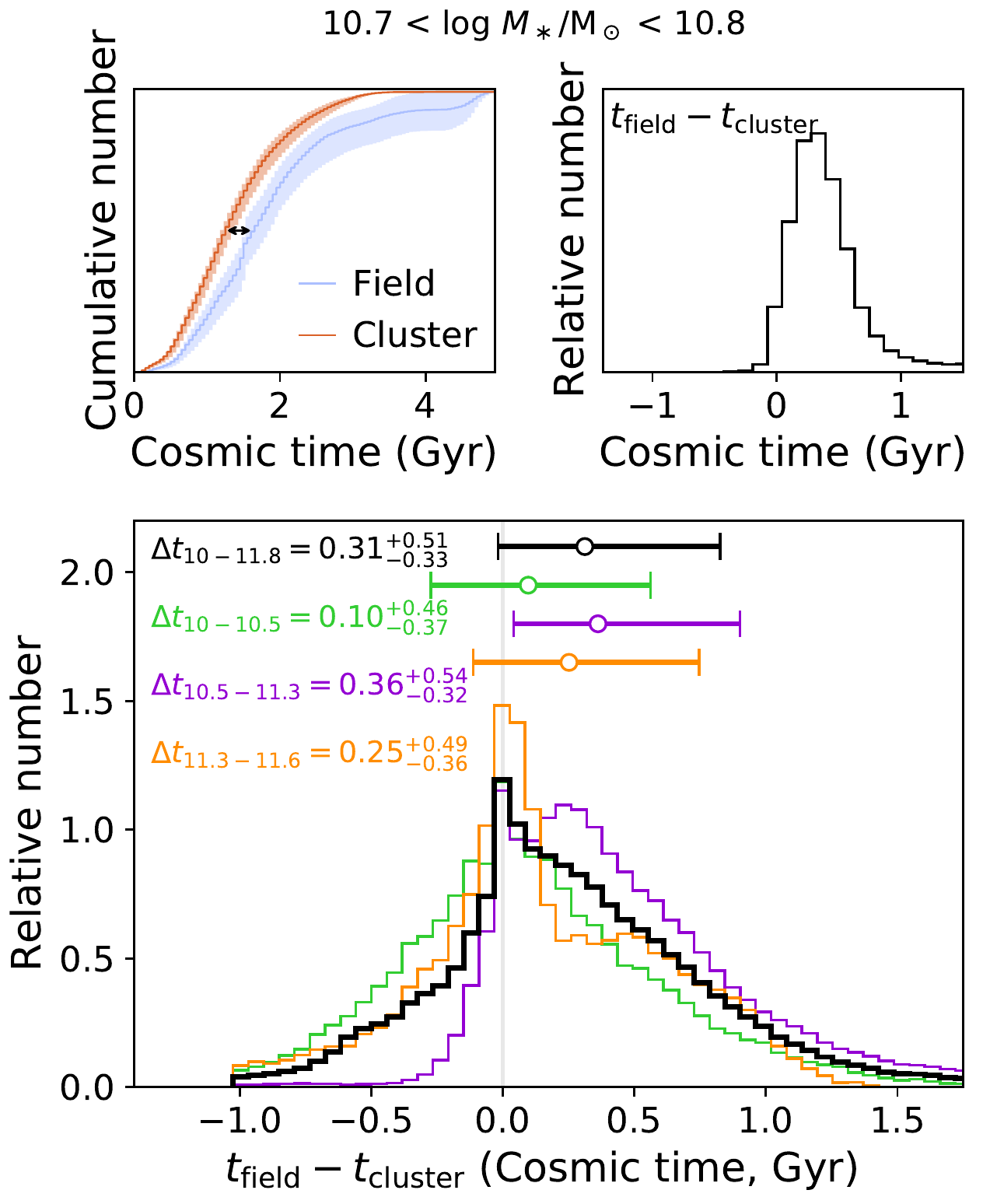}  
      \end{center}
      \caption{ 
      Distributions of offsets between cumulative age distributions of field and cluster galaxies (shown in Figure\,\ref{fig:diff_mwa} as non-cumulative histograms), in units of cosmic time. 
      The top row shows an example of this age comparison for galaxies with stellar masses $10.7{<}\log\,M_\ast/\rm{M}_{\sun}{<}10.8$. The cumulative mass-weighted age distributions for the field (blue) and cluster (red) galaxies is shown on the top left, where the samples have been bootstrapped and the variance is shown as a shaded region. The solid lines show the medians of the bootstrapped distributions. The corresponding offsets in the mass-weighted ages for interpolated bins spanning the cumulative distributions are shown in the top right plot.
      Galaxies are compared at fixed stellar mass (bins of 0.1\,dex) and combined, weighted by the integrated mass within the bins. 
      The combined distributions within the broader mass selections used throughout previous figures are included for reference: $10{<}\log\,M_\ast/\rm{M}_{\sun}{<}10.5$, green; $10.5{<}\log\,M_\ast/\rm{M}_{\sun}{<}11.3$, purple; $11.3{<}\log\,M_\ast/\rm{M}_{\sun}{<}11.8$, orange. The full mass range combined distribution is shown in black. The median age difference for each mass selection is labelled in the figure, which shows the age difference is within $0.31_{^{-0.33}}^{_{+0.51}}$\,Gyr. The median age difference is larger at lower stellar masses, and smaller for the highest mass galaxies. Error bars indicate the 68 per cent confidence range. 
      This quantitative comparison echoes the qualitative comparison shown in Figure\,\ref{fig:diff_mwa} in that the quiescent cluster galaxies are on average older than comparable field galaxies.
      } \label{fig:diff_mwa_cum}
    \end{figure}

In order to better quantify the difference in mass-weighted ages between field and cluster galaxies we compare the combined age distributions in a cumulative sense. This allows us to compare the cosmic time at which the two populations reach a given fraction of their mass-weighted age distribution. Within small (0.1\,dex) mass ranges we select field galaxies and cluster galaxies, calculate their respective combined age distributions, and interpolate the cumulative distributions to the same binning. Within a given mass bin, we include all portions of the posteriors that fall within the limits (i.e., we are not selecting based on the median mass). We then measure the horizontal offset (i.e., in units of time) between the distributions (field - cluster). An example of this procedure is shown in Figure\,\ref{fig:diff_mwa_cum}. The mass-selected comparisons are then combined, weighted by the number of samplings from the respective posteriors, and the overall age offset is determined. We bootstrap our galaxy sample 500 times to capture the true variance.

Figure\,\ref{fig:diff_mwa_cum} shows the cumulative-age-distribution comparisons combined into broader mass selections (coloured histograms), and for the full mass range of our sample (black). The median age difference is shown for each mass selection with error bars corresponding to the 68 per cent confidence region. Across the mass range of our sample, $10{<}\log\,M_\ast/\mathrm{M}_{\sun}{<}11.8$, the median age difference between field and cluster galaxies is $0.31_{^{-0.33}}^{_{+0.51}}$\,Gyr, in the sense that cluster galaxies are on average older than field galaxies. Interestingly, the age difference is slightly smaller for the lower and higher mass galaxies, and slightly larger for our moderate mass galaxies. Note that the lower mass selection is below the mass completeness limit of our sample, and is dominated by galaxies $z{<}1.2$. That is to say, the sample of galaxies below $10^{10.5}$\,M$_{\sun}$ is not a representative sample of the galaxy population, and the result is not as robust. Omitting the lower mass galaxies does not significantly change our result, however: the median age of the cluster galaxies is instead ${0.35}_{^{-0.32}}^{_{+0.51}}$\,Gyr older than that of field galaxies.

\begin{figure*}
      \begin{center}
        \includegraphics[width=\linewidth]{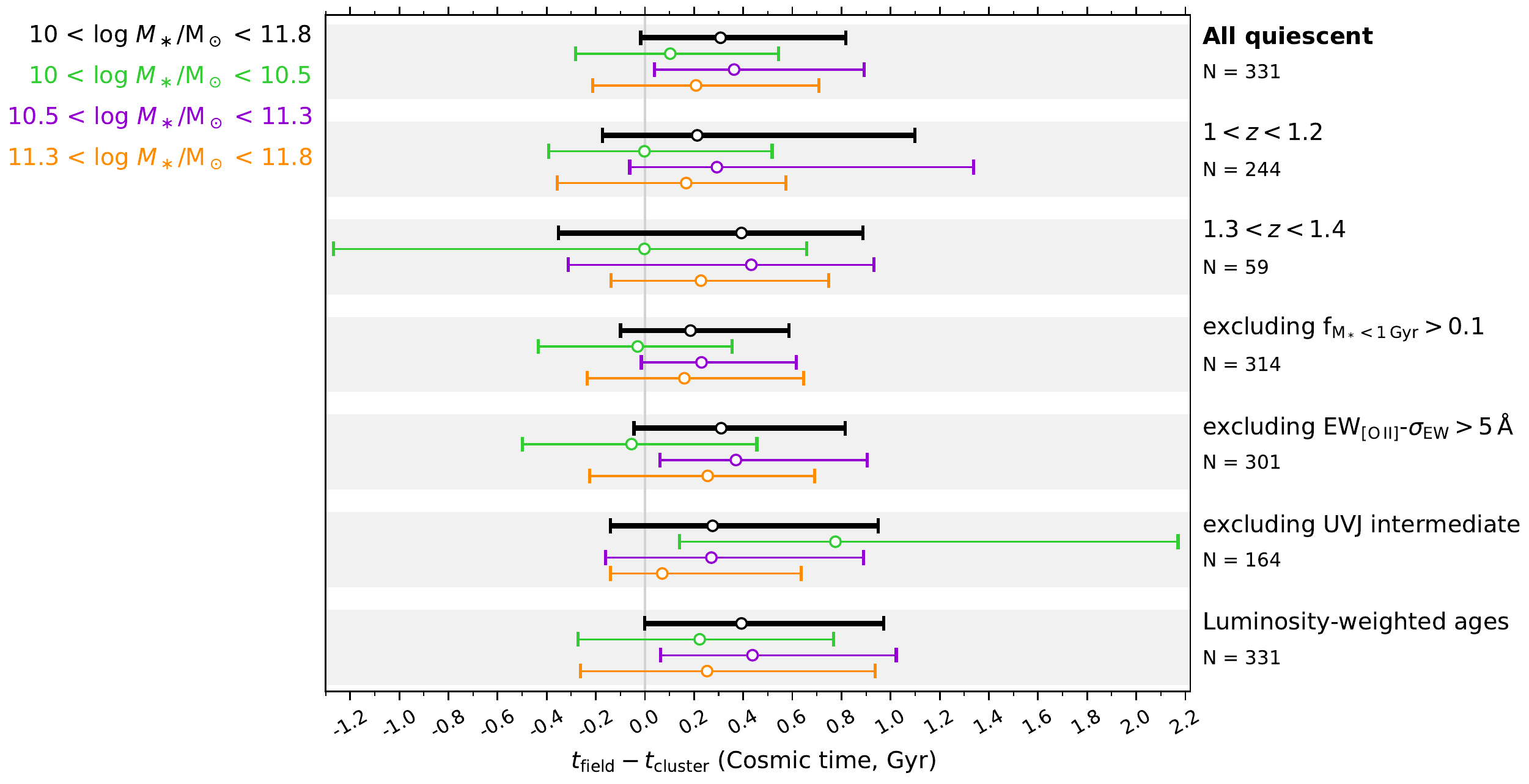}
      \end{center}
      \caption{Differences in cumulative mass-weighted age distributions between field and cluster galaxies for different selections of our sample of quiescent galaxies, as described in the text, in units of cosmic time. Ages are first compared within 0.1\,dex mass selections, and these comparisons are then combined, weighted by the number of posterior samplings in each selection (see Figure\,\ref{fig:diff_mwa_cum} for an example of this procedure). We show the comparisons in mass ranges of $10{<}\log {M}_\ast/\rm{M}_{\sun}{<}11.8$, black (i.e., the full mass range); $10.0{<}\log {M}_\ast/\rm{M}_{\sun}{<}10.5$, green; $10.5{<}\log {M}_\ast/\rm{M}_{\sun}{<}11.3$, purple; $11.3{<}\log {M}_\ast/\rm{M}_{\sun}{<}11.8$, orange. Note that the lowest mass bin is below our completeness limit. The median age difference for each mass selection is marked as a circle with error bars indicating the 68 per cent confidence range.
      The age comparison between luminosity-weighted ages is also shown, discussed in Appendix\,\ref{sec:lwas}, which predicts a slightly larger (by 0.1\,Gyr) age difference than mass-weighted ages for the lower and middle mass ranges.
      The age-comparison result for the full mass-range does not significantly change when excluding high-$z$, low-$z$, or `young' (\frejuv$>0.1$) galaxies, galaxies with \oii emission, or galaxies near the \textit{UVJ}-colour quiescent selection boundary (labelled \textit{UVJ} intermediate). The latter selections would reasonably exclude galaxies transitioning between star-forming and quiescence, or which have complex dust properties obscuring star forming populations. Overall the age difference between field and cluster galaxies is insensitive to recent star formation, unlike the low-redshift galaxies studied in \citet{thomas2010}.
      } \label{fig:diff_mwa_cum_all}
    \end{figure*}

We also consider the age comparison between galaxies at the lower end of our redshift selection, $1{<}z{<}1.2$, and find that the age difference is slightly smaller: ${0.21}_{^{-0.39}}^{_{+0.88}}$\,Gyr, but still consistent with out main result. On the other hand, galaxies at the higher end of our redshift selection, $1.3{<}z{<}1.4$, have a slightly larger age difference: ${0.39}_{^{-0.74}}^{_{+0.49}}$\,Gyr, but age consistent within the uncertainties. Figure\,\ref{fig:diff_mwa_cum_all} shows the mass-weighted age comparison for each mass and redshift selection of quiescent galaxies. We further test our result by identifying galaxies which are not necessarily passively evolving, discussed in the next Section.

\subsection{Recent star formation} \label{sec:rejuv} 

Our quiescent sample is selected based on \textit{UVJ}-colours. We have seen in Section\,\ref{sec:sfhs} that our \textit{UVJ} colour selection does not yield exclusively old galaxies with exponentially declining SFRs (cyan coloured SFHs in Figure\,\ref{fig:sfhs}, marked with diamonds in Figure\,\ref{fig:diff_mwa}). While four galaxies have fairly flat SFHs, most of these galaxies are `late-bloomers' with peaks in their sSFRs within the last 1\,Gyr \citep[similar to][]{dressler2018}. These galaxies are not necessarily `frosted' in the sense of \citealt{trager2000b}, or `rejuvenated' in the sense of \citet{thomas2010} or \citet{chauke2018}, given that these recent peaks account for a substantial fraction of the stellar mass. 

Given the breadth of the \textit{UVJ}-colour selection of these quiescent galaxies, it is conceivable that these galaxies are still in transition (the \textit{UVJ}-colour selection is discussed further below). In addition, some of our \textit{UVJ}-selected galaxies show significant \oii emission, which may be indicative of ongoing star formation. Both `young' and \oii-emitting galaxies are more frequent in our field sample \citep[similar to studies at lower redshifts, e.g.][]{treu1999, treu2001b, vandokkum2001a, vanderwel2004, bernardi2006}. We consider here if either population is the cause of the average mass-weighted age difference we find between field and cluster galaxies. 

We identify galaxies which are not intrinsically old by the fraction of stellar mass formed within the last 1\,Gyr,
\begin{equation}\label{eqn:rejuv}
f_{M_\ast{<}\,1\,\mathrm{Gyr}} = \frac{ \int_{{t}_\mathrm{obs}}^{{t}_\mathrm{obs}-1\,\mathrm{Gyr}} \mathrm{SFR}(t)\,\mathrm{d}t}{ \int^{0}_{{t}_\mathrm{obs}} \mathrm{SFR}(t)\,\mathrm{d}t}
\end{equation}
where we use $f_{M_\ast{<}1\,\mathrm{Gyr}}>0.1$ as the criteria (i.e., irrespective of \oii emission). This selects 18 (5\,per cent) galaxies in our total sample, based on the median \frejuv values. We note that four of these galaxies have $f_{M_\ast{<}1\,\mathrm{Gyr}}>0.85$ an no \oii emission, three of which are in clusters (one of which has particularly red \textit{UVJ} colours). The spectra of these four `young' galaxies are suggestive of recent star formation in the sense that they have relatively strong Balmer absorption lines, while two are particularly low SNR that their SFHs are not well constrained. 

\begin{figure}
      \begin{center}
        \includegraphics[width=0.96\linewidth]{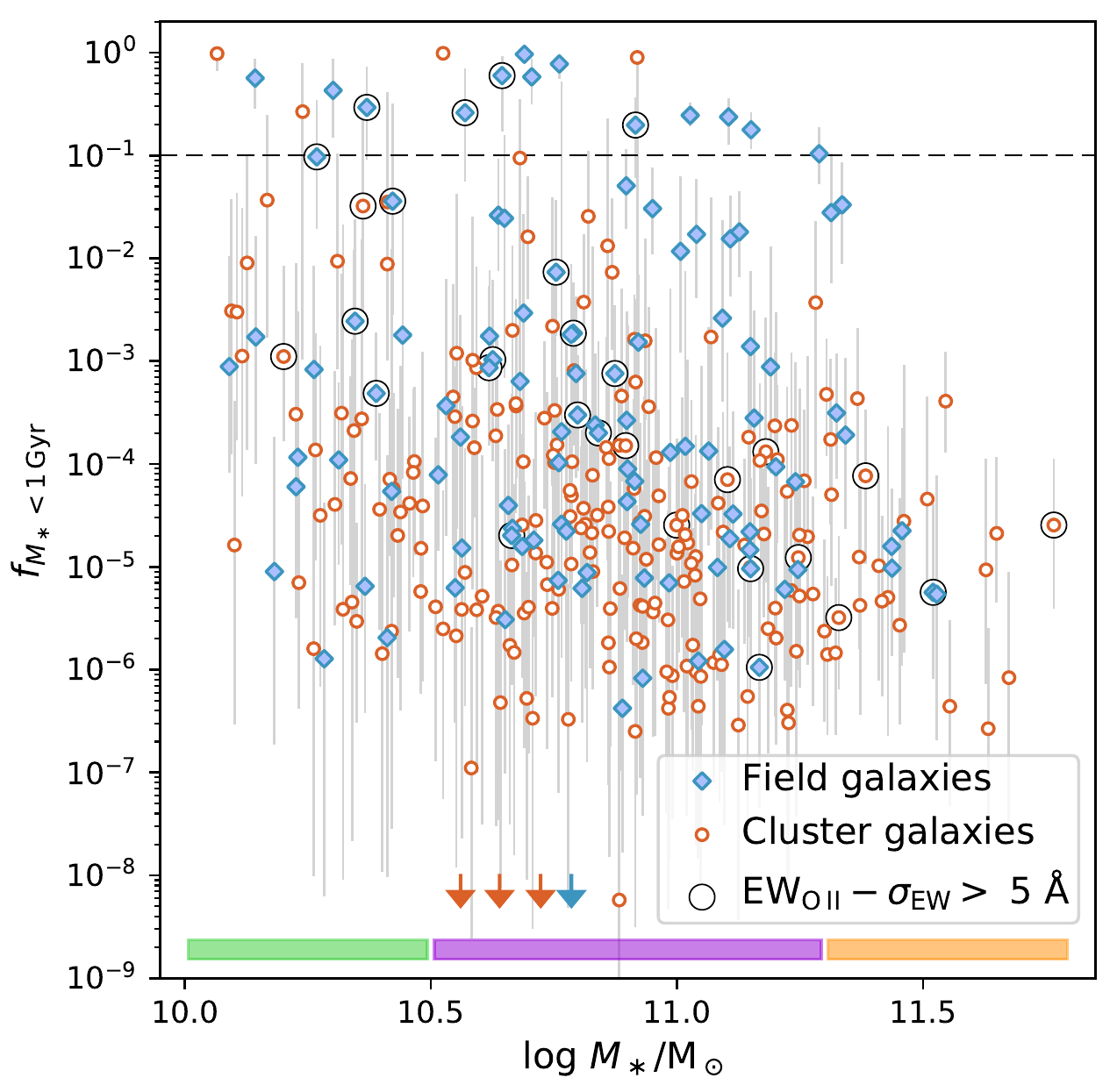}
      \end{center}
      \caption{The fraction of stellar mass formed within the last 1\,Gyr as a function of total stellar mass, for our \textit{UVJ}-selected sample of quiescent galaxies. Galaxies with EW(\oii){-}$\sigma_\mathrm{EW}>5$\,\AA\ are circled. Field galaxies are shown as blue diamonds, and cluster galaxies as orange circles. Arrows indicate points below the shown scale. 
      `Young' \frejuv$>0.1$ galaxies are more common among field galaxies, and at stellar masses ${<}10^{11.3}$\,M$_{\sun}$. There is no correlation between the presence of \oii emission and \frejuv$>0.1$, however. The robustness of age-comparison is tested by excluding this population of `young' galaxies, see Figure\,\ref{fig:diff_mwa_cum_all}.
      } \label{fig:rejuv}
    \end{figure}

Figure\,\ref{fig:rejuv} shows \frejuv as a function of stellar mass, separating cluster and field galaxies in colour, and galaxies which also have \oii emission are circled. Coloured boxes indicate the ranges of the three mass bins used throughout the paper. The number of galaxies which are `young' by this definition are labelled in Figure\,\ref{fig:sfhs} for each mass and environment subsample; 13 of these galaxies are in the field population, accounting for 16\,per cent (12\,per cent) of the lower (moderate) mass sample. Comparatively, the four `young' galaxies in our cluster sample account for 5\,per cent (2\,per cent) of the lower (moderate) mass samples. Although the relative fractions of these galaxies are higher in the field population, the overall fractions are still quite low. Indeed, the overall median SFHs shown in Figure\,\ref{fig:sfhs} are unchanged within the bootstrapped uncertainty when the `young' galaxies are excluded. 

The fraction of field galaxies in our sample with significant \oii emission (\,EW(\oii){-}$\sigma_\mathrm{EW}>5$\,\AA, cf Appendix\,\ref{sec:tracers}), 17\,per cent (19/109), is similarly larger than the 5\,per cent (11/222) of cluster galaxies. Moreover, as apparent in the co-added spectra discussed in Appendix\,\ref{sec:coadds}, the strength of \oii emission is higher for field galaxies. Similar to our results, \citet{rudnick2017} find that for a selection of intrinsically old galaxies the prevalence of \oii emission was higher for field galaxies, which they attributed to clusters (and groups) being sites where gas accretion onto massive galaxies (above $10^{10.4}$\,M$_{\sun}$) was shut off. Indeed, \oii emission can result from processes other than star formation \citep[AGN and/or LINER, e.g.][]{heckman1980, yan2006, singh2013}, and has complex dependence on ISM properties \citep{hogg1998}. Interestingly, the sites of \oii emission in our sample have different mass ranges between environments: for field galaxies the \oii emitting galaxies have masses ${<}10^{10.9}$\,M$_{\sun}$ for all but three galaxies, while the cluster galaxies have masses ${>}10^{10.9}$\,M$_{\sun}$ for all but three galaxies. We also note that only four of the \frejuv$>0.1$ galaxies also have \oii emission. 
    
\begin{figure}
      \begin{center}
        \includegraphics[width=0.96\linewidth]{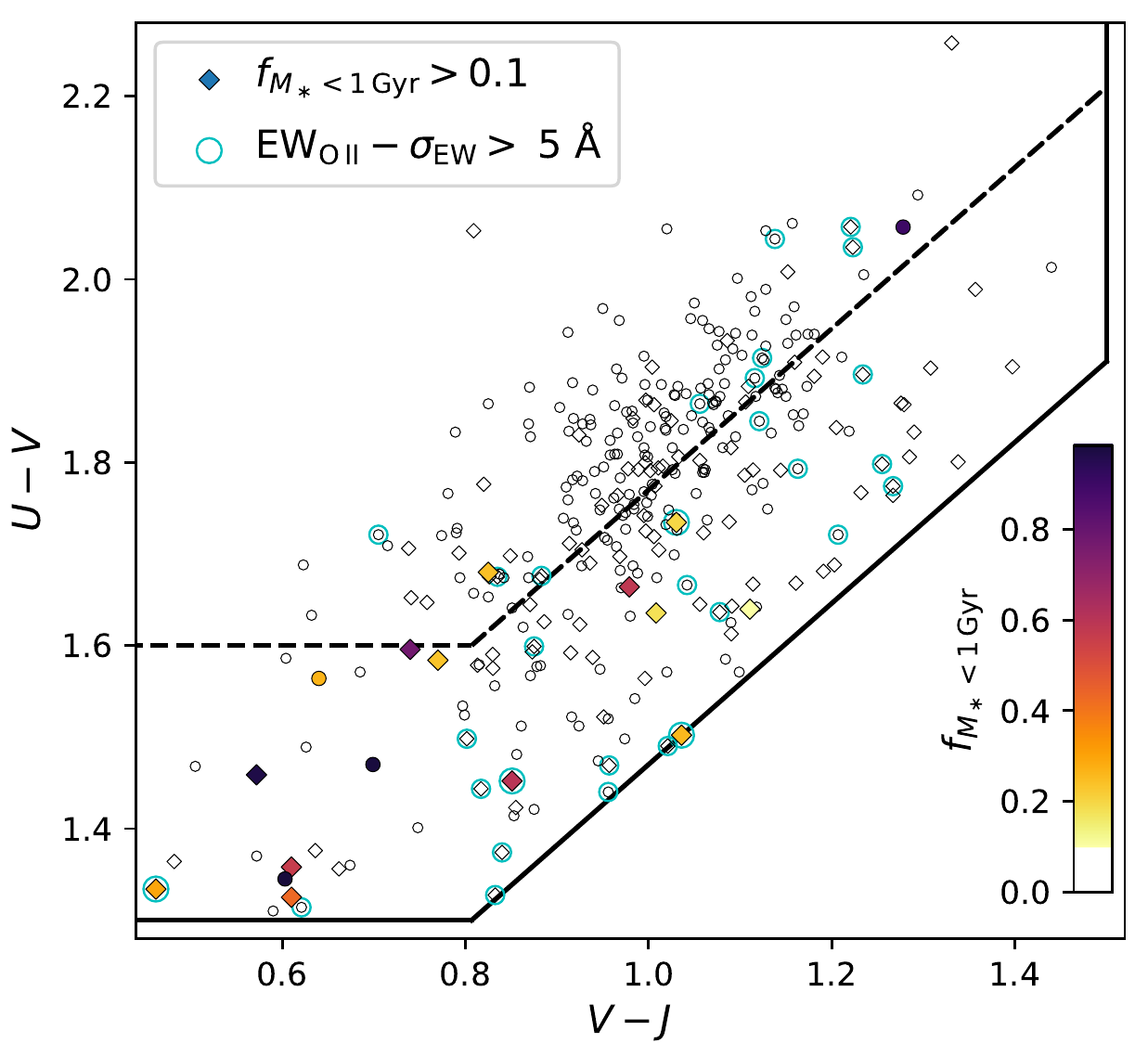}
      \end{center}
      \caption{Rest-frame \textit{UVJ} colours of the GOGREEN quiescent galaxies (plus marks), where \frejuv$>0.1$ galaxies are shown with colours according to the fraction of stellar mass formed within the last 1\,Gyr, \frejuv. Cluster galaxies are shown as circles, field galaxies as diamonds. Galaxies with significant \oii emission are circled. Only a few of the \frejuv$>0.1$ or \oii emitting galaxies occupy the densest region, ie. the `red clump'. We therefore test our age comparison for galaxies in this clump by increasing the $U{-}V$ selection by 0.3\,dex, shown as a dashed line (in other words, exclude the `\textit{UVJ} intermediate' galaxies), see Figure\,\ref{fig:diff_mwa_cum_all}.} \label{fig:rejuv_uvj}
    \end{figure}

Figure\,\ref{fig:rejuv_uvj} shows our quiescent sample in \textit{UVJ} colour space, where diamonds indicate \frejuv$>0.1$ galaxies, and galaxies with EW(\oii){-}$\sigma_\mathrm{EW}>$\,5\,{\AA} are circled. Interestingly, and perhaps as expected, the `young' galaxies occupy the bluer end of the \textit{UVJ} colours (except one galaxy), and both the `young' and \oii emitting galaxies preferentially occupy the colour space closer to the boundary of the quiescent selection. This region is below the dashed line in Figure\,\ref{fig:rejuv_uvj} where the \textit{U}-\textit{V} delimiter was increased by 0.3\,dex. 

We now repeat our measurement of the mass-weighted age difference between field and cluster galaxies, now excluding galaxies which are not intrinsically old. Figure\,\ref{fig:diff_mwa_cum_all} summarises the age comparisons for these various selections of quiescent galaxies, relative to the full sample. Our result does not significantly change when excluding `young' (\frejuv$>0.1$) galaxies, galaxies with \oii emission, or galaxies near the \textit{UVJ}-colour quiescent selection boundary (labelled \textit{UVJ} intermediate). The latter selection would reasonably exclude galaxies transitioning between star-forming and quiescence, or which have complex dust properties obscuring star forming populations. Saying that, the largest change comes from excluding the \frejuv$>0.1$ galaxies, particularly at lower stellar masses. On the other hand, excluding the \textit{UVJ} intermediate primarily increases the age difference between low mass galaxies, although the error bars are larger due to smaller numbers of galaxies. The exclusion of \oii emitting galaxies does not visibly affect the age difference at all except for the lower mass galaxies, decreasing the age difference. Overall the age difference between field and cluster galaxies is insensitive to recent star formation, unlike the low-redshift galaxies studied in \citet{thomas2010}. 

The age comparison between luminosity-weighted ages is also shown, discussed in Appendix\,\ref{sec:lwas}, which predicts a larger (by 0.1\,Gyr) age difference than mass-weighted ages, except for the highest mass galaxies. The luminosity-weighted age is more sensitive to recent star formation, so it is not unexpected that there is a mass dependence between \mwa and \lwa related to the mass-dependent SFHs.

 \section{Discussion}\label{sec:discussion} 
 
The main goal of this work is to compare quiescent galaxies in average density (field) and high density (galaxy cluster) environments, while accounting for any differences related to their stellar mass. We now discuss our result that the age difference is within $0.31_{^{-0.33}}^{_{+0.51}}$\,Gyr in the context of the literature. In Section\,\ref{sec:mass_discuss}, we discuss that our SFHs are consistent with `mass-dependent evolution', and the environmental dependence of the SFHs. In Section\,\ref{sec:env_discuss}, we compare the mass-weighted age measurements to similar results from the literature, and discuss the difference between mass-weighted ages as a function of environment. 

\noindent We then place the measured age difference in the context of two simple quenching models in Section\,\ref{sec:toy}, where environmental quenching is purely related to the time since infall, or where there 
is no environmental quenching but the field population forms later than cluster galaxies.

\subsection{Mass-dependent evolution of quiescent galaxies}\label{sec:mass_discuss}

Decades of work has shown that the bulk of star formation in massive ETGs occurred at high redshifts, and these galaxies have been passively evolving since. Studies connecting intermediate-redshift and local observations of 
the colour-magnitude relations \citep[e.g.][]{dressler1980, ellis1997, stanford1998}, 
the evolution of the luminosity function \citep[e.g.][]{depropris1999, toft2004},
the Fundamental Plane \citep[e.g.][]{vandokkum1998, kelson2000, cimatti2006, diseregoalighieri2006a,diseregoalighieri2006b, jorgensen2006,jorgensen2007, Beifiori2017, woodrum2017, saracco2020},
and absorption lines \citep[e.g,][]{bender1996, kelson2001, sanchez-blazquez2009}
suggest that ETGs have been evolving passively since $z{\sim}2$--3 (see \citealt{renzini2006} for a review). 
As large scale surveys became available \citep[e.g. SDSS;][]{york2000}, trends between the star formation histories and galaxy properties have increasingly been explored. A robust finding is that more massive galaxies form their stellar mass earlier and over shorter time scales than lower mass galaxies \citep[e.g.][]{gallazzi2014, heavens2004, jimenez2005, nelan2005, thomas2005, thomas2010, thomas2017, sanchez-blazquez2009}, i.e., `mass-dependent evolution'. This is similar to the concept of `downsizing' in the sense that there is mass-dependent decline in the SFRs of galaxies with time \citep[e.g.][]{cowie1996, bell2005, juneau2005}, or in the growth of the stellar mass function \citep[e.g.][]{cimatti2006, leitner2012}. This downsizing trend can be explained by the fact that more low mass galaxies are continuously (over time) being added to the quiescent population \citep{brammer2011, muzzin2013a, tomczak2014}. On the other hand, merger rates are mass dependent \citep{khochfar2009, emsellem2011}, and late time rejuvenated star formation is more common in low mass galaxies \citep{poggianti2008,poggianti2009, thomas2010, belli2015}.

The mass-dependence of SFHs has been confirmed at higher redshifts, where age indicators are more sensitive to older stellar populations. However, observations beyond $z{\sim}1$ are challenging. As a result, studies have been mainly limited to surveys of massive galaxies with 
small samples \citep[e.g.][]{vandokkum2010b, toft2012, vandesande2013, kriek2016, belli2015, belli2019, estrada-carpenter2019, saracco2020}, 
and rely on averaging photometric SFHs \citep[e.g.][]{rettura2011, snyder2012, strazzullo2013, pacifici2016, iyer2017}
or combining spectra \citep[e.g.][]{gobat2008, tanaka2013, whitaker2013, choi2014, onodera2015}. Only recently have large, high-redshift spectroscopic surveys been completed which allow more precise age estimates of individual galaxies. Notably, \citet{chauke2018} combine high resolution spectroscopy and photometry for more than 600 galaxies at $0.6{<}z{<}1$ from LEGA-C \citep{vanderwel2016, straatman2018} to show that galaxies with higher stellar velocity dispersions formed both earlier and faster, and that the majority of quiescent galaxies evolve passively since their main star forming epoch. \citet{carnall2019b} similarly use the VANDELS survey \citep{mclure2018, pentericci2018} to determine the SFHs for 75 massive quiescent galaxies at $1{<}z{<}1.3$, finding a trend between the average formation times of galaxies and their stellar mass of $1.48^{_{+0.34}}_{^{-0.39}}$\,Gyr per dex for $M_\ast{<}10^{11}$\,M$_{\sun}$.

The mass-dependent evolution in the GOGREEN quiescent galaxies is apparent in Figure\,\ref{fig:sfhs}, where we find the more massive galaxies to have sSFRs which are higher at earlier times, and decline at earlier times, than the lower mass galaxies, at fixed environment. The median mass-weighted ages are shown in Figure\,\ref{fig:diff_mwa} relative to stellar mass, where contours show the combined posteriors. While we see that the ages of lower mass galaxies are younger \textit{on average}, this appears to be driven by the fact that there is a broader distribution of ages among the lower mass galaxies. Indeed our trend between age and mass in our field sample is both flatter and offset towards older ages than found by \citet[][and references therein]{carnall2019b}. 

The ${\sim}$1\,Gyr offset in ages could be a result of differences in fitting procedure, where \citet{leja2019b} report that \prospector-$\alpha$\footnote{\prospector-$\alpha$ uses the \prospector framework, but includes additional parameters (such as dust emission, nebular emission, AGN emission).} predicts older ages and higher stellar masses than standard parametric modelling. \citet{carnall2019b} use a double-power-law form for their SFHs, however, which is more flexible than fiducial declining-exponential models, so the ages should be more similar than those reported by \citet{leja2019b}. Along the same lines, Forrest et al. (2020b; submitted) reconstruct the SFHs of ultra massive (${>}10^{11}$\,M$_{\sun}$) galaxies at $3{<}z{<}4$ and find that the bulk of star formation occurred between $4{<}z{<}6$ ( ${\lesssim}0.5$\,Gyr later than the median mass-weighted ages we measure), and the galaxies quenched several hundred Myr later, in some cases as early as $z{\sim}4$. Besides the difference in parameterization of the SFHs, an alternative explanation for the older ages we find is that it is a result of a lower metallicity in the best-fit model, since metallicities are strongly degenerate with ages. In fact, as discussed in Appendix\,\ref{sec:mzr}, our metallicities are systematically lower than other studies at intermediate redshifts \citep[e.g.][]{choi2014, estrada-carpenter2019, morishita2019}. 
An increase in metallicity by a factor of three (i.e, +0.5\,dex) would decrease the mass-weighted age by ${\sim}$0.5\,Gyr, which would account for most of the age difference. 

There is a stronger age difference between lower and higher mass galaxies at fixed environment, than between environments at fixed mass -- despite the fact that we find a flatter mass-dependence of the SFHs than other studies. For both the cluster and the field populations, the median difference in mass-weighted ages is ${\sim}0.7^{_{+0.3}}_{^{-0.6}}$\,Gyr between galaxies of mass $10^{10}$--$10^{10.5}$\,M$_{\sun}$ and $10^{11.3}$--$10^{11.8}$\,M$_{\sun}$, while the age differences between environments are ${<}0.4$\,Gyr (see Figure\,\ref{fig:diff_mwa_cum_all}), and are discussed further in the next Section. This result is consistent with the results of \citet{saglia2010} and \citet{woodrum2017}, where both measured the evolution of the M/L ratio between cluster and field galaxies at $z{\la}0.9$ and $z{<}1.2$, respectively, and found stronger differences between galaxies of different stellar mass than between environments. Similarly, \citet{raichoor2011} compared ETGs at  $z{\sim}1.3$ to conclude that the age difference between galaxies in cluster and field environments was less significant than between galaxies of different mass.

\subsection{Environment-dependent evolution of quiescent galaxies}\label{sec:env_discuss}

A number of recent studies find that field galaxies form over longer timescales than cluster galaxies, however, the exact timescales have been challenging to robustly quantify. Line strength studies of early type galaxies (ETGs) at low redshifts find that star formation in low density environments is delayed by 1--2\,Gyr \citep[e.g.][see also the review by \citealt{renzini2006}]{bernardi1998, balogh1999, thomas2005, clemens2006, sanchez-blazquez2006}. Using the Fundamental Plane, the evolution of the M/L between galaxies at $z{\la}1.2$ has shown that the slopes are steeper for galaxies in cluster environments, indicating that they formed at slightly higher redshifts than field galaxies \citep[e.g.][]{vandokkum2007, saglia2010, woodrum2017}. The M/L evolution can be interpreted as SFHs with models of simple stellar populations (SSPs), taking into account the structural evolution in the size of galaxies (and progenitor bias). \citet{vandokkum2007} infer that massive galaxies in clusters are ${\sim}$0.4\,Gyr older than field galaxies, \citet{saglia2010} estimate a ${\sim}$1.6\,Gyr age difference, while \citet{woodrum2017} estimate ${\la}0.3$\,Gyr difference for galaxies with low-velocity dispersions but ${\la}$1\,Gyr for high-velocity dispersions. Compared to luminosity-weighted ages derived from Balmer absorption lines, \citet{saglia2010} find consistent age estimates within their large uncertainties, while \citet{woodrum2017} find a larger age difference of 1--3\,Gyr. 

Measuring age differences at low redshifts does not necessarily reflect differences in star formation histories at early times, however. Late-stage environmental effects on galaxy evolution \citep[e.g.][]{thomas2010}, or progenitor effects, can obscure estimates of the ages of the oldest stellar populations; recent star formation can `outshine' older stars making age estimates from the integrated light difficult \citep{papovich2001}. Moreover, the population of ETGs has been in place since $z{\sim}$2 \citep[e.g.][]{bernardi1998, vandokkum2010a}, where the result is that galaxies older than ${\sim}$5\,Gyr have similar stellar spectra and are difficult to distinguish \citep{conroy2013a}. In order to explore whether environmental factors affected galaxy formation during the period where the galaxies assembled the majority of their mass requires higher redshift observations. 

At $z{\sim}1.2$, \citet{gobat2008} measured the ages of ETGs in a massive cluster relative to galaxies in the GOODS/CDF-S survey via SED fitting photometry and coadded spectroscopy, finding that cluster galaxies formed ${\sim}$0.5\,Gyr before field galaxies (particularly at ${<}10^{11}$\,M$_{\sun}$). On the other hand, for the same cluster \citet{rettura2010} independently compare the massive ETGs with equivalent galaxies in the GOODS survey, measuring ages from fitting photometry to SEDs (without spectroscopy), and conclude that there is no significant delay in formation epochs between the two environments within the typical uncertainty of ${\sim}$0.5\,Gyr. Two additional clusters are included in the comparison by \citet{rettura2011}, where again no difference was found in formation times within their average uncertainty, 0.5\,Gyr. At $z{\sim}1.3$, \citet{saracco2017} compare the median luminosity-weighted ages of elliptical galaxies in three clusters relative to the GOODS, COSMOS, and CANDELS fields. While they find that the structural properties of galaxies in cluster and field environments are consistent at fixed mass, and ${<}10^{11}$\,M$_{\sun}$, massive galaxies either assemble ${\sim}$0.3\,Gyr earlier or assemble more efficiently in clusters. 

Our results are fully consistent with these studies. We find that cluster galaxies are \textit{on average} $0.31_{^{-0.33}}^{_{+0.51}}$\,Gyr
older than field galaxies, at fixed stellar mass. While the age difference is largest for galaxies of masses $10^{10.5}$--$10^{11.3}$\,M$_{\sun}$, the age difference is positive (although sometimes consistent with zero) for all mass ranges. This result is robust when carefully removing galaxies which show recent star formation, \oii emission, or \textit{UVJ} colours outside of the red clump (see Section\,\ref{sec:rejuv}). 

\citet{muzzin2012} compare \dbreak values, as a proxy for stellar age, for quiescent\footnote{\citet{muzzin2012} select quiescent galaxies based on the lack of \oii emission, rather than \textit{UVJ} colours. See Appendix\,\ref{sec:tracers} for a comparison of these selections.} galaxies in the GCLASS survey; a subset of these clusters, and galaxy spectra, are included in GOGREEN. At fixed stellar mass, they find that \dbreak is independent of environment except perhaps for their lowest mass galaxies ${<}10^{10}$\,M$_{\sun}$. We compare the \dbreak of our spectra relative to \citet{muzzin2012} in Appendix\,\ref{sec:coadds}, where we find modestly larger differences between environments, consistent with our result of a small positive age difference. The GCLASS sample is dominated by $z{\sim}0.8$ clusters, however, particularly at low stellar masses. Thus, the small difference we observe may be a result of evolution.  

An important consideration when comparing to results from the literature is how the lower density sample is defined. Some studies separate galaxies in the cores and outskirts of clusters, or in higher- and lower-density regions within their sample, or carefully select for galaxies in clusters, groups, or in isolation. Our field sample is selected from the distant fore- and background of our clusters and is therefore expected to be representative of an average patch of the Universe. Comparing galaxies in clusters with those truly isolated in cosmic voids, or exclusively galaxies central to their halo, will likely have a larger contrast in properties than our results. Importantly, the `field' environment may be different at different mass scales; for example, more massive galaxies could be more likely to exist in cosmic overdensities (e.g. groups) than lower mass galaxies. Therefore, the comparison between galaxies of lower stellar masses could reflect different physical factors than between galaxies of higher mass. We leave a comparison of galaxies between different local densities to a future paper.   

A second consideration is the selection of quiescent galaxies: several studies classify quiescent galaxies based on morphology, or other star formation tracers than \textit{UVJ} colours. We do not expect this to significantly impact the relative age measurements, however, as long as the selection is consistent between environments. \citet{saracco2017} find that at $z{\sim}1.3$ elliptical galaxies have consistent structure and properties between field and cluster environments, however there are fewer large and massive elliptical galaxies in the field relative to clusters. Such differences between galaxy properties and environment could be important to the quiescent-selection in detail.

\subsection{Toy models of cluster galaxy evolution}\label{sec:toy}

It is well established that at low redshifts the fraction of quiescent galaxies is higher in denser environments \citep[e.g.][]{baldry2006}. Several studies also find a higher fraction of low-mass quiescent galaxies in denser environments 
\citep[e.g.][]{muzzin2012, woo2013}. \citet{peng2010} suggested that these two observations are consistent if galaxies in dense environments are subject to extra `environmental-quenching' which is independent of stellar mass, in addition to mass-dependent `self-quenching'. 

At $z{\gtrsim}1$ the situation is very different. While there is still an excess of quiescent galaxies in dense environments, the SMFs of quiescent galaxies are consistent between low- and high-density environments \citep[][]{nantais2016, vanderburg2020}. Moreover, the shapes of the SMFs for \textit{star forming} galaxies are also the same between cluster and field. We add to this picture the fact that there is a small, positive age difference between quiescent cluster and field galaxies. This is difficult to reconcile with the hypothesis that the higher fractions of quenched galaxies in galaxy clusters at this epoch result from the transformation of recently accreted, star-forming galaxies. 

We first consider whether a simple infall-based quenching model can be simultaneously compatible with both our measured average age difference, and the quenched fractions in cluster and field environments measured by \citet[][]{vanderburg2020}. We then consider an alternative model where cluster galaxies formed earlier than field galaxies, and infall-driven quenching is negligible. 

In order to determine the mass-weighted age evolution we need a prediction of the average SFH of star forming galaxies; we assume the SFRs evolve as defined in \citet{schreiber2015}, and that the SF is instantaneously truncated when the galaxy is `quenched'. We compare galaxies with final stellar masses between $10^{9.5}$-$10^{11.5}$\,M$_{\sun}$. As we are only interested in modelling the global properties of `average' galaxies we ignore any mass dependence in the data. Therefore we model the self-quenching efficiency using the same form as proposed by \citet{peng2010} (i.e., $\eta\propto \mathrm{SFR}/M^\ast$) using the SFR for an ${M}^\ast{=}10^{10.8}$\,M$_{\sun}$ galaxy. Given that \citet[][]{vanderburg2020} find the SMFs between star forming cluster and field populations to have the same shape, we require that the star forming SMFs in our model similarly do not evolve. Our toy model consists of tracking the number of star forming and quiescent galaxies from $z{=}10$ (when cluster galaxies are assumed to form) to $z{=}1.2$.\footnote{This toy model is qualitatively different than the mass-quenching model proposed by \citet{peng2010}, or as implemented by \citet[][]{vanderburg2020}. Furthermore, we neglect mergers. Including mergers, however, would only enhance the different galaxy properties between cluster and field environments.}


We acknowledge that this is a simple assumption for the evolution of field galaxies and may not be realistic \citep[e.g.][]{dressler2013, gladders2013, schawinski2014}. However, it serves as a useful starting point that characterizes the overall growth in the quiescent population with time. In future work we will consider more sophisticated models, in light of all the available GOGREEN data.   

\subsubsection{Post-infall environmental quenching and pre-processing}

For the infall-based quenching model we assume that all galaxies are subject to self-quenching, while in addition star forming galaxies that join clusters quench at a given time after infall (${t}_\mathrm{delay}$). The infall rate we assume follows the predictions of \citet{mcgee2009} for timescales of galaxies becoming satellites of larger halos (${>}10^{13}$\,M$_{\sun}$), based on the Millennium simulation \citep{springel2005} with additional prescriptions for halo assembly via merger trees \citep{helly2003, harker2006} and using the semi-analytic models of \citet{bower2006}\footnote{With updated modelling of strangulation, as per \citet{font2008}.}. This predicts that the rate at which galaxies join larger haloes is effectively constant in time. There are then two parameters in this model which determine the relative populations of star forming/quiescent and field/cluster galaxies: the normalization of the self-quenching efficiency, and $t_\mathrm{delay}$. Both of these parameters are constrained by observations of the quenched fractions at our fiducial stellar mass, measured for the GOGREEN sample to be ${f}_{\mathrm{Q,\,field}}(z{=}1.2)=0.3$ and ${f}_{\mathrm{Q,\,cluster}}(z{=}1.2)=0.65$ \citep[][]{vanderburg2020}. The self-quenching efficiency drives the quenched fraction in the field, while the delay time determines the additional quenching in clusters. We find that a delay time of $t_\mathrm{delay}{\sim}2.4$\,Gyr is required to match the observed quenched fractions. This is somewhat longer than expected from dynamical timescales at this redshift \citep[e.g.][]{balogh2016}; we caution that our toy model is merely illustrative (we ignore mass dependence and the mass-quenching rate is somewhat arbitrary), and this discrepancy does not significantly affect our conclusions here.

An important consequence of post-infall environmental quenching models is that quiescent galaxies in the field would be on average \textit{older} than quiescent galaxies in clusters at fixed mass (by $1.5^{_{+1.3}}_{^{-0.2}}$\,Gyr given the \tdelay and quenched fractions listed above). This is because the rate that recently-quenched galaxies are added to the quiescent population is higher at later times in the cluster, such that the overall population is younger. As we have constructed our model, environmental-quenching is stronger at later times \citep[e.g.][]{nantais2017}, while self-quenching dominates at early times. \citet{muzzin2012} come to a similar conclusion modelling the evolution of \dbreak for early-self-quenching late-environmental-quenching dominated efficiencies. Given that we find a small, but significant, average age difference between field and cluster galaxies in the opposite sense, we can exclude this model even for delay times approaching the age of the universe. 

One important simplification of this infall-based quenching model is that we have neglected the role of pre-possessing in the field population. That is, galaxies which quenched in locally overdense clumps (i.e., groups or filaments) prior to joining clusters \citep[e.g.][]{dressler1980, fujita2004, moran2007}. The infall rate we use predicts the number of galaxies which become satellites of haloes with masses ${>}10^{13}$\,M$_{\sun}$ within a given time, and we have considered all such structures `clusters' when realistically some fraction makes up the `field'. Secondly, some fraction of these pre-processed groups will eventually accrete onto clusters. \citet{mcgee2009}'s model predicts that at $z{=}1.5$ around 20\,per cent of galaxies were in halos of mass $10^{13}$--$10^{14}$\,M$_{\sun}$\,$h^{-1}$ prior to becoming a member of their final halo. Along the same lines \citet{poggianti2006} use the fraction of cluster galaxies with \oii emission to constrain the fraction of galaxies which were `primordially quenched' at high redshifts, or experienced environmental quenching in haloes above $10^{14}$\,M$_{\sun}$. They find that $z{\sim}1.5$ marks a turnover between these two populations, where only galaxies in haloes with high velocity dispersions have appreciable numbers of `quenched' galaxies. \citet{delucia2012} build on the model of \citet{mcgee2009} to show that the accretion history of satellites onto clusters is stellar mass dependent, where lower stellar mass galaxies are more likely to be satellite of a smaller structure when joining a cluster. Moreover, if the groups which accrete onto clusters represent a biased sample (e.g. the oldest groups) this would make the cluster quenched population older on average, and the field younger. In this scenario it is no longer clear that the field quenched population is necessarily older than quiescent cluster galaxies. The exact age differences are difficult to predict, however, as they depend on the distribution of galaxies in groups between field and cluster environments. Lastly, even considering the fact that some galaxies may be part of smaller substructures prior to joining clusters, whether or not they are quenched in such environments likely depends on the halo mass and how long they have been satellites. If the dominant environmental-quenching processes are only relevant over long timescales, the effect of pre-processing at high redshifts may not be significant. We leave a more complete analysis to a future paper.

\subsubsection{Delayed formation of field galaxies}

\begin{figure}
      \begin{center}
        \includegraphics[width=\linewidth]{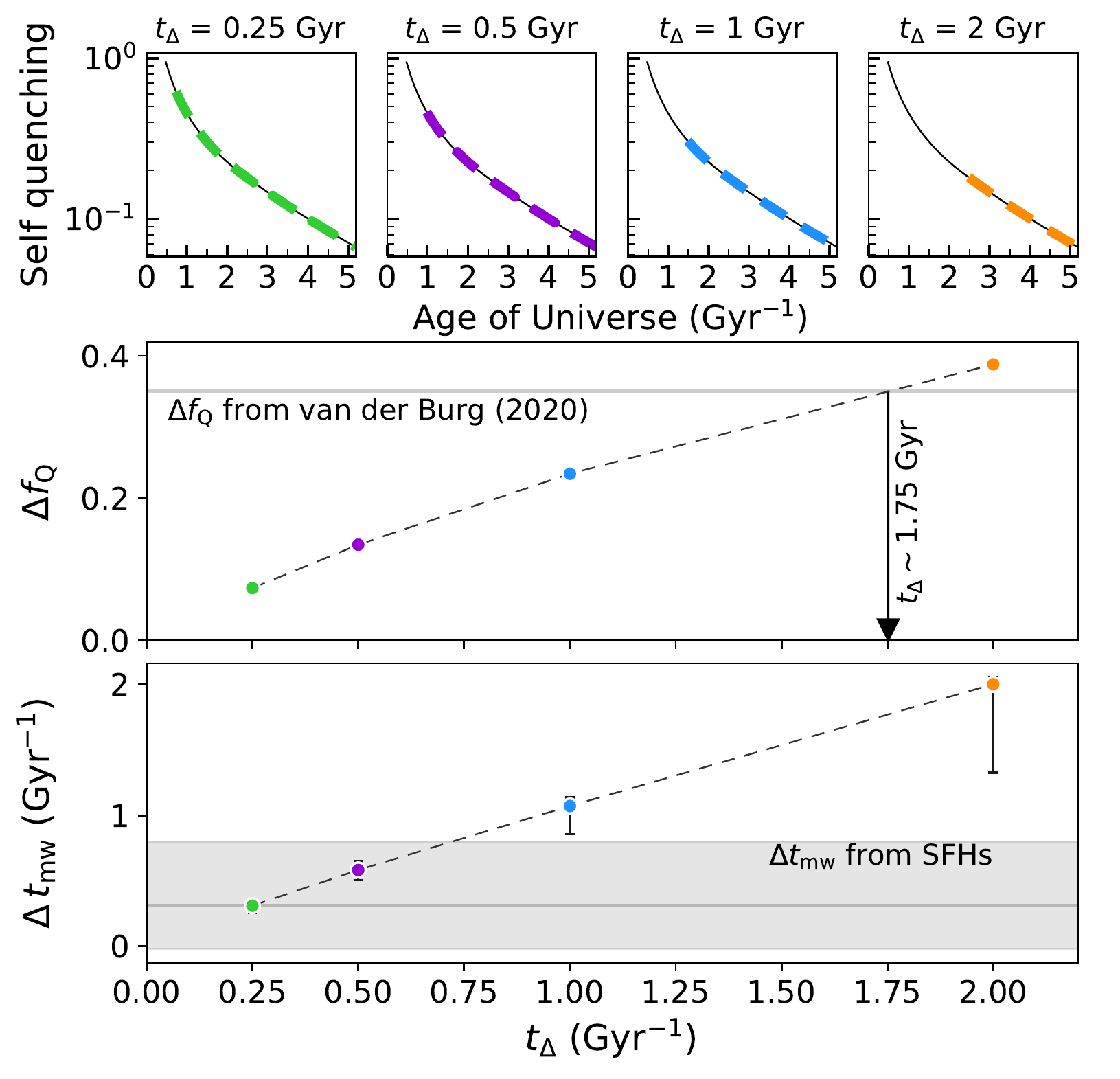} \\
      \end{center}
      \caption{Toy model of the expected difference in quenched fractions and \mwa given an offset in the formation of the field population of $t_\Delta$. Top: quenching rate for four values of \tdelta: 0.25\,Gyr, 0.5\,Gyr, 1\,Gyr, and 2\,Gyr. The cluster quenching rate is shown as a black line starting at $z_\mathrm{form}{=}10$, and field quenching rate as a coloured (according to \tdelta) dashed line. Middle: The difference in quenched fractions, $\Delta\mathrm{f}_\mathrm{Q}$, for fixed stellar mass at $z{=}1.2$, for the four \tdelta models shown. Larger offsets in the formation of field galaxies corresponds to larger $\Delta\mathrm{f}_\mathrm{Q}$. A horizontal line indicates the measured difference in quenched fractions from \citet[][]{vanderburg2020}. Bottom: Average difference in cumulative \mwa distributions, $\Delta{\mathrm{t}_\mathrm{mw}}$, between field and cluster galaxies, with error bars indicating the 68 per cent spread. The grey shaded region indicates the measured average age difference, see Section\,\ref{sec:mwas}. Larger offsets in the formation of field galaxies corresponds to larger $\Delta{\mathrm{t}_\mathrm{mw}}$. In the context of this simple model, \tdelta${<}$0.75\,Gyr is consistent with our observations, but is inconsistent with the time derived by the difference in quenched fractions.  }
     \label{fig:toy}
    \end{figure}

Motivated by this challenge for the simplistic post-infall quenching model to explain our results, we now turn to a model where the self-quenching of cluster galaxies gets a head start relative to the average field. Figure\,\ref{fig:toy} illustrates this toy model of delayed formation times between cluster and field galaxies. Here the only quenching is self-quenching, which is shown in the top row for cluster galaxies (thin black line, starting at $z{=}10$) normalised such that ${f}_{\mathrm{Q,\,cluster}}(z{=}1.2)=0.65$, and for the field galaxies (coloured dashed lines). Galaxies in the two environments quench through the same processes; however, field galaxies form and quench later, starting at a time offset from the cluster, labelled as \tdelta\footnote{We note that in the simple model of \citet{peng2010} (their Sec\,6) the formation of field galaxies is delayed ${\sim}$1\,Gyr ($z_\mathrm{form}{=}4$)}.
For the four delay times, (\tdelta {=} 0.25\,Gyr, 0.5\,Gyr, 1\,Gyr, and 2\,Gyr), the difference in quenched fractions between cluster and field galaxies, $\Delta{f}_\mathrm{Q}$, and the median cumulative mass-weighted age difference, $\Delta{{t}_\mathrm{mw}}$, are calculated and shown in the second and third plots, respectively. The error bars on $\Delta{{t}_\mathrm{mw}}$ correspond to the 68 per cent region of the age comparison, and the grey region indicates the measured age difference discussed in Section\,\ref{sec:mwas}.

Figure\,\ref{fig:toy} shows that any delay time $\gtrsim$1\,Gyr would result in a mass-weighted age difference that is excluded by our data. To match the observed $\Delta\mathrm{f}_\mathrm{Q}{\sim}0.35$, however, would require \tdelta${\sim}$1.75\,Gyr in our simple model. This is larger than the delay in formation time predicted to match the quenched fractions in the toy model of \citet[][]{vanderburg2020} of ${\sim}$1\,Gyr, which likely is due to different assumptions of the growth of the SMF and mass-dependent self-quenching. In either model, such a long formation delay time would result in a mass-weighted aged difference of ${>}$1\,Gyr, strongly excluded by our observations.

\hspace{\linewidth}

Neither the simple post-infall environment quenching model nor the delayed-formation model can fully explain the difference in galaxy properties between high and low density environments at $z{=}1$. In principle a combination of the two models can, even without pre-processing. For example, with a delay between infall and environmental quenching of ${\sim}$2.8\,Gyr (which is still long), and delaying the formation of field galaxies by 1\,Gyr relative to cluster galaxies, this simple model can simultaneously match both the observed quenched fraction and age difference. Before concluding that such a hybrid model is successful, however, it will be important to test the stellar mass and halo-mass dependence of the predictions. We leave this work to a future paper.

\section{Conclusions}\label{sec:conclusions}

In this work we determined the SFHs for 331 quiescent galaxies in 11 GOGREEN clusters and field galaxies at $1{<}z{<}1.5$ based on rest-frame optical spectroscopy and multi-band photometry fit to SED templates with the Bayesian inference code \prospector. The following summarises our comparison of the quiescent field and cluster galaxies:

\begin{enumerate}

\item Comparing SFHs between galaxies of different mass we found that more massive galaxies form earlier, and over shorter timescales, than lower mass galaxies (see Section\,\ref{sec:sfhs} and the bottom row of Figure\,\ref{fig:sfhs}). This picture is consistent with the `mass-dependent evolution' scenario. Comparing SFHs between galaxies in cluster and field environments, we conclude that below $10^{11.3}$\,M$_{\sun}$ the SFRs declined earlier and more rapidly for galaxies in denser environments, at fixed mass (see the right-hand column of Figure\,\ref{fig:sfhs}). 

\item From the SFHs we calculate posteriors for mass-weighted ages for each galaxy, shown in Figure\,\ref{fig:diff_mwa} relative to stellar mass. Overall, 90\,per cent of all galaxies have formed half their stellar mass by $z{\sim}2.2$. The majority (${>}50$\,per cent) of galaxies with masses $10^{11.3}$--$10^{11.8}$\,M$_{\sun}$ have formed at least half of their stellar mass by $z{\sim}5.4$, while the same is true for galaxies with masses $10^{10}$--$10^{10.5}$\,M$_{\sun}$ at $z{\sim}3.4$. The formation times we estimate are older than similar ages in the literature; this may be a consequence of the age-metallicity degeneracy and the fact that our fits prefer somewhat lower metallicities than other studies (see discussion in Sec\,\ref{sec:env_discuss}). Future telescopes, such as the James Webb Space Telescope (JWST), can directly observe galaxies at these redshifts and will be able to confirm our predictions. 

\item We compare the mass-weighted ages for galaxies of similar stellar mass between the two environments (see Figure\,\ref{fig:diff_mwa_cum}). The distribution of ages for field galaxies is broader than for cluster galaxies, where the field population has a higher relative fraction of young galaxies. As a result, the mass-weighted age difference between field and cluster galaxies with stellar masses between $10^{10}$--$10^{11.8}$\,M$_{\sun}$ is within $0.31_{^{-0.33}}^{_{+0.51}}$\,Gyr, in the sense that cluster galaxies are older on average. This result holds when we exclude galaxies which have formed more than 10\,per cent of the stellar mass within the last 1\,Gyr, have significant \oii emission, or have \textit{UVJ} colours closer to the star formation population (see discussion in Sec\,\ref{sec:rejuv}). 

\item We consider two simple models consistent with the higher fraction of quenched galaxies in clusters, and show neither one is consistent with our age measurements. If the environmentally-quenched population is built up entirely through post-infall quenching processes (without pre-processing), we predict field galaxies would be \textit{older} than cluster galaxies -- in contrast to our results. On the other hand, if quenching in cluster environments gets a head start, this needs to be ${>}$1\,Gyr to explain the difference in quenched fraction, which results in an average mass-weighted age difference that is much larger than we observe.

\end{enumerate}

This work builds on previous evidence \citep{balogh2016,vanderburg2020} that the substantial quenched population in galaxy clusters at $z{>}1$ has been built up in a fundamentally different way from clusters at $z{=}0$. In particular, the infall-based environmental quenching models of \citet{peng2010}, \citet{wetzel2012}, and others, that are so successful at matching local observations, are not able to account for the properties of the GOGREEN cluster sample. The data seem to require that a substantial population of protocluster galaxies are quenched at early times, $z{>}3$, via a process which is accelerated but otherwise indistinguishable from the mass-quenching that affects all galaxies. Evidence of quenched populations of massive galaxies at this epoch is growing \citep[e.g.][Forrest et al. 2020b; submitted, McConachie et al. 2020; in prep]{glazebrook2017,schreiber2018a, tanaka2019, forrest2020, tanaka2020, valentino2020}. Environmental quenching must still play a role, but it may only become dominant at $z{<}1$.
In future work we will use the stellar-mass and halo-mass dependence of these observations to further constrain these toy models; comparison with simulations and semianalytic models will be important to help identify the physical origin of the quenching mechanisms postulated here (e.g., Kukstas et al. 2020; in prep). Finally, these data indicate that much of the quenching activity responsible for building up galaxy clusters occurred in the protocluster environment at $z{>}3$; data from JWST will be crucial for understanding the nature and cause of this phenomenon.


\section*{Acknowledgements}

We thank the native Hawaiians for the use of Mauna Kea, as observations from Gemini, CFHT, and Subaru were all used as part of our survey. We acknowledge the helpful advice from the referee.

This research is supported by the following grants: National Sciences and Engineering Reserach Council of Canada (NSERC) CGS-D award (KW), Discovery grants (MLB and AM), and USRA award (KB); European Space Agency (ESA) Research Fellowship (LJO); National Science Foundation (NSF) grants AST-1517863 (GW), AST-1517815 (GHR); NASA, through grants GO-15294 (GW), GO-15294.012-A (GHR) and the Astrophysics Data Anaylsis Program (ADAP) 80NSSC17K0019 (GW), and 80NSSC19K0592 (GHR).
JN acknowledges support from the UNAB Internal Project number DI-12-19/R. 
BV acknowledges financial contribution from the grant PRIN MIUR 2017 n.20173ML3WW\_001 (PI Cimatti) and from the INAF main-stream funding programme (PI Vulcani). 
RD gratefully acknowledges support from the Chilean Centro de Excelencia en Astrof\'isica y Tecnolog\'ias Afines (CATA) BASAL grant AFB-170002 
PC acknowledges the support from the ALMA-CONICYT grant no 31180051.
Support for program number GO-15294 was provided by NASA through a grant from the Space Telescope Science Institute, which is operated by the Association of Universities for Research in Astronomy, Incorporated, under NASA contract NAS5-26555. GHR also acknowledges the support of the International Space Sciences Institute for sponsoring a team meeting and for the European Southern Observatory for supporting him with a visiting fellowship for summer research. This work was made possible by the facilities of the Shared Hierarchical Academic Research Computing Network (SHARCNET:www.sharcnet.ca) and Compute/Calcul Canada. 

This paper includes data gathered with the Gemini Observatory, which is operated by the Association of Universities for Research in Astronomy, Inc., under a cooperative agreement with the NSF on behalf of the Gemini partnership: the National Science Foundation (United States), the National Research Council (Canada), CONICYT (Chile), Ministerio de Ciencia, Tecnologa e Innovacin Productiva (Argentina), and Ministrio da Cilncia, Tecnologia e Inovao (Brazil); the 6.5 metre Magellan Telescopes located at Las Campanas Observatory, Chile; the Canada-France-Hawaii Telescope (CFHT) which is operated by the National Research Council of Canada, the Institut National des Sciences de l’Univers of the Centre National de la Recherche Scientifique of France, and the University of Hawaii; MegaPrime/MegaCam, a joint project of CFHT and CEA/DAPNIA; Subaru Telescope, which is operated by the National Astronomical Observatory of Japan; and the ESO Telescopes at the La Silla Paranal Observatory under programme ID 097.A-0734. This work was based on data products from observations made with ESO Telescopes at the La Silla Paranal Observatory under ESO programme ID 179.A-2005 and on data products produced by CALET and the Cambridge Astronomy Survey Unit on behalf of the UltraVISTA consortium. This work made use of {\sc Prospector} \citep{leja2017, johnson2019}, {\sc python-fsps} \citep{foreman-mackey2013}, {\sc FSPS} \citep{fsps}, {\sc Astropy} \citep{astropy1,astropy2}, {\sc matplotlib} \citep{matplotlib}, {\sc scipy} \citep{scipy}, {\sc ipython} \citep{ipython}, {\sc numpy} \citep{numpy}, {\sc pandas} \citep{pandas}, and {\sc emcee} \citep{foreman-mackey2013}.

\section*{Data Availability}

The data underlying this article are available from the Canadian Advanced Network for Astronomical Research (CANFAR), at {https://www.canfar.net/citation/landing?doi=20.0009}, DOI:10.11570/20.0009. The GOGREEN Public data release (DR1) is available at {https://www.canfar.net/storage/list/GOGREEN/DR1}, as described in Balogh et al. (2020; submitted). More information on this release is available at {http://gogreensurvey.ca/data-releases/data-packages/gogreen-and-gclass-first-data-release/}.


\bibliographystyle{mnras}
\bibliography{paper}

\begin{thebibliography}{}
\makeatletter
\relax
\def\mn@urlcharsother{\let\do\@makeother \do\$\do\&\do\#\do\^\do\_\do\%\do\~}
\def\mn@doi{\begingroup\mn@urlcharsother \@ifnextchar [ {\mn@doi@}
  {\mn@doi@[]}}
\def\mn@doi@[#1]#2{\def\@tempa{#1}\ifx\@tempa\@empty \href
  {http://dx.doi.org/#2} {doi:#2}\else \href {http://dx.doi.org/#2} {#1}\fi
  \endgroup}
\def\mn@eprint#1#2{\mn@eprint@#1:#2::\@nil}
\def\mn@eprint@arXiv#1{\href {http://arxiv.org/abs/#1} {{\tt arXiv:#1}}}
\def\mn@eprint@dblp#1{\href {http://dblp.uni-trier.de/rec/bibtex/#1.xml}
  {dblp:#1}}
\def\mn@eprint@#1:#2:#3:#4\@nil{\def\@tempa {#1}\def\@tempb {#2}\def\@tempc
  {#3}\ifx \@tempc \@empty \let \@tempc \@tempb \let \@tempb \@tempa \fi \ifx
  \@tempb \@empty \def\@tempb {arXiv}\fi \@ifundefined
  {mn@eprint@\@tempb}{\@tempb:\@tempc}{\expandafter \expandafter \csname
  mn@eprint@\@tempb\endcsname \expandafter{\@tempc}}}

\bibitem[\protect\citeauthoryear{Baldry, Balogh, Bower, Glazebrook, Nichol,
  Bamford  \& Budavari}{Baldry et~al.}{2006}]{baldry2006}
Baldry I.~K.,  Balogh M.~L.,  Bower R.~G.,  Glazebrook K.,  Nichol R.~C.,
  Bamford S.~P.,   Budavari T.,  2006, \mn@doi [\mnras]
  {10.1111/j.1365-2966.2006.11081.x}, 373, 469

\bibitem[\protect\citeauthoryear{Balogh, Morris, Yee, Carlberg  \&
  Ellingson}{Balogh et~al.}{1999}]{balogh1999}
Balogh M.~L.,  Morris S.~L.,  Yee H. K.~C.,  Carlberg R.~G.,   Ellingson E.,
  1999, \mn@doi [\apj] {10.1086/308056}, 527, 54

\bibitem[\protect\citeauthoryear{Balogh, Navarro  \& Morris}{Balogh
  et~al.}{2000}]{balogh2000}
Balogh M.~L.,  Navarro J.~F.,   Morris S.~L.,  2000, \mn@doi [\apj]
  {10.1086/309323}, 540, 113

\bibitem[\protect\citeauthoryear{Balogh et~al.,}{Balogh
  et~al.}{2014}]{balogh2014}
Balogh M.~L.,  et~al., 2014, \mn@doi [\mnras] {10.1093/mnras/stu1332}, 443,
  2679

\bibitem[\protect\citeauthoryear{Balogh et~al.,}{Balogh
  et~al.}{2016}]{balogh2016}
Balogh M.~L.,  et~al., 2016, \mn@doi [\mnras] {10.1093/mnras/stv2949}, 456,
  4364

\bibitem[\protect\citeauthoryear{Balogh et~al.,}{Balogh
  et~al.}{2017}]{balogh2017}
Balogh M.~L.,  et~al., 2017, \mn@doi [\mnras] {10.1093/mnras/stx1370}, 470,
  4168

\bibitem[\protect\citeauthoryear{Barro et~al.,}{Barro et~al.}{2017}]{barro2017}
Barro G.,  et~al., 2017, \mn@doi [\apj] {10.3847/1538-4357/aa6b05}, 840, 47

\bibitem[\protect\citeauthoryear{Beifiori et~al.,}{Beifiori
  et~al.}{2017}]{Beifiori2017}
Beifiori A.,  et~al., 2017, \mn@doi [\apj] {10.3847/1538-4357/aa8368}, 846, 120

\bibitem[\protect\citeauthoryear{Bell et~al.,}{Bell et~al.}{2005}]{bell2005}
Bell E.~F.,  et~al., 2005, \mn@doi [\apj] {10.1086/429552}, 625, 23

\bibitem[\protect\citeauthoryear{Belli, Newman  \& Ellis}{Belli
  et~al.}{2015}]{belli2015}
Belli S.,  Newman A.~B.,   Ellis R.~S.,  2015, \mn@doi [\apj]
  {10.1088/0004-637X/799/2/206}, 799, 206

\bibitem[\protect\citeauthoryear{Belli, Newman  \& Ellis}{Belli
  et~al.}{2019}]{belli2019}
Belli S.,  Newman A.~B.,   Ellis R.~S.,  2019, \mn@doi [\apj]
  {10.3847/1538-4357/ab07af}, 874, 17

\bibitem[\protect\citeauthoryear{Bender, Ziegler  \& Bruzual}{Bender
  et~al.}{1996}]{bender1996}
Bender R.,  Ziegler B.,   Bruzual G.,  1996, \mn@doi [\apj] {10.1086/310071},
  463, L51

\bibitem[\protect\citeauthoryear{Bernardi, Renzini, {da Costa}, Wegner, Alonso,
  Pellegrini, Rit{\'e}  \& Willmer}{Bernardi et~al.}{1998}]{bernardi1998}
Bernardi M.,  Renzini A.,  {da Costa} L.~N.,  Wegner G.,  Alonso M.~V.,
  Pellegrini P.~S.,  Rit{\'e} C.,   Willmer C. N.~A.,  1998, \mn@doi [\apj]
  {10.1086/311742}, 508, L143

\bibitem[\protect\citeauthoryear{Bernardi, Nichol, Sheth, Miller  \&
  Brinkmann}{Bernardi et~al.}{2006}]{bernardi2006}
Bernardi M.,  Nichol R.~C.,  Sheth R.~K.,  Miller C.~J.,   Brinkmann J.,  2006,
  \mn@doi [\aj] {10.1086/499522}, 131, 1288

\bibitem[\protect\citeauthoryear{Bower, Benson, Malbon, Helly, Frenk, Baugh,
  Cole  \& Lacey}{Bower et~al.}{2006}]{bower2006}
Bower R.~G.,  Benson A.~J.,  Malbon R.,  Helly J.~C.,  Frenk C.~S.,  Baugh
  C.~M.,  Cole S.,   Lacey C.~G.,  2006, \mn@doi [\mnras]
  {10.1111/j.1365-2966.2006.10519.x}, 370, 645

\bibitem[\protect\citeauthoryear{Bower, Benson  \& Crain}{Bower
  et~al.}{2012}]{bower2012}
Bower R.~G.,  Benson A.~J.,   Crain R.~A.,  2012, \mn@doi [\mnras]
  {10.1111/j.1365-2966.2012.20516.x}, 422, 2816

\bibitem[\protect\citeauthoryear{Brammer, {van Dokkum}  \& Coppi}{Brammer
  et~al.}{2008}]{brammer2008}
Brammer G.~B.,  {van Dokkum} P.~G.,   Coppi P.,  2008, \mn@doi [\apj]
  {10.1086/591786}, 686, 1503

\bibitem[\protect\citeauthoryear{Brammer et~al.,}{Brammer
  et~al.}{2011}]{brammer2011}
Brammer G.~B.,  et~al., 2011, \mn@doi [\apj] {10.1088/0004-637X/739/1/24}, 739,
  24

\bibitem[\protect\citeauthoryear{Brinchmann, Charlot, White, Tremonti,
  Kauffmann, Heckman  \& Brinkmann}{Brinchmann et~al.}{2004}]{brinchmann2004}
Brinchmann J.,  Charlot S.,  White S. D.~M.,  Tremonti C.,  Kauffmann G.,
  Heckman T.,   Brinkmann J.,  2004, \mn@doi [\mnras]
  {10.1111/j.1365-2966.2004.07881.x}, 351, 1151

\bibitem[\protect\citeauthoryear{Brodwin et~al.,}{Brodwin
  et~al.}{2010}]{brodwin2010}
Brodwin M.,  et~al., 2010, \mn@doi [\apj] {10.1088/0004-637X/721/1/90}, 721, 90

\bibitem[\protect\citeauthoryear{Bruzual \& Charlot}{Bruzual \&
  Charlot}{2003}]{bruzual2003}
Bruzual G.,  Charlot S.,  2003, \mn@doi [\mnras]
  {10.1046/j.1365-8711.2003.06897.x}, 344, 1000

\bibitem[\protect\citeauthoryear{Calzetti, Armus, Bohlin, Kinney, Koornneef  \&
  Storchi-Bergmann}{Calzetti et~al.}{2000}]{calzetti2000}
Calzetti D.,  Armus L.,  Bohlin R.~C.,  Kinney A.~L.,  Koornneef J.,
  Storchi-Bergmann T.,  2000, \mn@doi [\apj] {10.1086/308692}, 533, 682

\bibitem[\protect\citeauthoryear{Cardelli, Clayton  \& Mathis}{Cardelli
  et~al.}{1989}]{cardelli1989}
Cardelli J.~A.,  Clayton G.~C.,   Mathis J.~S.,  1989, \mn@doi [\apj]
  {10.1086/167900}, 345, 245

\bibitem[\protect\citeauthoryear{Carnall, McLure, Dunlop  \& Dav{\'e}}{Carnall
  et~al.}{2018}]{carnall2018}
Carnall A.~C.,  McLure R.~J.,  Dunlop J.~S.,   Dav{\'e} R.,  2018, \mn@doi
  [\mnras] {10.1093/mnras/sty2169}, 480, 4379

\bibitem[\protect\citeauthoryear{Carnall et~al.,}{Carnall
  et~al.}{2019a}]{carnall2019b}
Carnall A.~C.,  et~al., 2019a, \mn@doi [\mnras] {10.1093/mnras/stz2544}, 490,
  417

\bibitem[\protect\citeauthoryear{Carnall, Leja, Johnson, McLure, Dunlop  \&
  Conroy}{Carnall et~al.}{2019b}]{carnall2019a}
Carnall A.~C.,  Leja J.,  Johnson B.~D.,  McLure R.~J.,  Dunlop J.~S.,   Conroy
  C.,  2019b, \mn@doi [\apj] {10.3847/1538-4357/ab04a2}, 873, 44

\bibitem[\protect\citeauthoryear{Chabrier}{Chabrier}{2003}]{chabrier2003}
Chabrier G.,  2003, \mn@doi [\pasp] {10.1086/376392}, 115, 763

\bibitem[\protect\citeauthoryear{Chan et~al.,}{Chan et~al.}{2019}]{chan2019}
Chan J. C.~C.,  et~al., 2019, arXiv e-prints

\bibitem[\protect\citeauthoryear{Chauke et~al.,}{Chauke
  et~al.}{2018}]{chauke2018}
Chauke P.,  et~al., 2018, \mn@doi [\apj] {10.3847/1538-4357/aac324}, 861, 13

\bibitem[\protect\citeauthoryear{Choi, Conroy, Moustakas, Graves, Holden,
  Brodwin, Brown  \& {van Dokkum}}{Choi et~al.}{2014}]{choi2014}
Choi J.,  Conroy C.,  Moustakas J.,  Graves G.~J.,  Holden B.~P.,  Brodwin M.,
  Brown M. J.~I.,   {van Dokkum} P.~G.,  2014, \mn@doi [\apj]
  {10.1088/0004-637X/792/2/95}, 792, 95

\bibitem[\protect\citeauthoryear{Choi, Dotter, Conroy, Cantiello, Paxton  \&
  Johnson}{Choi et~al.}{2016}]{choi2016}
Choi J.,  Dotter A.,  Conroy C.,  Cantiello M.,  Paxton B.,   Johnson B.~D.,
  2016, \mn@doi [\apj] {10.3847/0004-637X/823/2/102}, 823, 102

\bibitem[\protect\citeauthoryear{Cimatti, Daddi  \& Renzini}{Cimatti
  et~al.}{2006}]{cimatti2006}
Cimatti A.,  Daddi E.,   Renzini A.,  2006, \mn@doi [\aap]
  {10.1051/0004-6361:20065155}, 453, L29

\bibitem[\protect\citeauthoryear{Citro, Pozzetti, Moresco  \& Cimatti}{Citro
  et~al.}{2016}]{citro2016}
Citro A.,  Pozzetti L.,  Moresco M.,   Cimatti A.,  2016, \mn@doi [\aap]
  {10.1051/0004-6361/201527772}, 592, A19

\bibitem[\protect\citeauthoryear{Clemens, Bressan, Nikolic, Alexander, Annibali
   \& Rampazzo}{Clemens et~al.}{2006}]{clemens2006}
Clemens M.~S.,  Bressan A.,  Nikolic B.,  Alexander P.,  Annibali F.,
  Rampazzo R.,  2006, \mn@doi [\mnras] {10.1111/j.1365-2966.2006.10530.x}, 370,
  702

\bibitem[\protect\citeauthoryear{Conroy}{Conroy}{2013}]{conroy2013a}
Conroy C.,  2013, \mn@doi [\araa] {10.1146/annurev-astro-082812-141017}, 51,
  393

\bibitem[\protect\citeauthoryear{Conroy \& Gunn}{Conroy \& Gunn}{2010}]{fsps}
Conroy C.,  Gunn J.~E.,  2010, {{FSPS}}: {{Flexible}} Stellar Population
  Synthesis, \url {https://ui.adsabs.harvard.edu/abs/2010ascl.soft10043C}

\bibitem[\protect\citeauthoryear{Conroy, Gunn  \& White}{Conroy
  et~al.}{2009}]{conroy2009}
Conroy C.,  Gunn J.~E.,   White M.,  2009, \mn@doi [\apj]
  {10.1088/0004-637X/699/1/486}, 699, 486

\bibitem[\protect\citeauthoryear{Conroy, Graves  \& {van Dokkum}}{Conroy
  et~al.}{2013}]{conroy2013b}
Conroy C.,  Graves G.~J.,   {van Dokkum} P.~G.,  2013, \mn@doi [\apj]
  {10.1088/0004-637X/780/1/33}, 780, 33

\bibitem[\protect\citeauthoryear{Cooper, Gallazzi, Newman  \& Yan}{Cooper
  et~al.}{2010}]{cooper2010}
Cooper M.~C.,  Gallazzi A.,  Newman J.~A.,   Yan R.,  2010, \mn@doi [\mnras]
  {10.1111/j.1365-2966.2009.16020.x}, 402, 1942

\bibitem[\protect\citeauthoryear{Cowie, Songaila, Hu  \& Cohen}{Cowie
  et~al.}{1996}]{cowie1996}
Cowie L.~L.,  Songaila A.,  Hu E.~M.,   Cohen J.~G.,  1996, \mn@doi [\aj]
  {10.1086/118058}, 112, 839

\bibitem[\protect\citeauthoryear{Croton et~al.,}{Croton
  et~al.}{2006}]{croton2006}
Croton D.~J.,  et~al., 2006, \mn@doi [\mnras]
  {10.1111/j.1365-2966.2005.09675.x}, 365, 11

\bibitem[\protect\citeauthoryear{Darvish, Mobasher, Sobral, Rettura, Scoville,
  Faisst  \& Capak}{Darvish et~al.}{2016}]{darvish2016}
Darvish B.,  Mobasher B.,  Sobral D.,  Rettura A.,  Scoville N.,  Faisst A.,
  Capak P.,  2016, \mn@doi [\apj] {10.3847/0004-637X/825/2/113}, 825, 113

\bibitem[\protect\citeauthoryear{De~Lucia, Weinmann, Poggianti,
  {Arag{\'o}n-Salamanca}  \& Zaritsky}{De~Lucia et~al.}{2012}]{delucia2012}
De~Lucia G.,  Weinmann S.,  Poggianti B.~M.,  {Arag{\'o}n-Salamanca} A.,
  Zaritsky D.,  2012, \mn@doi [\mnras] {10.1111/j.1365-2966.2012.20983.x}, 423,
  1277

\bibitem[\protect\citeauthoryear{De~Propris, Stanford, Eisenhardt, Dickinson
  \& Elston}{De~Propris et~al.}{1999}]{depropris1999}
De~Propris R.,  Stanford S.~A.,  Eisenhardt P.~R.,  Dickinson M.,   Elston R.,
  1999, \mn@doi [\aj] {10.1086/300978}, 118, 719

\bibitem[\protect\citeauthoryear{Demarco et~al.,}{Demarco
  et~al.}{2010}]{demarco2010}
Demarco R.,  et~al., 2010, \mn@doi [\apj] {10.1088/0004-637X/711/2/1185}, 711,
  1185

\bibitem[\protect\citeauthoryear{Dotter}{Dotter}{2016}]{dotter2016}
Dotter A.,  2016, \mn@doi [\apjs] {10.3847/0067-0049/222/1/8}, 222, 8

\bibitem[\protect\citeauthoryear{Dressler}{Dressler}{1980}]{dressler1980}
Dressler A.,  1980, \mn@doi [\apj] {10.1086/157753}, 236, 351

\bibitem[\protect\citeauthoryear{Dressler, Oemler, Poggianti, Gladders,
  Abramson  \& Vulcani}{Dressler et~al.}{2013}]{dressler2013}
Dressler A.,  Oemler A.,  Poggianti B.~M.,  Gladders M.~D.,  Abramson L.,
  Vulcani B.,  2013, \mn@doi [\apj] {10.1088/0004-637X/770/1/62}, 770, 62

\bibitem[\protect\citeauthoryear{Dressler, Kelson  \& E.~Abramson}{Dressler
  et~al.}{2018}]{dressler2018}
Dressler A.,  Kelson D.~D.,   E.~Abramson L.,  2018, \mn@doi [ApJ]
  {10.3847/1538-4357/aaedbe}, 869, 152

\bibitem[\protect\citeauthoryear{Ellis, Smail, Dressler, Couch, Oemler, Butcher
   \& Sharples}{Ellis et~al.}{1997}]{ellis1997}
Ellis R.~S.,  Smail I.,  Dressler A.,  Couch W.~J.,  Oemler Jr. A.,  Butcher
  H.,   Sharples R.~M.,  1997, \mn@doi [\apj] {10.1086/304261}, 483, 582

\bibitem[\protect\citeauthoryear{Emsellem et~al.,}{Emsellem
  et~al.}{2011}]{emsellem2011}
Emsellem E.,  et~al., 2011, \mn@doi [\mnras]
  {10.1111/j.1365-2966.2011.18496.x}, 414, 888

\bibitem[\protect\citeauthoryear{{Estrada-Carpenter}
  et~al.,}{{Estrada-Carpenter} et~al.}{2019}]{estrada-carpenter2019}
{Estrada-Carpenter} V.,  et~al., 2019, \mn@doi [\apj]
  {10.3847/1538-4357/aaf22e}, 870, 133

\bibitem[\protect\citeauthoryear{{Estrada-Carpenter}
  et~al.,}{{Estrada-Carpenter} et~al.}{2020}]{estrada-carpenter2020}
{Estrada-Carpenter} V.,  et~al., 2020, arXiv e-prints, p. arXiv:2005.12289

\bibitem[\protect\citeauthoryear{Faber et~al.,}{Faber et~al.}{2007}]{faber2007}
Faber S.~M.,  et~al., 2007, \mn@doi [\apj] {10.1086/519294}, 665, 265

\bibitem[\protect\citeauthoryear{Ferreras et~al.,}{Ferreras
  et~al.}{2019}]{ferreras2019}
Ferreras I.,  et~al., 2019, \mn@doi [\mnras] {10.1093/mnras/stz849}, 486, 1358

\bibitem[\protect\citeauthoryear{Foley et~al.,}{Foley et~al.}{2011}]{foley2011}
Foley R.~J.,  et~al., 2011, \mn@doi [\apj] {10.1088/0004-637X/731/2/86}, 731,
  86

\bibitem[\protect\citeauthoryear{Font et~al.,}{Font et~al.}{2008}]{font2008}
Font A.~S.,  et~al., 2008, \mn@doi [\mnras] {10.1111/j.1365-2966.2008.13698.x},
  389, 1619

\bibitem[\protect\citeauthoryear{Fontana et~al.,}{Fontana
  et~al.}{2004}]{fontana2004}
Fontana A.,  et~al., 2004, \mn@doi [\aap] {10.1051/0004-6361:20035626}, 424, 23

\bibitem[\protect\citeauthoryear{{Foreman-Mackey} et~al.,}{{Foreman-Mackey}
  et~al.}{2013}]{foreman-mackey2013}
{Foreman-Mackey} D.,  et~al., 2013, Emcee: {{The MCMC}} Hammer, \url
  {https://ui.adsabs.harvard.edu/abs/2013ascl.soft03002F}

\bibitem[\protect\citeauthoryear{Forrest et~al.,}{Forrest
  et~al.}{2020}]{forrest2020}
Forrest B.,  et~al., 2020, \mn@doi [\apj] {10.3847/2041-8213/ab5b9f}, 890, L1

\bibitem[\protect\citeauthoryear{Fujita}{Fujita}{2004}]{fujita2004}
Fujita Y.,  2004, \mn@doi [\pasj] {10.1093/pasj/56.1.29}, 56, 29

\bibitem[\protect\citeauthoryear{Galametz et~al.,}{Galametz
  et~al.}{2013}]{galametz2013}
Galametz A.,  et~al., 2013, \mn@doi [\apjs] {10.1088/0067-0049/206/2/10}, 206,
  10

\bibitem[\protect\citeauthoryear{Gallazzi, Charlot, White  \&
  Brinchmann}{Gallazzi et~al.}{2004}]{gallazzi2004}
Gallazzi A.,  Charlot S.,  White S. D.~M.,   Brinchmann J.,  2004, \mn@doi
  [IAU] {10.1017/S1743921304001036}, 2004

\bibitem[\protect\citeauthoryear{Gallazzi, Charlot, Brinchmann, White  \&
  Tremonti}{Gallazzi et~al.}{2005}]{gallazzi2005}
Gallazzi A.,  Charlot S.,  Brinchmann J.,  White S. D.~M.,   Tremonti C.~A.,
  2005, \mn@doi [\mnras] {10.1111/j.1365-2966.2005.09321.x}, 362, 41

\bibitem[\protect\citeauthoryear{Gallazzi, Bell, Zibetti, Brinchmann  \&
  Kelson}{Gallazzi et~al.}{2014}]{gallazzi2014}
Gallazzi A.,  Bell E.~F.,  Zibetti S.,  Brinchmann J.,   Kelson D.~D.,  2014,
  \mn@doi [\apj] {10.1088/0004-637X/788/1/72}, 788, 72

\bibitem[\protect\citeauthoryear{Gladders, Oemler, Dressler, Poggianti, Vulcani
   \& Abramson}{Gladders et~al.}{2013}]{gladders2013}
Gladders M.~D.,  Oemler A.,  Dressler A.,  Poggianti B.,  Vulcani B.,
  Abramson L.,  2013, \mn@doi [\apj] {10.1088/0004-637X/770/1/64}, 770, 64

\bibitem[\protect\citeauthoryear{Glazebrook et~al.,}{Glazebrook
  et~al.}{2004}]{glazebrook2004}
Glazebrook K.,  et~al., 2004, \mn@doi [\nat] {10.1038/nature02667}, 430, 181

\bibitem[\protect\citeauthoryear{Glazebrook et~al.,}{Glazebrook
  et~al.}{2017}]{glazebrook2017}
Glazebrook K.,  et~al., 2017, \mn@doi [\nat] {10.1038/nature21680}, 544, 71

\bibitem[\protect\citeauthoryear{Gobat, Rosati, Strazzullo, Rettura, Demarco
  \& Nonino}{Gobat et~al.}{2008}]{gobat2008}
Gobat R.,  Rosati P.,  Strazzullo V.,  Rettura A.,  Demarco R.,   Nonino M.,
  2008, \mn@doi [\aap] {10.1051/0004-6361:200809531}, 488, 853

\bibitem[\protect\citeauthoryear{Guglielmo et~al.,}{Guglielmo
  et~al.}{2019}]{guglielmo2019}
Guglielmo V.,  et~al., 2019, \mn@doi [\aap] {10.1051/0004-6361/201834970}, 625,
  A112

\bibitem[\protect\citeauthoryear{Gunn \& Gott}{Gunn \& Gott}{1972}]{gunn1972}
Gunn J.~E.,  Gott III J.~R.,  1972, \mn@doi [\apj] {10.1086/151605}, 176, 1

\bibitem[\protect\citeauthoryear{Han \& Han}{Han \& Han}{2018}]{han2018}
Han Y.,  Han Z.,  2018, \mn@doi [\apjs] {10.3847/1538-4365/aaeffa}, 240, 3

\bibitem[\protect\citeauthoryear{Harker, Cole, Helly, Frenk  \& Jenkins}{Harker
  et~al.}{2006}]{harker2006}
Harker G.,  Cole S.,  Helly J.,  Frenk C.,   Jenkins A.,  2006, \mn@doi
  [\mnras] {10.1111/j.1365-2966.2006.10022.x}, 367, 1039

\bibitem[\protect\citeauthoryear{Heavens, Jimenez  \& Lahav}{Heavens
  et~al.}{2000}]{heavens2000}
Heavens A.~F.,  Jimenez R.,   Lahav O.,  2000, \mn@doi [\mnras]
  {10.1046/j.1365-8711.2000.03692.x}, 317, 965

\bibitem[\protect\citeauthoryear{Heavens, Panter, Jimenez  \& Dunlop}{Heavens
  et~al.}{2004}]{heavens2004}
Heavens A.,  Panter B.,  Jimenez R.,   Dunlop J.,  2004, \mn@doi [\nat]
  {10.1038/nature02474}, 428, 625

\bibitem[\protect\citeauthoryear{Heckman}{Heckman}{1980}]{heckman1980}
Heckman T.~M.,  1980, \aap, 500, 187

\bibitem[\protect\citeauthoryear{Helly, Cole, Frenk, Baugh, Benson  \&
  Lacey}{Helly et~al.}{2003}]{helly2003}
Helly J.~C.,  Cole S.,  Frenk C.~S.,  Baugh C.~M.,  Benson A.,   Lacey C.,
  2003, \mn@doi [\mnras] {10.1046/j.1365-8711.2003.06151.x}, 338, 903

\bibitem[\protect\citeauthoryear{Hinton, Davis, Lidman, Glazebrook  \&
  Lewis}{Hinton et~al.}{2016}]{hinton2016}
Hinton S.~R.,  Davis T.~M.,  Lidman C.,  Glazebrook K.,   Lewis G.~F.,  2016,
  \mn@doi [Astronomy and Computing] {10.1016/j.ascom.2016.03.001}, 15, 61

\bibitem[\protect\citeauthoryear{Hirschmann, De~Lucia  \& Fontanot}{Hirschmann
  et~al.}{2016}]{hirschmann2016}
Hirschmann M.,  De~Lucia G.,   Fontanot F.,  2016, \mn@doi [\mnras]
  {10.1093/mnras/stw1318}, 461, 1760

\bibitem[\protect\citeauthoryear{Hogg, Cohen, Blandford  \& Pahre}{Hogg
  et~al.}{1998}]{hogg1998}
Hogg D.~W.,  Cohen J.~G.,  Blandford R.,   Pahre M.~A.,  1998, \mn@doi [\apj]
  {10.1086/306122}, 504, 622

\bibitem[\protect\citeauthoryear{Hunter}{Hunter}{2007}]{matplotlib}
Hunter J.~D.,  2007, \mn@doi [Computing in Science \& Engineering]
  {10.1109/MCSE.2007.55}, 9, 90

\bibitem[\protect\citeauthoryear{Iyer \& Gawiser}{Iyer \&
  Gawiser}{2017}]{iyer2017}
Iyer K.,  Gawiser E.,  2017, \mn@doi [\apj] {10.3847/1538-4357/aa63f0}, 838,
  127

\bibitem[\protect\citeauthoryear{Jaff{\'e} et~al.,}{Jaff{\'e}
  et~al.}{2016}]{jaffe2016}
Jaff{\'e} Y.~L.,  et~al., 2016, \mn@doi [\mnras] {10.1093/mnras/stw984}, 461,
  1202

\bibitem[\protect\citeauthoryear{Jian et~al.,}{Jian et~al.}{2017}]{jian2017}
Jian H.-Y.,  et~al., 2017, \mn@doi [\apj] {10.3847/1538-4357/aa7de2}, 845, 74

\bibitem[\protect\citeauthoryear{Jimenez, Panter, Heavens  \& Verde}{Jimenez
  et~al.}{2005}]{jimenez2005}
Jimenez R.,  Panter B.,  Heavens A.~F.,   Verde L.,  2005, \mn@doi [\mnras]
  {10.1111/j.1365-2966.2004.08469.x}, 356, 495

\bibitem[\protect\citeauthoryear{Johnson, Leja, Conroy  \& Speagle}{Johnson
  et~al.}{2019}]{johnson2019}
Johnson B.~D.,  Leja J.~L.,  Conroy C.,   Speagle J.~S.,  2019, Prospector:
  {{Stellar}} Population Inference from Spectra and {{SEDs}}, \url
  {https://ui.adsabs.harvard.edu/abs/2019ascl.soft05025J}

\bibitem[\protect\citeauthoryear{J{\o}rgensen, Chiboucas, Flint, Bergmann, Barr
   \& Davies}{J{\o}rgensen et~al.}{2006}]{jorgensen2006}
J{\o}rgensen I.,  Chiboucas K.,  Flint K.,  Bergmann M.,  Barr J.,   Davies R.,
   2006, \mn@doi [\apj] {10.1086/501348}, 639, L9

\bibitem[\protect\citeauthoryear{J{\o}rgensen, Chiboucas, Flint, Bergmann, Barr
   \& Davies}{J{\o}rgensen et~al.}{2007}]{jorgensen2007}
J{\o}rgensen I.,  Chiboucas K.,  Flint K.,  Bergmann M.,  Barr J.,   Davies R.,
   2007, \mn@doi [\apj] {10.1086/511010}, 654, L179

\bibitem[\protect\citeauthoryear{J{\o}rgensen, Chiboucas, Berkson, Smith,
  Takamiya  \& Villaume}{J{\o}rgensen et~al.}{2017}]{jorgensen2017}
J{\o}rgensen I.,  Chiboucas K.,  Berkson E.,  Smith O.,  Takamiya M.,
  Villaume A.,  2017, \mn@doi [\aj] {10.3847/1538-3881/aa96a3}, 154, 251

\bibitem[\protect\citeauthoryear{J{\o}rgensen, Chiboucas, Webb  \&
  Woodrum}{J{\o}rgensen et~al.}{2018}]{jorgensen2018}
J{\o}rgensen I.,  Chiboucas K.,  Webb K.,   Woodrum C.,  2018, \mn@doi [\aj]
  {10.3847/1538-3881/aae522}, 156, 224

\bibitem[\protect\citeauthoryear{Juneau et~al.,}{Juneau
  et~al.}{2005}]{juneau2005}
Juneau S.,  et~al., 2005, \mn@doi [\apj] {10.1086/427937}, 619, L135

\bibitem[\protect\citeauthoryear{Kauffmann et~al.,}{Kauffmann
  et~al.}{2003}]{kauffmann2003b}
Kauffmann G.,  et~al., 2003, \mn@doi [\mnras]
  {10.1046/j.1365-8711.2003.06292.x}, 341, 54

\bibitem[\protect\citeauthoryear{Kauffmann, White, Heckman, M{\'e}nard,
  Brinchmann, Charlot, Tremonti  \& Brinkmann}{Kauffmann
  et~al.}{2004}]{kauffmann2004}
Kauffmann G.,  White S. D.~M.,  Heckman T.~M.,  M{\'e}nard B.,  Brinchmann J.,
  Charlot S.,  Tremonti C.,   Brinkmann J.,  2004, \mn@doi [\mnras]
  {10.1111/j.1365-2966.2004.08117.x}, 353, 713

\bibitem[\protect\citeauthoryear{Kelson, Illingworth, {van Dokkum}  \&
  Franx}{Kelson et~al.}{2000}]{kelson2000}
Kelson D.~D.,  Illingworth G.~D.,  {van Dokkum} P.~G.,   Franx M.,  2000,
  \mn@doi [\apj] {10.1086/308440}, 531, 184

\bibitem[\protect\citeauthoryear{Kelson, Illingworth, Franx  \& {van
  Dokkum}}{Kelson et~al.}{2001}]{kelson2001}
Kelson D.~D.,  Illingworth G.~D.,  Franx M.,   {van Dokkum} P.~G.,  2001,
  \mn@doi [\apj] {10.1086/320252}, 552, L17

\bibitem[\protect\citeauthoryear{Khochfar \& Silk}{Khochfar \&
  Silk}{2009}]{khochfar2009}
Khochfar S.,  Silk J.,  2009, \mn@doi [\mnras]
  {10.1111/j.1365-2966.2009.14958.x}, 397, 506

\bibitem[\protect\citeauthoryear{Kimm et~al.,}{Kimm et~al.}{2009}]{kimm2009}
Kimm T.,  et~al., 2009, \mn@doi [\mnras] {10.1111/j.1365-2966.2009.14414.x},
  394, 1131

\bibitem[\protect\citeauthoryear{Kodama et~al.,}{Kodama
  et~al.}{2004}]{kodama2004}
Kodama T.,  et~al., 2004, \mn@doi [\mnras] {10.1111/j.1365-2966.2004.07711.x},
  350, 1005

\bibitem[\protect\citeauthoryear{Koyama et~al.,}{Koyama
  et~al.}{2013}]{koyama2013}
Koyama Y.,  et~al., 2013, \mn@doi [\mnras] {10.1093/mnras/stt1035}, 434, 423

\bibitem[\protect\citeauthoryear{Kriek, {van Dokkum}, Labb{\'e}, Franx,
  Illingworth, Marchesini  \& Quadri}{Kriek et~al.}{2009}]{kriek2009}
Kriek M.,  {van Dokkum} P.~G.,  Labb{\'e} I.,  Franx M.,  Illingworth G.~D.,
  Marchesini D.,   Quadri R.~F.,  2009, \mn@doi [\apj]
  {10.1088/0004-637X/700/1/221}, 700, 221

\bibitem[\protect\citeauthoryear{Kriek et~al.,}{Kriek et~al.}{2016}]{kriek2016}
Kriek M.,  et~al., 2016, \mn@doi [\nat] {10.1038/nature20570}, 540, 248

\bibitem[\protect\citeauthoryear{Kriek et~al.,}{Kriek et~al.}{2019}]{kriek2019}
Kriek M.,  et~al., 2019, \mn@doi [\apj] {10.3847/2041-8213/ab2e75}, 880, L31

\bibitem[\protect\citeauthoryear{Labb{\'e} et~al.,}{Labb{\'e}
  et~al.}{2005}]{labbe2005}
Labb{\'e} I.,  et~al., 2005, \mn@doi [\apj] {10.1086/430700}, 624, L81

\bibitem[\protect\citeauthoryear{Larson, Tinsley  \& Caldwell}{Larson
  et~al.}{1980}]{larson1980}
Larson R.~B.,  Tinsley B.~M.,   Caldwell C.~N.,  1980, \mn@doi [\apj]
  {10.1086/157917}, 237, 692

\bibitem[\protect\citeauthoryear{{Lee-Brown} et~al.,}{{Lee-Brown}
  et~al.}{2017}]{lee-brown2017}
{Lee-Brown} D.~B.,  et~al., 2017, \mn@doi [\apj] {10.3847/1538-4357/aa7948},
  844, 43

\bibitem[\protect\citeauthoryear{Leethochawalit, Kirby, Moran, Ellis  \&
  Treu}{Leethochawalit et~al.}{2018}]{leethochawalit2018}
Leethochawalit N.,  Kirby E.~N.,  Moran S.~M.,  Ellis R.~S.,   Treu T.,  2018,
  \mn@doi [\apj] {10.3847/1538-4357/aab26a}, 856, 15

\bibitem[\protect\citeauthoryear{Leitner}{Leitner}{2012}]{leitner2012}
Leitner S.~N.,  2012, \mn@doi [\apj] {10.1088/0004-637X/745/2/149}, 745, 149

\bibitem[\protect\citeauthoryear{Leja, Johnson, Conroy, van Dokkum  \&
  Byler}{Leja et~al.}{2017}]{leja2017}
Leja J.,  Johnson B.~D.,  Conroy C.,  van Dokkum P.~G.,   Byler N.,  2017,
  \mn@doi [\apj] {10.3847/1538-4357/aa5ffe}, 837, 170

\bibitem[\protect\citeauthoryear{Leja, Carnall, Johnson, Conroy  \&
  Speagle}{Leja et~al.}{2019a}]{leja2019a}
Leja J.,  Carnall A.~C.,  Johnson B.~D.,  Conroy C.,   Speagle J.~S.,  2019a,
  \mn@doi [\apj] {10.3847/1538-4357/ab133c}, 876, 3

\bibitem[\protect\citeauthoryear{Leja et~al.,}{Leja et~al.}{2019b}]{leja2019b}
Leja J.,  et~al., 2019b, \mn@doi [\apj] {10.3847/1538-4357/ab1d5a}, 877, 140

\bibitem[\protect\citeauthoryear{Lin et~al.,}{Lin et~al.}{2014}]{lin2014}
Lin L.,  et~al., 2014, \mn@doi [\apj] {10.1088/0004-637X/782/1/33}, 782, 33

\bibitem[\protect\citeauthoryear{Loh, Ellingson, Yee, Gilbank, Gladders  \&
  Barrientos}{Loh et~al.}{2008}]{loh2008}
Loh Y.-S.,  Ellingson E.,  Yee H. K.~C.,  Gilbank D.~G.,  Gladders M.~D.,
  Barrientos L.~F.,  2008, \mn@doi [\apj] {10.1086/587830}, 680, 214

\bibitem[\protect\citeauthoryear{Lonsdale et~al.,}{Lonsdale
  et~al.}{2003}]{lonsdale2003}
Lonsdale C.~J.,  et~al., 2003, \mn@doi [\pasp] {10.1086/376850}, 115, 897

\bibitem[\protect\citeauthoryear{MacArthur, Gonz{\'a}lez  \&
  Courteau}{MacArthur et~al.}{2009}]{macarthur2009}
MacArthur L.~A.,  Gonz{\'a}lez J.~J.,   Courteau S.,  2009, \mn@doi [\mnras]
  {10.1111/j.1365-2966.2009.14519.x}, 395, 28

\bibitem[\protect\citeauthoryear{Maier et~al.,}{Maier et~al.}{2016}]{maier2016}
Maier C.,  et~al., 2016, \mn@doi [\aap] {10.1051/0004-6361/201628223}, 590,
  A108

\bibitem[\protect\citeauthoryear{Maraston}{Maraston}{2005}]{maraston2005}
Maraston C.,  2005, \mn@doi [\mnras] {10.1111/j.1365-2966.2005.09270.x}, 362,
  799

\bibitem[\protect\citeauthoryear{Mauduit et~al.,}{Mauduit
  et~al.}{2012}]{mauduit2012}
Mauduit J.-C.,  et~al., 2012, \mn@doi [\pasp] {10.1086/666945}, 124, 714

\bibitem[\protect\citeauthoryear{McDermid et~al.,}{McDermid
  et~al.}{2015}]{mcdermid2015}
McDermid R.~M.,  et~al., 2015, \mn@doi [\mnras] {10.1093/mnras/stv105}, 448,
  3484

\bibitem[\protect\citeauthoryear{McGee, Balogh, Bower, Font  \& McCarthy}{McGee
  et~al.}{2009}]{mcgee2009}
McGee S.~L.,  Balogh M.~L.,  Bower R.~G.,  Font A.~S.,   McCarthy I.~G.,  2009,
  \mn@doi [\mnras] {10.1111/j.1365-2966.2009.15507.x}, 400, 937

\bibitem[\protect\citeauthoryear{McGee, Balogh, Wilman, Bower, Mulchaey, Parker
   \& Oemler}{McGee et~al.}{2011}]{mcgee2011}
McGee S.~L.,  Balogh M.~L.,  Wilman D.~J.,  Bower R.~G.,  Mulchaey J.~S.,
  Parker L.~C.,   Oemler A.,  2011, \mn@doi [\mnras]
  {10.1111/j.1365-2966.2010.18189.x}, 413, 996

\bibitem[\protect\citeauthoryear{McGee, Bower  \& Balogh}{McGee
  et~al.}{2014}]{mcgee2014}
McGee S.~L.,  Bower R.~G.,   Balogh M.~L.,  2014, \mn@doi [\mnras]
  {10.1093/mnrasl/slu066}, 442, L105

\bibitem[\protect\citeauthoryear{McKinney}{McKinney}{2010}]{pandas}
McKinney W.,  2010, in {van der Walt} S.,  Millman J.,  eds, Proceedings of the
  9th {{Python}} in {{Science Conference}}. pp 56--61,
  \mn@doi{10.25080/Majora-92bf1922-00a}, \url
  {https://doi.org/10.25080/Majora-92bf1922-00a}

\bibitem[\protect\citeauthoryear{McLure et~al.,}{McLure
  et~al.}{2018}]{mclure2018}
McLure R.~J.,  et~al., 2018, \mn@doi [\mnras] {10.1093/mnras/sty1213}

\bibitem[\protect\citeauthoryear{Mehta et~al.,}{Mehta et~al.}{2018}]{mehta2018}
Mehta V.,  et~al., 2018, \mn@doi [\apjs] {10.3847/1538-4365/aab60c}, 235, 36

\bibitem[\protect\citeauthoryear{Merritt}{Merritt}{1983}]{merritt1983}
Merritt D.,  1983, \mn@doi [\apj] {10.1086/160571}, 264, 24

\bibitem[\protect\citeauthoryear{Moore, Katz, Lake, Dressler  \& Oemler}{Moore
  et~al.}{1996}]{moore1996}
Moore B.,  Katz N.,  Lake G.,  Dressler A.,   Oemler A.,  1996, \mn@doi [\nat]
  {10.1038/379613a0}, 379, 613

\bibitem[\protect\citeauthoryear{Moran, Ellis, Treu, Smith, Rich  \&
  Smail}{Moran et~al.}{2007}]{moran2007}
Moran S.~M.,  Ellis R.~S.,  Treu T.,  Smith G.~P.,  Rich R.~M.,   Smail I.,
  2007, \mn@doi [\apj] {10.1086/522303}, 671, 1503

\bibitem[\protect\citeauthoryear{Morishita et~al.,}{Morishita
  et~al.}{2018}]{morishita2018}
Morishita T.,  et~al., 2018, \mn@doi [\apj] {10.3847/2041-8213/aab493}, 856, L4

\bibitem[\protect\citeauthoryear{Morishita et~al.,}{Morishita
  et~al.}{2019}]{morishita2019}
Morishita T.,  et~al., 2019, \mn@doi [\apj] {10.3847/1538-4357/ab1d53}, 877,
  141

\bibitem[\protect\citeauthoryear{Muzzin, Marchesini, {van Dokkum}, Labb{\'e},
  Kriek  \& Franx}{Muzzin et~al.}{2009}]{muzzin2009}
Muzzin A.,  Marchesini D.,  {van Dokkum} P.~G.,  Labb{\'e} I.,  Kriek M.,
  Franx M.,  2009, \mn@doi [\apj] {10.1088/0004-637X/701/2/1839}, 701, 1839

\bibitem[\protect\citeauthoryear{Muzzin et~al.,}{Muzzin
  et~al.}{2012}]{muzzin2012}
Muzzin A.,  et~al., 2012, \mn@doi [\apj] {10.1088/0004-637X/746/2/188}, 746,
  188

\bibitem[\protect\citeauthoryear{Muzzin et~al.,}{Muzzin
  et~al.}{2013a}]{muzzin2013b}
Muzzin A.,  et~al., 2013a, \mn@doi [\apjs] {10.1088/0067-0049/206/1/8}, 206, 8

\bibitem[\protect\citeauthoryear{Muzzin, Wilson, Demarco, Lidman, Nantais,
  Hoekstra, Yee  \& Rettura}{Muzzin et~al.}{2013b}]{muzzin2013a}
Muzzin A.,  Wilson G.,  Demarco R.,  Lidman C.,  Nantais J.,  Hoekstra H.,  Yee
  H. K.~C.,   Rettura A.,  2013b, \mn@doi [\apj] {10.1088/0004-637X/767/1/39},
  767, 39

\bibitem[\protect\citeauthoryear{Muzzin et~al.,}{Muzzin
  et~al.}{2013c}]{muzzin2013c}
Muzzin A.,  et~al., 2013c, \mn@doi [\apj] {10.1088/0004-637X/777/1/18}, 777, 18

\bibitem[\protect\citeauthoryear{Muzzin et~al.,}{Muzzin
  et~al.}{2014}]{muzzin2014}
Muzzin A.,  et~al., 2014, \mn@doi [\apj] {10.1088/0004-637X/796/1/65}, 796, 65

\bibitem[\protect\citeauthoryear{Nantais et~al.,}{Nantais
  et~al.}{2016}]{nantais2016}
Nantais J.~B.,  et~al., 2016, \mn@doi [\aap] {10.1051/0004-6361/201628663},
  592, A161

\bibitem[\protect\citeauthoryear{Nantais et~al.,}{Nantais
  et~al.}{2017}]{nantais2017}
Nantais J.~B.,  et~al., 2017, \mn@doi [\mnras] {10.1093/mnrasl/slw224}, 465,
  L104

\bibitem[\protect\citeauthoryear{Neistein, {van den Bosch}  \& Dekel}{Neistein
  et~al.}{2006}]{neistein2006}
Neistein E.,  {van den Bosch} F.~C.,   Dekel A.,  2006, \mn@doi [\mnras]
  {10.1111/j.1365-2966.2006.10918.x}, 372, 933

\bibitem[\protect\citeauthoryear{Nelan, Smith, Hudson, Wegner, Lucey, Moore,
  Quinney  \& Suntzeff}{Nelan et~al.}{2005}]{nelan2005}
Nelan J.~E.,  Smith R.~J.,  Hudson M.~J.,  Wegner G.~A.,  Lucey J.~R.,  Moore
  S. A.~W.,  Quinney S.~J.,   Suntzeff N.~B.,  2005, \mn@doi [\apj]
  {10.1086/431962}, 632, 137

\bibitem[\protect\citeauthoryear{Newman, Ellis, Andreon, Treu, Raichoor  \&
  Trinchieri}{Newman et~al.}{2014}]{newman2014}
Newman A.~B.,  Ellis R.~S.,  Andreon S.,  Treu T.,  Raichoor A.,   Trinchieri
  G.,  2014, \mn@doi [\apj] {10.1088/0004-637X/788/1/51}, 788, 51

\bibitem[\protect\citeauthoryear{Odekon et~al.,}{Odekon
  et~al.}{2016}]{odekon2016}
Odekon M.~C.,  et~al., 2016, \mn@doi [\apj] {10.3847/0004-637X/824/2/110}, 824,
  110

\bibitem[\protect\citeauthoryear{Old et~al.,}{Old et~al.}{2020}]{old2020}
Old L.~J.,  et~al., 2020, \mn@doi [\mnras] {10.1093/mnras/staa579}, 493, 5987

\bibitem[\protect\citeauthoryear{Onodera et~al.,}{Onodera
  et~al.}{2012}]{onodera2012}
Onodera M.,  et~al., 2012, \mn@doi [\apj] {10.1088/0004-637X/755/1/26}, 755, 26

\bibitem[\protect\citeauthoryear{Onodera et~al.,}{Onodera
  et~al.}{2015}]{onodera2015}
Onodera M.,  et~al., 2015, \mn@doi [\apj] {10.1088/0004-637X/808/2/161}, 808,
  161

\bibitem[\protect\citeauthoryear{Pacifici et~al.,}{Pacifici
  et~al.}{2016}]{pacifici2016}
Pacifici C.,  et~al., 2016, \mn@doi [\apj] {10.3847/0004-637X/832/1/79}, 832,
  79

\bibitem[\protect\citeauthoryear{Pallero, G{\'o}mez, Padilla, {Torres-Flores},
  Demarco, Cerulo  \& {Olave-Rojas}}{Pallero et~al.}{2019}]{pallero2019}
Pallero D.,  G{\'o}mez F.~A.,  Padilla N.~D.,  {Torres-Flores} S.,  Demarco R.,
   Cerulo P.,   {Olave-Rojas} D.,  2019, \mn@doi [\mnras]
  {10.1093/mnras/stz1745}, 488, 847

\bibitem[\protect\citeauthoryear{Panter, Heavens  \& Jimenez}{Panter
  et~al.}{2003}]{panter2003}
Panter B.,  Heavens A.~F.,   Jimenez R.,  2003, \mn@doi [\mnras]
  {10.1046/j.1365-8711.2003.06722.x}, 343, 1145

\bibitem[\protect\citeauthoryear{Panter, Jimenez, Heavens  \& Charlot}{Panter
  et~al.}{2007}]{panter2007}
Panter B.,  Jimenez R.,  Heavens A.~F.,   Charlot S.,  2007, \mn@doi [\mnras]
  {10.1111/j.1365-2966.2007.11909.x}, 378, 1550

\bibitem[\protect\citeauthoryear{Panter, Jimenez, Heavens  \& Charlot}{Panter
  et~al.}{2008}]{panter2008}
Panter B.,  Jimenez R.,  Heavens A.~F.,   Charlot S.,  2008, \mn@doi [\mnras]
  {10.1111/j.1365-2966.2008.13981.x}, 391, 1117

\bibitem[\protect\citeauthoryear{Papovich, Dickinson  \& Ferguson}{Papovich
  et~al.}{2001}]{papovich2001}
Papovich C.,  Dickinson M.,   Ferguson H.~C.,  2001, \mn@doi [\apj]
  {10.1086/322412}, 559, 620

\bibitem[\protect\citeauthoryear{Patel, Holden, Kelson, Franx, {van der Wel}
  \& Illingworth}{Patel et~al.}{2012}]{patel2012}
Patel S.~G.,  Holden B.~P.,  Kelson D.~D.,  Franx M.,  {van der Wel} A.,
  Illingworth G.~D.,  2012, \mn@doi [\apj] {10.1088/2041-8205/748/2/L27}, 748,
  L27

\bibitem[\protect\citeauthoryear{{Paulino-Afonso}, Sobral, Darvish, Ribeiro,
  Smail, Best, Stroe  \& Cairns}{{Paulino-Afonso}
  et~al.}{2020}]{paulino-afonso2020}
{Paulino-Afonso} A.,  Sobral D.,  Darvish B.,  Ribeiro B.,  Smail I.,  Best P.,
   Stroe A.,   Cairns J.,  2020, \mn@doi [\aap] {10.1051/0004-6361/201834244},
  633, A70

\bibitem[\protect\citeauthoryear{Paxton, Bildsten, Dotter, Herwig, Lesaffre  \&
  Timmes}{Paxton et~al.}{2011}]{paxton2011}
Paxton B.,  Bildsten L.,  Dotter A.,  Herwig F.,  Lesaffre P.,   Timmes F.,
  2011, \mn@doi [\apjs] {10.1088/0067-0049/192/1/3}, 192, 3

\bibitem[\protect\citeauthoryear{Paxton et~al.,}{Paxton
  et~al.}{2013}]{paxton2013}
Paxton B.,  et~al., 2013, \mn@doi [\apjs] {10.1088/0067-0049/208/1/4}, 208, 4

\bibitem[\protect\citeauthoryear{Paxton et~al.,}{Paxton
  et~al.}{2015}]{paxton2015}
Paxton B.,  et~al., 2015, \mn@doi [\apjs] {10.1088/0067-0049/220/1/15}, 220, 15

\bibitem[\protect\citeauthoryear{Paxton et~al.,}{Paxton
  et~al.}{2018}]{paxton2018}
Paxton B.,  et~al., 2018, \mn@doi [\apjs] {10.3847/1538-4365/aaa5a8}, 234, 34

\bibitem[\protect\citeauthoryear{Peng et~al.,}{Peng et~al.}{2010}]{peng2010}
Peng Y.-j.,  et~al., 2010, \mn@doi [\apj] {10.1088/0004-637X/721/1/193}, 721,
  193

\bibitem[\protect\citeauthoryear{Peng, Lilly, Renzini  \& Carollo}{Peng
  et~al.}{2012}]{peng2012}
Peng Y.-j.,  Lilly S.~J.,  Renzini A.,   Carollo M.,  2012, \mn@doi [\apj]
  {10.1088/0004-637X/757/1/4}, 757, 4

\bibitem[\protect\citeauthoryear{Peng, Maiolino  \& Cochrane}{Peng
  et~al.}{2015}]{peng2015}
Peng Y.,  Maiolino R.,   Cochrane R.,  2015, \mn@doi [\nat]
  {10.1038/nature14439}, 521, 192

\bibitem[\protect\citeauthoryear{Pentericci et~al.,}{Pentericci
  et~al.}{2018}]{pentericci2018}
Pentericci L.,  et~al., 2018, \mn@doi [\aap] {10.1051/0004-6361/201833047},
  616, A174

\bibitem[\protect\citeauthoryear{P{\'e}rez \& Granger}{P{\'e}rez \&
  Granger}{2007}]{ipython}
P{\'e}rez F.,  Granger B.~E.,  2007, \mn@doi [Computing in Science \&
  Engineering] {10.1109/MCSE.2007.53}, 9, 21

\bibitem[\protect\citeauthoryear{{Pintos-Castro}, Yee, Muzzin, Old  \&
  Wilson}{{Pintos-Castro} et~al.}{2019}]{pintos-castro2019}
{Pintos-Castro} I.,  Yee H. K.~C.,  Muzzin A.,  Old L.,   Wilson G.,  2019,
  \mn@doi [\apj] {10.3847/1538-4357/ab14ee}, 876, 40

\bibitem[\protect\citeauthoryear{Poggianti, Smail, Dressler, Couch, Barger,
  Butcher, Ellis  \& Oemler}{Poggianti et~al.}{1999}]{poggianti1999}
Poggianti B.~M.,  Smail I.,  Dressler A.,  Couch W.~J.,  Barger A.~J.,  Butcher
  H.,  Ellis R.~S.,   Oemler Jr. A.,  1999, \mn@doi [\apj] {10.1086/307322},
  518, 576

\bibitem[\protect\citeauthoryear{Poggianti et~al.,}{Poggianti
  et~al.}{2006}]{poggianti2006}
Poggianti B.~M.,  et~al., 2006, \mn@doi [\apj] {10.1086/500666}, 642, 188

\bibitem[\protect\citeauthoryear{Poggianti et~al.,}{Poggianti
  et~al.}{2008}]{poggianti2008}
Poggianti B.~M.,  et~al., 2008, \mn@doi [\apj] {10.1086/589936}, 684, 888

\bibitem[\protect\citeauthoryear{Poggianti et~al.,}{Poggianti
  et~al.}{2009}]{poggianti2009}
Poggianti B.~M.,  et~al., 2009, \mn@doi [\apj] {10.1088/0004-637X/693/1/112},
  693, 112

\bibitem[\protect\citeauthoryear{Popesso et~al.,}{Popesso
  et~al.}{2011}]{popesso2011}
Popesso P.,  et~al., 2011, \mn@doi [\aap] {10.1051/0004-6361/201015672}, 532,
  A145

\bibitem[\protect\citeauthoryear{Raichoor et~al.,}{Raichoor
  et~al.}{2011}]{raichoor2011}
Raichoor A.,  et~al., 2011, \mn@doi [\apj] {10.1088/0004-637X/732/1/12}, 732,
  12

\bibitem[\protect\citeauthoryear{Renzini}{Renzini}{2006}]{renzini2006}
Renzini A.,  2006, \mn@doi [\araa] {10.1146/annurev.astro.44.051905.092450},
  44, 141

\bibitem[\protect\citeauthoryear{Renzini}{Renzini}{2016}]{renzini2016}
Renzini A.,  2016, \mn@doi [\mnras] {10.1093/mnrasl/slw066}, 460, L45

\bibitem[\protect\citeauthoryear{Rettura et~al.,}{Rettura
  et~al.}{2010}]{rettura2010}
Rettura A.,  et~al., 2010, \mn@doi [\apj] {10.1088/0004-637X/709/1/512}, 709,
  512

\bibitem[\protect\citeauthoryear{Rettura et~al.,}{Rettura
  et~al.}{2011}]{rettura2011}
Rettura A.,  et~al., 2011, \mn@doi [\apj] {10.1088/0004-637X/732/2/94}, 732, 94

\bibitem[\protect\citeauthoryear{Rudnick et~al.,}{Rudnick
  et~al.}{2017}]{rudnick2017}
Rudnick G.,  et~al., 2017, \mn@doi [\apj] {10.3847/1538-4357/aa866c}, 850, 181

\bibitem[\protect\citeauthoryear{Saglia et~al.,}{Saglia
  et~al.}{2010}]{saglia2010}
Saglia R.~P.,  et~al., 2010, \mn@doi [\aap] {10.1051/0004-6361/201014703}, 524,
  A6

\bibitem[\protect\citeauthoryear{{S{\'a}nchez-Bl{\'a}zquez}
  et~al.,}{{S{\'a}nchez-Bl{\'a}zquez} et~al.}{2009}]{sanchez-blazquez2009}
{S{\'a}nchez-Bl{\'a}zquez} P.,  et~al., 2009, \mn@doi [\aap]
  {10.1051/0004-6361/200811355}, 499, 47

\bibitem[\protect\citeauthoryear{{S{\'a}nchez-Bl{\'a}zquez}, Ocvirk, Gibson,
  P{\'e}rez  \& Peletier}{{S{\'a}nchez-Bl{\'a}zquez}
  et~al.}{2011}]{sanchez-blazquez2011}
{S{\'a}nchez-Bl{\'a}zquez} P.,  Ocvirk P.,  Gibson B.~K.,  P{\'e}rez I.,
  Peletier R.~F.,  2011, \mn@doi [\mnras] {10.1111/j.1365-2966.2011.18749.x},
  415, 709

\bibitem[\protect\citeauthoryear{{S{\'a}nchez{\textendash}Bl{\'a}zquez},
  Gorgas, Cardiel  \& Gonz{\'a}lez}{{S{\'a}nchez{\textendash}Bl{\'a}zquez}
  et~al.}{2006}]{sanchez-blazquez2006}
{S{\'a}nchez{\textendash}Bl{\'a}zquez} P.,  Gorgas J.,  Cardiel N.,
  Gonz{\'a}lez J.~J.,  2006, \mn@doi [\aap] {10.1051/0004-6361:20064842}, 457,
  787

\bibitem[\protect\citeauthoryear{Sanders et~al.,}{Sanders
  et~al.}{2007}]{sanders2007}
Sanders D.~B.,  et~al., 2007, \mn@doi [\apjs] {10.1086/517885}, 172, 86

\bibitem[\protect\citeauthoryear{Saracco, Gargiulo, Ciocca  \&
  Marchesini}{Saracco et~al.}{2017}]{saracco2017}
Saracco P.,  Gargiulo A.,  Ciocca F.,   Marchesini D.,  2017, \mn@doi [\aap]
  {10.1051/0004-6361/201628866}, 597, A122

\bibitem[\protect\citeauthoryear{Saracco, Gargiulo, La~Barbera, Annunziatella
  \& Marchesini}{Saracco et~al.}{2020}]{saracco2020}
Saracco P.,  Gargiulo A.,  La~Barbera F.,  Annunziatella M.,   Marchesini D.,
  2020, \mn@doi [\mnras] {10.1093/mnras/stz3109}, 491, 1777

\bibitem[\protect\citeauthoryear{Schawinski et~al.,}{Schawinski
  et~al.}{2014}]{schawinski2014}
Schawinski K.,  et~al., 2014, \mn@doi [\mnras] {10.1093/mnras/stu327}, 440, 889

\bibitem[\protect\citeauthoryear{Schreiber et~al.,}{Schreiber
  et~al.}{2015}]{schreiber2015}
Schreiber C.,  et~al., 2015, \mn@doi [\aap] {10.1051/0004-6361/201425017}, 575,
  A74

\bibitem[\protect\citeauthoryear{Schreiber et~al.,}{Schreiber
  et~al.}{2018a}]{schreiber2018a}
Schreiber C.,  et~al., 2018a, \mn@doi [\aap] {10.1051/0004-6361/201731917},
  611, A22

\bibitem[\protect\citeauthoryear{Schreiber et~al.,}{Schreiber
  et~al.}{2018b}]{schreiber2018b}
Schreiber C.,  et~al., 2018b, \mn@doi [\aap] {10.1051/0004-6361/201833070},
  618, A85

\bibitem[\protect\citeauthoryear{Singh et~al.,}{Singh et~al.}{2013}]{singh2013}
Singh R.,  et~al., 2013, \mn@doi [\aap] {10.1051/0004-6361/201322062}, 558, A43

\bibitem[\protect\citeauthoryear{Snyder et~al.,}{Snyder
  et~al.}{2012}]{snyder2012}
Snyder G.~F.,  et~al., 2012, \mn@doi [\apj] {10.1088/0004-637X/756/2/114}, 756,
  114

\bibitem[\protect\citeauthoryear{Springel}{Springel}{2005}]{springel2005}
Springel V.,  2005, \mn@doi [\mnras] {10.1111/j.1365-2966.2005.09655.x}, 364,
  1105

\bibitem[\protect\citeauthoryear{Stalder et~al.,}{Stalder
  et~al.}{2013}]{stalder2013}
Stalder B.,  et~al., 2013, \mn@doi [\apj] {10.1088/0004-637X/763/2/93}, 763, 93

\bibitem[\protect\citeauthoryear{Stanford, Eisenhardt  \& Dickinson}{Stanford
  et~al.}{1998}]{stanford1998}
Stanford S.~A.,  Eisenhardt P.~R.,   Dickinson M.,  1998, \mn@doi [\apj]
  {10.1086/305050}, 492, 461

\bibitem[\protect\citeauthoryear{Straatman et~al.,}{Straatman
  et~al.}{2014}]{straatman2014}
Straatman C. M.~S.,  et~al., 2014, \mn@doi [\apj]
  {10.1088/2041-8205/783/1/L14}, 783, L14

\bibitem[\protect\citeauthoryear{Straatman et~al.,}{Straatman
  et~al.}{2018}]{straatman2018}
Straatman C. M.~S.,  et~al., 2018, \mn@doi [\apjs] {10.3847/1538-4365/aae37a},
  239, 27

\bibitem[\protect\citeauthoryear{Strazzullo et~al.,}{Strazzullo
  et~al.}{2013}]{strazzullo2013}
Strazzullo V.,  et~al., 2013, \mn@doi [\apj] {10.1088/0004-637X/772/2/118},
  772, 118

\bibitem[\protect\citeauthoryear{Tanaka et~al.,}{Tanaka
  et~al.}{2013}]{tanaka2013}
Tanaka M.,  et~al., 2013, \mn@doi [\apj] {10.1088/0004-637X/772/2/113}, 772,
  113

\bibitem[\protect\citeauthoryear{Tanaka et~al.,}{Tanaka
  et~al.}{2019}]{tanaka2019}
Tanaka M.,  et~al., 2019, \mn@doi [\apj] {10.3847/2041-8213/ab4ff3}, 885, L34

\bibitem[\protect\citeauthoryear{Tanaka et~al.,}{Tanaka
  et~al.}{2020}]{tanaka2020}
Tanaka M.,  et~al., 2020, \mn@doi [\apj] {10.3847/2041-8213/ab8b5a}, 894, L13

\bibitem[\protect\citeauthoryear{{The Astropy Collaboration} et~al.,}{{The
  Astropy Collaboration} et~al.}{2013}]{astropy2}
{The Astropy Collaboration} et~al., 2013, \mn@doi [\aap]
  {10.1051/0004-6361/201322068}, 558, A33

\bibitem[\protect\citeauthoryear{{The Astropy Collaboration} et~al.,}{{The
  Astropy Collaboration} et~al.}{2018}]{astropy1}
{The Astropy Collaboration} et~al., 2018, \mn@doi [\aj]
  {10.3847/1538-3881/aabc4f}, 156, 123

\bibitem[\protect\citeauthoryear{Thomas, Maraston, Bender  \& {de
  Oliveira}}{Thomas et~al.}{2005}]{thomas2005}
Thomas D.,  Maraston C.,  Bender R.,   {de Oliveira} C.~M.,  2005, \mn@doi
  [\apj] {10.1086/426932}, 621, 673

\bibitem[\protect\citeauthoryear{Thomas, Maraston, Schawinski, Sarzi  \&
  Silk}{Thomas et~al.}{2010}]{thomas2010}
Thomas D.,  Maraston C.,  Schawinski K.,  Sarzi M.,   Silk J.,  2010, \mn@doi
  [\mnras] {10.1111/j.1365-2966.2010.16427.x}

\bibitem[\protect\citeauthoryear{Thomas et~al.,}{Thomas
  et~al.}{2017}]{thomas2017}
Thomas R.,  et~al., 2017, \mn@doi [\aap] {10.1051/0004-6361/201628141}, 602,
  A35

\bibitem[\protect\citeauthoryear{Toft, Mainieri, Rosati, Lidman, Demarco,
  Nonino  \& Stanford}{Toft et~al.}{2004}]{toft2004}
Toft S.,  Mainieri V.,  Rosati P.,  Lidman C.,  Demarco R.,  Nonino M.,
  Stanford S.~A.,  2004, \mn@doi [\aap] {10.1051/0004-6361:20030621}, 422, 29

\bibitem[\protect\citeauthoryear{Toft, Gallazzi, Zirm, Wold, Zibetti, Grillo
  \& Man}{Toft et~al.}{2012}]{toft2012}
Toft S.,  Gallazzi A.,  Zirm A.,  Wold M.,  Zibetti S.,  Grillo C.,   Man A.,
  2012, \mn@doi [\apj] {10.1088/0004-637X/754/1/3}, 754, 3

\bibitem[\protect\citeauthoryear{Tomczak et~al.,}{Tomczak
  et~al.}{2014}]{tomczak2014}
Tomczak A.~R.,  et~al., 2014, \mn@doi [\apj] {10.1088/0004-637X/783/2/85}, 783,
  85

\bibitem[\protect\citeauthoryear{Trager, Faber, Worthey  \&
  Gonz{\'a}lez}{Trager et~al.}{2000a}]{trager2000a}
Trager S.~C.,  Faber S.~M.,  Worthey G.,   Gonz{\'a}lez J.~J.,  2000a, \mn@doi
  [\aj] {10.1086/301299}, 119, 1645

\bibitem[\protect\citeauthoryear{Trager, Faber, Worthey  \&
  Gonz{\'a}lez}{Trager et~al.}{2000b}]{trager2000b}
Trager S.~C.,  Faber S.~M.,  Worthey G.,   Gonz{\'a}lez J.~J.,  2000b, \mn@doi
  [\aj] {10.1086/301442}, 120, 165

\bibitem[\protect\citeauthoryear{Tremonti et~al.,}{Tremonti
  et~al.}{2004}]{tremonti2004}
Tremonti C.~A.,  et~al., 2004, \mn@doi [\apj] {10.1086/423264}, 613, 898

\bibitem[\protect\citeauthoryear{Treu, Stiavelli, Casertano, Moller  \&
  Bertin}{Treu et~al.}{1999}]{treu1999}
Treu T.,  Stiavelli M.,  Casertano S.,  Moller P.,   Bertin G.,  1999, \mn@doi
  [\mnras] {10.1046/j.1365-8711.1999.02794.x}, 308, 1037

\bibitem[\protect\citeauthoryear{Treu, Stiavelli, Bertin, Casertano  \&
  M{\o}ller}{Treu et~al.}{2001}]{treu2001b}
Treu T.,  Stiavelli M.,  Bertin G.,  Casertano S.,   M{\o}ller P.,  2001,
  \mn@doi [\mnras] {10.1046/j.1365-8711.2001.04720.x}, 326, 237

\bibitem[\protect\citeauthoryear{Treu, Ellis, Liao  \& {van Dokkum}}{Treu
  et~al.}{2005a}]{treu2005a}
Treu T.,  Ellis R.~S.,  Liao T.~X.,   {van Dokkum} P.~G.,  2005a, \mn@doi
  [\apj] {10.1086/429374}, 622, L5

\bibitem[\protect\citeauthoryear{Treu et~al.,}{Treu et~al.}{2005b}]{treu2005b}
Treu T.,  et~al., 2005b, \mn@doi [\apj] {10.1086/444585}, 633, 174

\bibitem[\protect\citeauthoryear{Valentino et~al.,}{Valentino
  et~al.}{2020}]{valentino2020}
Valentino F.,  et~al., 2020, \mn@doi [\apj] {10.3847/1538-4357/ab64dc}, 889, 93

\bibitem[\protect\citeauthoryear{Vazdekis}{Vazdekis}{1999}]{vazdekis1999}
Vazdekis A.,  1999, \mn@doi [\apj] {10.1086/306843}, 513, 224

\bibitem[\protect\citeauthoryear{Vazdekis et~al.,}{Vazdekis
  et~al.}{2015}]{vazdekis2015}
Vazdekis A.,  et~al., 2015, \mn@doi [\mnras] {10.1093/mnras/stv151}, 449, 1177

\bibitem[\protect\citeauthoryear{Virtanen et~al.,}{Virtanen
  et~al.}{2020}]{scipy}
Virtanen P.,  et~al., 2020, \mn@doi [Nature Methods]
  {10.1038/s41592-019-0686-2}, 17, 261

\bibitem[\protect\citeauthoryear{Vulcani, Poggianti, Finn, Rudnick, Desai  \&
  Bamford}{Vulcani et~al.}{2010}]{vulcani2010}
Vulcani B.,  Poggianti B.~M.,  Finn R.~A.,  Rudnick G.,  Desai V.,   Bamford
  S.,  2010, \mn@doi [\apj] {10.1088/2041-8205/710/1/L1}, 710, L1

\bibitem[\protect\citeauthoryear{Vulcani et~al.,}{Vulcani
  et~al.}{2011}]{vulcani2011}
Vulcani B.,  et~al., 2011, \mn@doi [\mnras] {10.1111/j.1365-2966.2010.17904.x},
  412, 246

\bibitem[\protect\citeauthoryear{Vulcani et~al.,}{Vulcani
  et~al.}{2013}]{vulcani2013}
Vulcani B.,  et~al., 2013, \mn@doi [\aap] {10.1051/0004-6361/201118388}, 550,
  A58

\bibitem[\protect\citeauthoryear{Weinmann, {van den Bosch}, Yang  \&
  Mo}{Weinmann et~al.}{2006}]{weinmann2006}
Weinmann S.~M.,  {van den Bosch} F.~C.,  Yang X.,   Mo H.~J.,  2006, \mn@doi
  [\mnras] {10.1111/j.1365-2966.2005.09865.x}, 366, 2

\bibitem[\protect\citeauthoryear{Weinmann, Pasquali, Oppenheimer, Finlator,
  Mendel, Crain  \& Macci{\`o}}{Weinmann et~al.}{2012}]{weinmann2012}
Weinmann S.~M.,  Pasquali A.,  Oppenheimer B.~D.,  Finlator K.,  Mendel J.~T.,
  Crain R.~A.,   Macci{\`o} A.~V.,  2012, \mn@doi [\mnras]
  {10.1111/j.1365-2966.2012.21931.x}, 426, 2797

\bibitem[\protect\citeauthoryear{Wetzel, Tinker  \& Conroy}{Wetzel
  et~al.}{2012}]{wetzel2012}
Wetzel A.~R.,  Tinker J.~L.,   Conroy C.,  2012, \mn@doi [\mnras]
  {10.1111/j.1365-2966.2012.21188.x}, 424, 232

\bibitem[\protect\citeauthoryear{Wetzel, Tinker, Conroy  \& {van den
  Bosch}}{Wetzel et~al.}{2013}]{wetzel2013}
Wetzel A.~R.,  Tinker J.~L.,  Conroy C.,   {van den Bosch} F.~C.,  2013,
  \mn@doi [\mnras] {10.1093/mnras/stt469}, 432, 336

\bibitem[\protect\citeauthoryear{Whitaker, {van Dokkum}, Brammer  \&
  Franx}{Whitaker et~al.}{2012}]{whitaker2012b}
Whitaker K.~E.,  {van Dokkum} P.~G.,  Brammer G.,   Franx M.,  2012, \mn@doi
  [\apj] {10.1088/2041-8205/754/2/L29}, 754, L29

\bibitem[\protect\citeauthoryear{Whitaker et~al.,}{Whitaker
  et~al.}{2013}]{whitaker2013}
Whitaker K.~E.,  et~al., 2013, \mn@doi [\apj] {10.1088/2041-8205/770/2/L39},
  770, L39

\bibitem[\protect\citeauthoryear{Williams, Quadri, Franx, {van Dokkum}  \&
  Labb{\'e}}{Williams et~al.}{2009}]{williams2009}
Williams R.~J.,  Quadri R.~F.,  Franx M.,  {van Dokkum} P.,   Labb{\'e} I.,
  2009, \mn@doi [\apj] {10.1088/0004-637X/691/2/1879}, 691, 1879

\bibitem[\protect\citeauthoryear{Williams, Quadri, Franx, {van Dokkum}, Toft,
  Kriek  \& Labb{\'e}}{Williams et~al.}{2010}]{williams2010}
Williams R.~J.,  Quadri R.~F.,  Franx M.,  {van Dokkum} P.,  Toft S.,  Kriek
  M.,   Labb{\'e} I.,  2010, \mn@doi [\apj] {10.1088/0004-637X/713/2/738}, 713,
  738

\bibitem[\protect\citeauthoryear{Wilman, Zibetti  \& Budav{\'a}ri}{Wilman
  et~al.}{2010}]{wilman2010}
Wilman D.~J.,  Zibetti S.,   Budav{\'a}ri T.,  2010, \mn@doi [\mnras]
  {10.1111/j.1365-2966.2010.16845.x}, pp no--no

\bibitem[\protect\citeauthoryear{Wilson et~al.,}{Wilson
  et~al.}{2009}]{wilson2009}
Wilson G.,  et~al., 2009, \mn@doi [\apj] {10.1088/0004-637X/698/2/1943}, 698,
  1943

\bibitem[\protect\citeauthoryear{Woo et~al.,}{Woo et~al.}{2013}]{woo2013}
Woo J.,  et~al., 2013, \mn@doi [\mnras] {10.1093/mnras/sts274}, 428, 3306

\bibitem[\protect\citeauthoryear{Woodrum, J{\o}rgensen, Fisher, Oberhelman,
  Demarco, Contreras  \& Bieker}{Woodrum et~al.}{2017}]{woodrum2017}
Woodrum C.,  J{\o}rgensen I.,  Fisher R.~S.,  Oberhelman L.,  Demarco R.,
  Contreras T.,   Bieker J.,  2017, \mn@doi [\apj] {10.3847/1538-4357/aa8871},
  847, 20

\bibitem[\protect\citeauthoryear{Worthey}{Worthey}{1994}]{worthey1994}
Worthey G.,  1994, \mn@doi [\apjs] {10.1086/192096}, 95, 107

\bibitem[\protect\citeauthoryear{Wuyts et~al.,}{Wuyts et~al.}{2007}]{wuyts2007}
Wuyts S.,  et~al., 2007, \mn@doi [\apj] {10.1086/509708}, 655, 51

\bibitem[\protect\citeauthoryear{Xie, De~Lucia, Hirschmann  \& Fontanot}{Xie
  et~al.}{2020}]{xie2020}
Xie L.,  De~Lucia G.,  Hirschmann M.,   Fontanot F.,  2020, arXiv e-prints

\bibitem[\protect\citeauthoryear{Yan, Newman, Faber, Konidaris, Koo  \&
  Davis}{Yan et~al.}{2006}]{yan2006}
Yan R.,  Newman J.~A.,  Faber S.~M.,  Konidaris N.,  Koo D.,   Davis M.,  2006,
  \mn@doi [\apj] {10.1086/505629}, 648, 281

\bibitem[\protect\citeauthoryear{York et~al.,}{York et~al.}{2000}]{york2000}
York D.~G.,  et~al., 2000, \mn@doi [\aj] {10.1086/301513}, 120, 1579

\bibitem[\protect\citeauthoryear{Zabludoff \& Mulchaey}{Zabludoff \&
  Mulchaey}{1998}]{zabludoff1998}
Zabludoff A.~I.,  Mulchaey J.~S.,  1998, \mn@doi [\apj] {10.1086/305355}, 496,
  39

\bibitem[\protect\citeauthoryear{Zhang, Zaritsky, Behroozi  \& Werk}{Zhang
  et~al.}{2019}]{zhang2019}
Zhang H.,  Zaritsky D.,  Behroozi P.,   Werk J.,  2019, \mn@doi [\apj]
  {10.3847/1538-4357/ab2761}, 880, 28

\bibitem[\protect\citeauthoryear{{di Serego Alighieri}, Lanzoni  \&
  J{\o}rgensen}{{di Serego Alighieri} et~al.}{2006a}]{diseregoalighieri2006a}
{di Serego Alighieri} S.,  Lanzoni B.,   J{\o}rgensen I.,  2006a, \mn@doi
  [\apj] {10.1086/507582}, 647, L99

\bibitem[\protect\citeauthoryear{{di Serego Alighieri}, Lanzoni  \&
  J{\o}rgensen}{{di Serego Alighieri} et~al.}{2006b}]{diseregoalighieri2006b}
{di Serego Alighieri} S.,  Lanzoni B.,   J{\o}rgensen I.,  2006b, \mn@doi
  [\apj] {10.1086/510409}, 652, L145

\bibitem[\protect\citeauthoryear{{van Dokkum} \& Brammer}{{van Dokkum} \&
  Brammer}{2010}]{vandokkum2010b}
{van Dokkum} P.~G.,  Brammer G.,  2010, \mn@doi [\apj]
  {10.1088/2041-8205/718/2/L73}, 718, L73

\bibitem[\protect\citeauthoryear{{van Dokkum} \& Stanford}{{van Dokkum} \&
  Stanford}{2003}]{vandokkum2003}
{van Dokkum} P.~G.,  Stanford S.~A.,  2003, \mn@doi [\apj] {10.1086/345989},
  585, 78

\bibitem[\protect\citeauthoryear{{van Dokkum} \& {van der Marel}}{{van Dokkum}
  \& {van der Marel}}{2007}]{vandokkum2007}
{van Dokkum} P.~G.,  {van der Marel} R.~P.,  2007, \mn@doi [\apj]
  {10.1086/509633}, 655, 30

\bibitem[\protect\citeauthoryear{{van Dokkum}, Franx, Kelson  \&
  Illingworth}{{van Dokkum} et~al.}{1998}]{vandokkum1998}
{van Dokkum} P.~G.,  Franx M.,  Kelson D.~D.,   Illingworth G.~D.,  1998,
  \mn@doi [\apj] {10.1086/311567}, 504, L17

\bibitem[\protect\citeauthoryear{{van Dokkum}, Franx, Kelson  \&
  Illingworth}{{van Dokkum} et~al.}{2001}]{vandokkum2001a}
{van Dokkum} P.~G.,  Franx M.,  Kelson D.~D.,   Illingworth G.~D.,  2001,
  \mn@doi [\apj] {10.1086/320502}, 553, L39

\bibitem[\protect\citeauthoryear{{van Dokkum} et~al.,}{{van Dokkum}
  et~al.}{2010}]{vandokkum2010a}
{van Dokkum} P.~G.,  et~al., 2010, \mn@doi [\apj]
  {10.1088/0004-637X/709/2/1018}, 709, 1018

\bibitem[\protect\citeauthoryear{{van de Sande} et~al.,}{{van de Sande}
  et~al.}{2013}]{vandesande2013}
{van de Sande} J.,  et~al., 2013, \mn@doi [\apj] {10.1088/0004-637X/771/2/85},
  771, 85

\bibitem[\protect\citeauthoryear{{van der Burg} et~al.,}{{van der Burg}
  et~al.}{2013}]{vanderburg2013}
{van der Burg} R. F.~J.,  et~al., 2013, \mn@doi [\aap]
  {10.1051/0004-6361/201321237}, 557, A15

\bibitem[\protect\citeauthoryear{{van der Burg} et~al.,}{{van der Burg}
  et~al.}{2020}]{vanderburg2020}
{van der Burg} R. F.~J.,  et~al., 2020, \mn@doi [\aap]
  {10.1051/0004-6361/202037754}, 638, A112

\bibitem[\protect\citeauthoryear{van~der Walt, Colbert  \& Varoquaux}{van~der
  Walt et~al.}{2011}]{numpy}
van~der Walt S.,  Colbert S.~C.,   Varoquaux G.,  2011, \mn@doi [Computing in
  Science \& Engineering] {10.1109/MCSE.2011.37}, 13, 22

\bibitem[\protect\citeauthoryear{{van der Wel}, Franx, {van Dokkum}  \&
  Rix}{{van der Wel} et~al.}{2004}]{vanderwel2004}
{van der Wel} A.,  Franx M.,  {van Dokkum} P.~G.,   Rix H.-W.,  2004, \mn@doi
  [\apj] {10.1086/381887}, 601, L5

\bibitem[\protect\citeauthoryear{{van der Wel} et~al.,}{{van der Wel}
  et~al.}{2016}]{vanderwel2016}
{van der Wel} A.,  et~al., 2016, The Messenger, 164, 36

\makeatother
\end{thebibliography}

\section*{Affiliations}

$^{1}$Department of Physics and Astronomy, University of Waterloo, Waterloo, Ontario N2L 3G1, Canada \\
$^{2}$Waterloo Centre for Astrophysics, University of Waterloo, Waterloo, Ontario, N2L3G1, Canada \\
$^{3}$Department of Astronomy and Astrophysics, Pennsylvania State University, University Park, PA 16802 \\
$^{4}$European Southern Observatory, Karl-Schwarzschild-Str. 2, 85748, Garching, Germany \\
$^{5}$Department of Physics and Astronomy, The University of Kansas, 1251 Wescoe Hall Drive, Lawrence, KS 66045, USA \\
$^{6}$Department of Physics and Astronomy, York University, 4700 Keele Street, Toronto, Ontario, ON MJ3 1P3, Canada \\
$^{7}$Departamento de Ingenier\'ia Inform\'atica y Ciencias de la Computaci\'on, Universidad de Concepci\'on, Concepci\'on, Chile \\
$^{8}$South African Astronomical Observatory, P.O. Box 9, Observatory 7935 Cape Town, South Africa \\
$^{9}$Centre for Space Research, North-West University, Potchefstroom 2520 Cape Town, South Africa \\
$^{10}$Research School of Astronomy and Astrophysics, The Australian National University, ACT 2601, Australia \\
$^{11}$Centre for Gravitational Astrophysics, College of Science, The Australian National University, ACT 2601, Australia \\
$^{12}$European Space Agency (ESA), European Space Astronomy Centre, Villanueva de la Ca\~{n}ada, E-28691 Madrid, Spain \\
$^{13}$Department of Astronomy \& Astrophysics, University of Toronto, Toronto, Canada \\
$^{14}$School of Physics and Astronomy, University of Birmingham, Edgbaston, Birmingham B15 2TT, England \\
$^{15}$Department of Physics, McGill University, 3600 rue University, Montr\'{e}al, Qu\'{e}bec, H3P 1T3, Canada \\
$^{16}$INAF - Osservatorio Astronomico di Trieste, via G. B. Tiepolo 11, I-34143 Trieste, Italy \\
$^{17}$IFPU - Institute for Fundamental Physics of the Universe, via Beirut 2, 34014 Trieste, Italy \\
$^{18}$Department of Physics and Astronomy, University of California, Riverside, 900 University Avenue, Riverside, CA 92521, USA \\
$^{19}$Department of Physics and Astronomy, University of California, Irvine, 4129 Frederick Reines Hall, Irvine, CA 92697, USA \\
$^{20}$Departamento de Astronom\'ia, Facultad de Ciencias F\'isicas y Matem\'aticas, Universidad de Concepci\'on, Concepci\'on, Chile \\
$^{21}$Laboratoire d'astrophysique, \'Ecole Polytechnique F\'ed\'erale de Lausanne (EPFL), 1290 Sauverny, Switzerland  \\
$^{22}$GEPI, Observatoire de Paris, Universit\'e PSL, CNRS, Place Jules Janssen, F-92190 Meudon, France \\
$^{23}$Astrophysics Research Institute, Liverpool John Moores University, 146 Brownlow Hill, Liverpool L3 5RF, UK \\
$^{24}$Departamento de Ciencias F\'{i}sicas, Universidad Andres Bello, Fernandez Concha 700, Las Condes 7591538, Santiago, Regi\'{o}n Metropolitana, Chile \\
$^{25}$Arizona State University, School of Earth and Space Exploration, Tempe, AZ 871404, USA \\
$^{26}$MIT Kavli Institute for Astrophysics and Space Research, 70 Vassar St, Cambridge, MA 02109, USA \\
$^{27}$INAF - Osservatorio astronomico di Padova, Vicolo Osservatorio 5, IT-35122 Padova, Italy \\
$^{28}$Steward Observatory and Department of Astronomy, University of Arizona, Tucson, AZ 85721, USA \\


\appendix

\section{Quiescent indicators}\label{sec:tracers}

    \begin{figure*}
      \begin{center}
        \includegraphics[width=\linewidth]{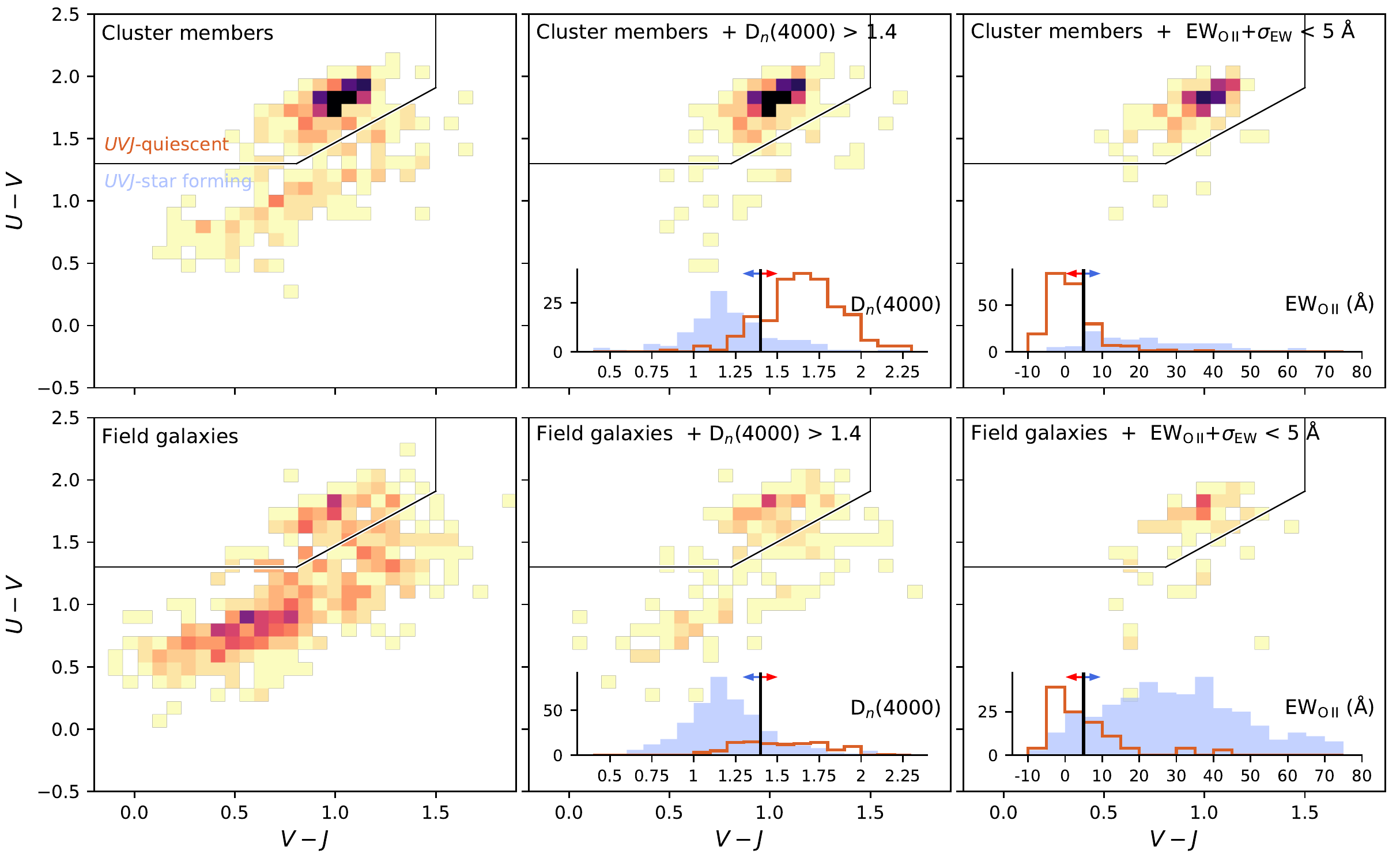}
      \end{center}
      \caption{2D histograms of the spectroscopic sample in \textit{UVJ} colour space. The selection criteria of \textit{UVJ}-quiescent galaxies is shown as a black line with arrows indicating the region of quiescent (red) or star forming (blue) galaxies. In the top (bottom) row, the cluster (field) galaxies are shown. The middle column compares the \textit{UVJ} selection against \dbreak${>}$1.4, where the \dbreak threshold was chosen based on the bimodality of the \textit{UVJ} selection relative to \dbreak shown in the inset histogram. The right-hand column compare the \textit{UVJ} selection against EW(\oii){+}$\sigma_\mathrm{EW}<5$\,\AA, where the threshold was chosen based on the bimodality of the \textit{UVJ} selection relative to EW(\oii) shown in the inset histogram. This comparison shows that for our sample, \textit{UVJ} colours are broadly consistent with both \dbreak and EW(\oii) tracers for quiescent galaxies. } \label{fig:uvj_binned_selections}
    \end{figure*}

In this work we selected quiescent galaxies by their position in rest-frame \textit{UVJ} colour space. However, there are several other tracers of SFR that could have been used instead. The \dbreak has been used as a proxy for the age of a stellar population \citep[][]{balogh1999,kauffmann2003b,muzzin2012} as the strength of the break increases with the fraction of old stars (but also with metallicity). The flux of the \oii emission line is sensitive to recent excitations in the ISM from young stars -- although indirectly, and is also dependent on the metallicity of the gas. Galaxies selected by each tracer as quiescent are shown in the \textit{UVJ} plane in Figure\,\ref{fig:uvj_binned_selections}. The first columns show the 2D histograms of the GOGREEN spectroscopic sample in \textit{UVJ} colour space, with galaxies in clusters shown in the first row and galaxies in the field in the second row. The separation of quiescent and star forming galaxies is shown as a black line. 
    
The positions of galaxies in \textit{UVJ} colour space are then shown for galaxies which satisfy alternative indicators of passive evolution: \dbreak${>}1.4$ in the middle column, and EW(\oii){+}$\sigma_\mathrm{EW}<5$\,{\AA} in the right-hand column. Among the cluster galaxies, the highest density of galaxies selected by \dbreak or \oii is predominantly in the \textit{UVJ}-quiescent region. A much larger fraction of the `quenched' galaxies in the field are \textit{UVJ}-star forming. 

The distribution of \textit{UVJ}-quiescent (red) or \textit{UVJ}-star forming (blue) according to \dbreak (EW(\oii)) is shown in the inset histogram in the second (third) columns. The selection of quiescent galaxies from \dbreak or EW(\oii) is determined by the break in the \textit{UVJ}-quiescent and \textit{UVJ}-star forming distributions, corresponding to \dbreak${\sim}$1.4 and EW(\oii){+}$\sigma_\mathrm{EW}<5$\,\AA. Both \dbreak and \oii emission select the majority of the \textit{UVJ}-selected quiescent sample -- 79\,per cent and 63\,per cent, respectively. While only 32\,per cent (24\,per cent) are \textit{UVJ}-quiescent among the \dbreak-quiescent (\oii-quiescent), the contamination of \textit{UVJ}-star forming galaxies is only 13\,per cent (3\,per cent). This brief comparison shows that these tracers are broadly consistent, and using \dbreak or EW(\oii) instead to select quiescent galaxies would not qualitatively change our conclusions. 

\section{Mass-metallicity relation} \label{sec:mzr}

    \begin{figure*}
      \begin{center}
        \includegraphics[width=\linewidth]{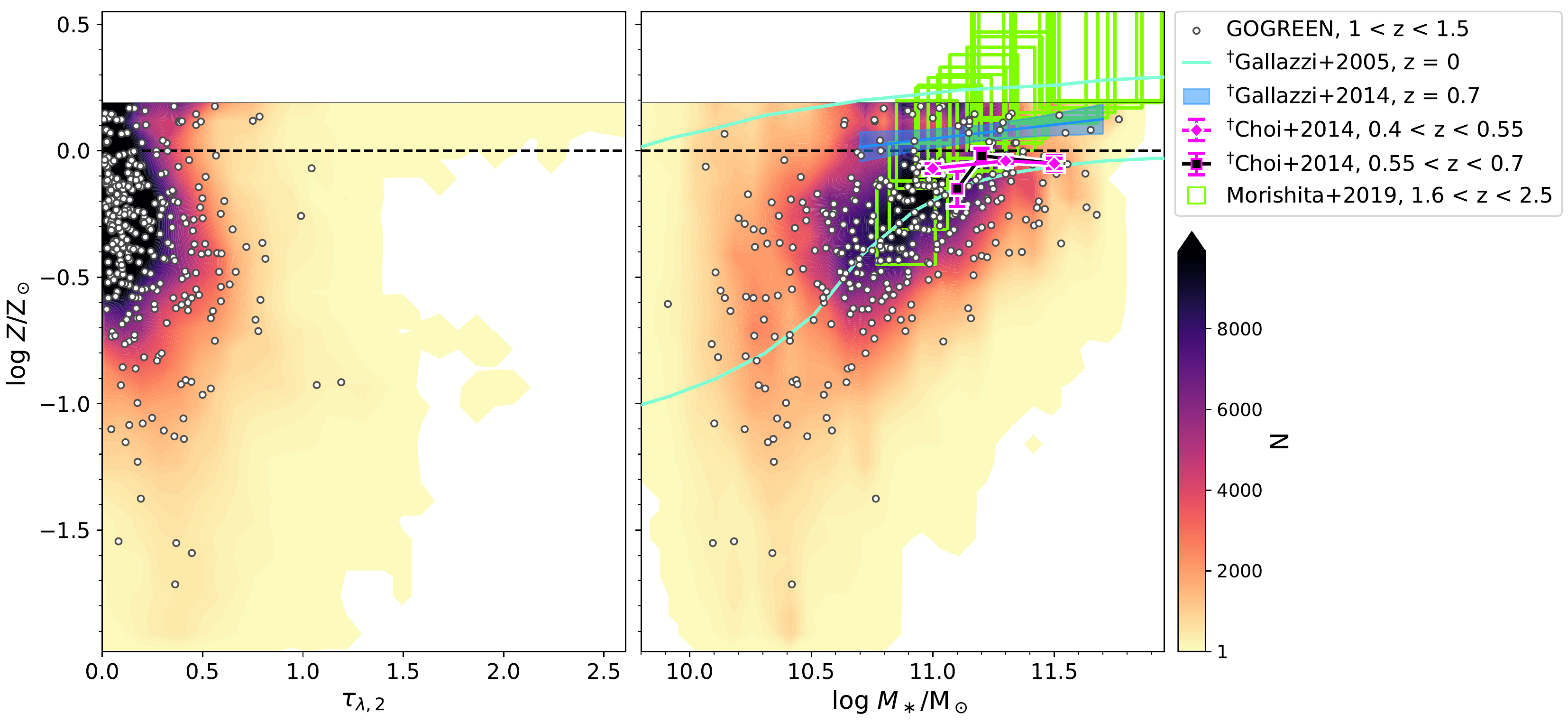}
      \end{center}
      \caption{Metallicity as a function of diffuse dust optical depth (left) and stellar mass (right) for the GOGREEN \textit{UVJ}-quiescent sample. A dashed line indicates solar metallicity, and a solid line indicates the maximum metallicity allowed by the {\sc MILES} spectral templates. The local mass-metallicity relation (MZR) for early type galaxies from \citet{gallazzi2004} at $z{\la}0.22$ is shown with two cyan lines indicating the lower and upper limits of the reported 68 per cent confidence region. This relation was used as a prior in our SFH fitting procedure. The MZR for $z{\sim}0.7$ quiescent galaxies is shown as a blue line with a shaded region indicating the uncertainty region, from \citet{gallazzi2014}.
      A selection of moderate-redshift quiescent galaxies at $0.4{<}z{<}0.55$ and $0.55{<}z{<}0.7$ from full continuum-normalised spectral fits from \citet{choi2014} are shown, without correction for differences in $\alpha$-abundance. A high-redshift sample of massive quiescent galaxies from \citet{morishita2019} are also included, shown in green. Daggers denote where data have been adapted from the relevant study to compensate for difference in stellar mass estimates. The colour scale shows the density of the combined posteriors in the GOGREEN data, with white circles indicating the median values of the individual posteriors. 
      } \label{fig:mzr}
    \end{figure*}
    
Stellar mass, dust, and metallicity are correlated throughout a galaxy's evolution, and the relation between the two has been well studied in the local universe \citep{gallazzi2005, gallazzi2014, tremonti2004, panter2008, choi2014}. Observables used to estimate the ages of stellar populations, such as colours and spectral lines, can be strongly degenerate with dust and metallicity. Understanding such degeneracies at $z{>}1$ is challenging, especially given that most studies are limited to small numbers of massive galaxies \citep{onodera2012, onodera2015, newman2014, kriek2016, lee-brown2017, morishita2018,morishita2019, estrada-carpenter2019}. Moreover, without high-resolution spectroscopy it is difficult to accurately model the complex behaviour of these parameters. Given the limited wavelength coverage in our spectra, and typically low SNR, we do not tightly constrain metallicity in our fits -- however, is important to consider the average metallicity we fit, as a function of mass and environment, because of its degeneracy with age. For instance, we find that a difference in metallicity of a factor of three (${\sim}$0.5\,dex) can change the mass-weighted age estimate by ${\sim}$0.5\,Gyr.

The {\sc MIST} isochrones cover an extended range of metallicities (-4$<$[Z/H]$<$0.5), while the {\sc MILES} templates are limited to [Z/H]${<}$0.19. We also impose an additional limit of [Z/H]${>}$-2 to avoid extrapolating the templates to less well constrained parameter space. Although updated isochrones libraries include variation of $\alpha$-abundances, the current version of {\sc FSPS} includes only scaled-solar abundances. Studies of high SNR spectra of passive galaxies show that \aFe scales with galaxy properties (eg., velocity dispersion, stellar mass), and a number of old massive galaxies with super-solar $\alpha$-abundances have been discovered \citep{thomas2005, choi2014, conroy2013b, onodera2015, kriek2016, kriek2019, jorgensen2017, jorgensen2018}. Underestimating $\alpha$-abundance affects the slope of the UV-NIR continuum, where \citet{vazdekis2015} show differences of 10\,per cent in optical colours, or 40\,per cent in flux within a bandpass, between solar \aFe and +0.4 albeit for galaxies much older than included in our study. 

We explored the sensitivity of the metallicity measurements in our fits through the stellar mass--metallicity relation (MZR) and relative to the diffuse dust optical depth. Figure\,\ref{fig:mzr} shows the posteriors of metallicity and dust (left) and stellar masses (right) for the galaxies in our sample, with circles showing the medians of individual posteriors. The GOGREEN measurements are shown relative to the local (field) relation for quiescent SDSS galaxies from \citet{gallazzi2005}, marked as cyan lines corresponding to the 16\thh and 84\thh percentiles of the reported trend. Note that this relation was used as a prior in our fitting procedure. We also include the $z{\sim}0.7$ MZR for quiescent galaxies from \citet{gallazzi2014} as a blue region. The MZR for quiescent galaxies reported by \citet{choi2014} at $0.4{<}z{<}0.55$ is shown as pink points, and $0.55{<}z{<}0.7$ as black points with pink error bars. Lastly, we show the $1\,\sigma$ region of individual measurements of $1.6{<}z{<}2.5$ massive galaxies from \citet{morishita2019} as green boxes. The \citet{gallazzi2005}, \citet{gallazzi2014}, and \citet{choi2014} data are shown corrected for differences in stellar mass estimates (i.e., +0.2\,dex, see Appendix\,\ref{sec:param_nonparam}), but not corrected for differences in definitions of solar metallicity or $\alpha$-abundance. \citet{choi2014} incorporated $\alpha$-abundance corrections in their continuum-normalised spectral fitting. \citet{morishita2019} used a higher limit on metallicity, as they use the updated {\sc MIST} isochrones which extend to [Z/H]${<}$0.5. 

We note that these studies all use different methodologies: \citet{gallazzi2005} and \citet{gallazzi2014} relied on line indices, \citet{choi2014} use full spectrum SPS modelling for continuum-corrected co-added spectra, while \citet{morishita2019} use full spectrum SPS modelling of spectroscopy and photometry, more similar to our own procedure. Although not shown in Figure\,\ref{fig:mzr}, \citet{leethochawalit2018} study the MZR with respect to [Fe/H] for quiescent galaxies at $z{\sim}0.4$ using spectral modelling, and recover values consistent with the highest density (purple) region in our plot (see their figure\,7). Interestingly, \citet{kriek2019} measure the metallicity of three massive quiescent galaxies at $z{\sim}1.4$, using high-resolution spectroscopy to measure absorption lines, and find that the [Fe/H] values are ${\sim}$0.2\,dex lower than the $z{<}0.7$ relation. \citet{jorgensen2007} similarly find evidence of evolution of cluster galaxies since $z{\sim}1$. On the other hand, \citet{onodera2015} find the [Z/H] of 24 massive quiescent galaxies at $z{\sim}1.6$ to be well in line with the local relation, based on a similar line index analysis.

While our metallicities are lower than reported by similar studies, as long as the mass-metallicity relation does not have a strong environmental dependence, the \textit{relative} comparison of cluster and field galaxy ages will not be sensitive to our model metallicities. Indeed, we find no difference in the MZR between field and cluster galaxies from our fits. \citet{peng2015} compared the stellar metallicities of galaxies in SDSS, and found no significant difference between satellite and central galaxies above $10^{10}$\,M$_{\sun}$. Tangentially, \citet{maier2016} measured enhanced \textit{gas-phase} metallicities of accreted star-forming cluster galaxies relative to comparable field galaxies at $z{\sim}0.4$ for ${<}10^{10.5}$\,M$_{\sun}$, but no significant difference at higher masses. 

As mentioned above, there is a degeneracy between age, metallicity, and dust. For completeness we show the combined posteriors of metallicity and the diffuse dust optical depth in the left plot of Figure\,\ref{fig:mzr}. The majority of galaxies have very little dust, $\tau_{\lambda,2}{<}0.5$, even the galaxies with very low metallicities. This perhaps suggests that the dust model we have assumed (i.e., Milky Way extinction curve; \citealt{cardelli1989}) is insufficient.

\section{Prospector nonparametric vs FAST parametric models}\label{sec:param_nonparam}

\begin{figure}
      \begin{center}
        \includegraphics[width=\linewidth]{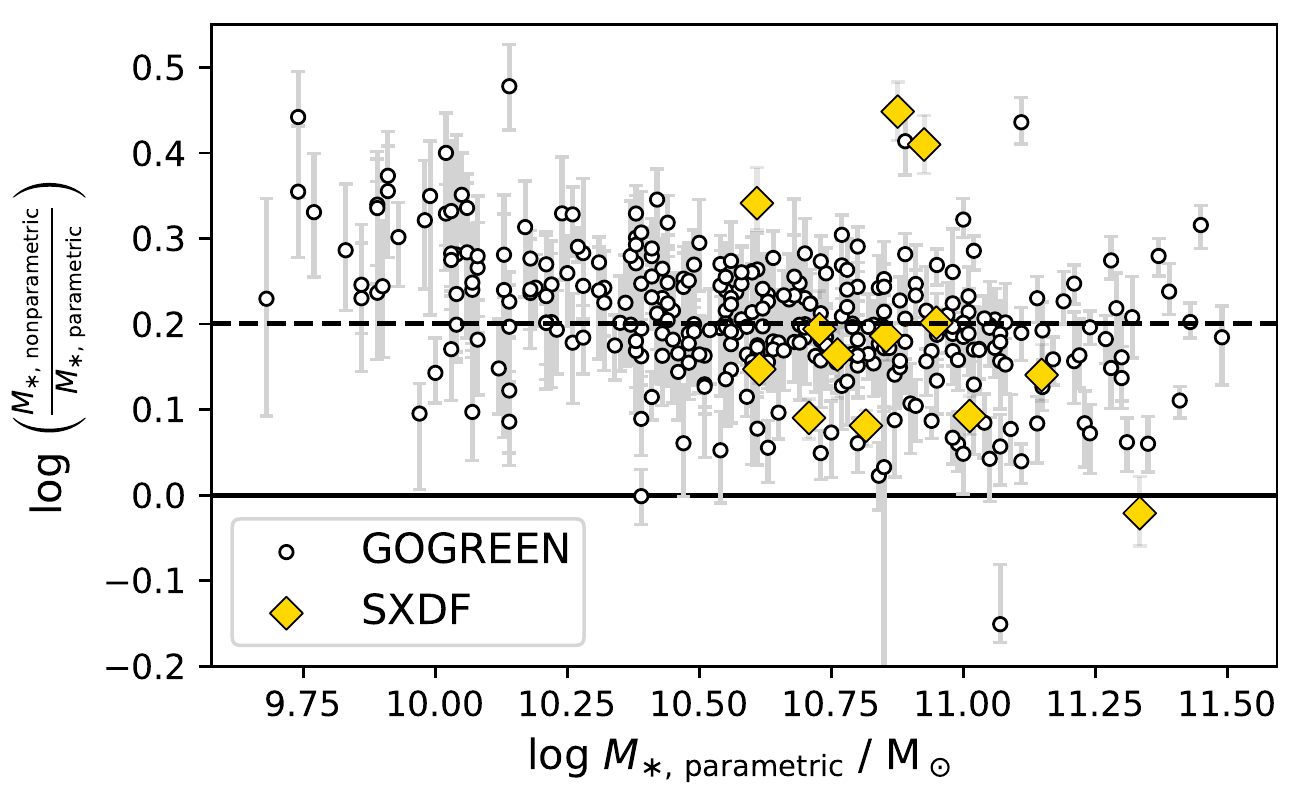}
      \end{center}
      \caption{Comparison of {\sc FAST} (parametric) and \prospector (nonparametric) derived stellar masses. We confirm the systematic offset reported by \citet{leja2019b} that nonparametric SFHs yield larger masses, by ${\sim}$0.2\,dex (shown as a dashed line), with a mild mass dependence. Yellow diamonds indicate SXDF galaxies which have parametric stellar masses from \citet{mehta2018}. } \label{fig:masses}
    \end{figure}

We confirm the systematic offset between parametric-SFH derived stellar masses using {\sc FAST} \citep{kriek2009} with non-parametric-SFH derived stellar masses using \prospector reported by other studies \citep[e.g.][]{leja2019b}. Our comparison is shown in Figure\,\ref{fig:masses}, where nonparametric-SFH masses are on average 1.6$\times$ (0.2\,dex) higher. Stellar masses were derived with {\sc FAST} for the same SFH as was used to measure the rest-frame colours with {\sc EZGAL} (see Section\,\ref{sec:photo_data} -- a declining exponential SFR). The SXDF galaxies are marked as yellow diamonds in Figure\,\ref{fig:masses} as their fiducial masses were not derived from {\sc FAST} but from similar template fitting with \citet{bruzual2003} models described in \citet{mehta2018}. The stellar masses used in \citet{old2020} and \citet{vanderburg2020}, as well as in the upcoming data release (Balogh et al. 2020, in prep) are based on {\sc FAST} masses, and therefore will differ from the stellar masses in this paper.

\section{Average spectral characteristics}\label{sec:coadds}

    \begin{figure*}
      \begin{center}
        \includegraphics[width=\linewidth]{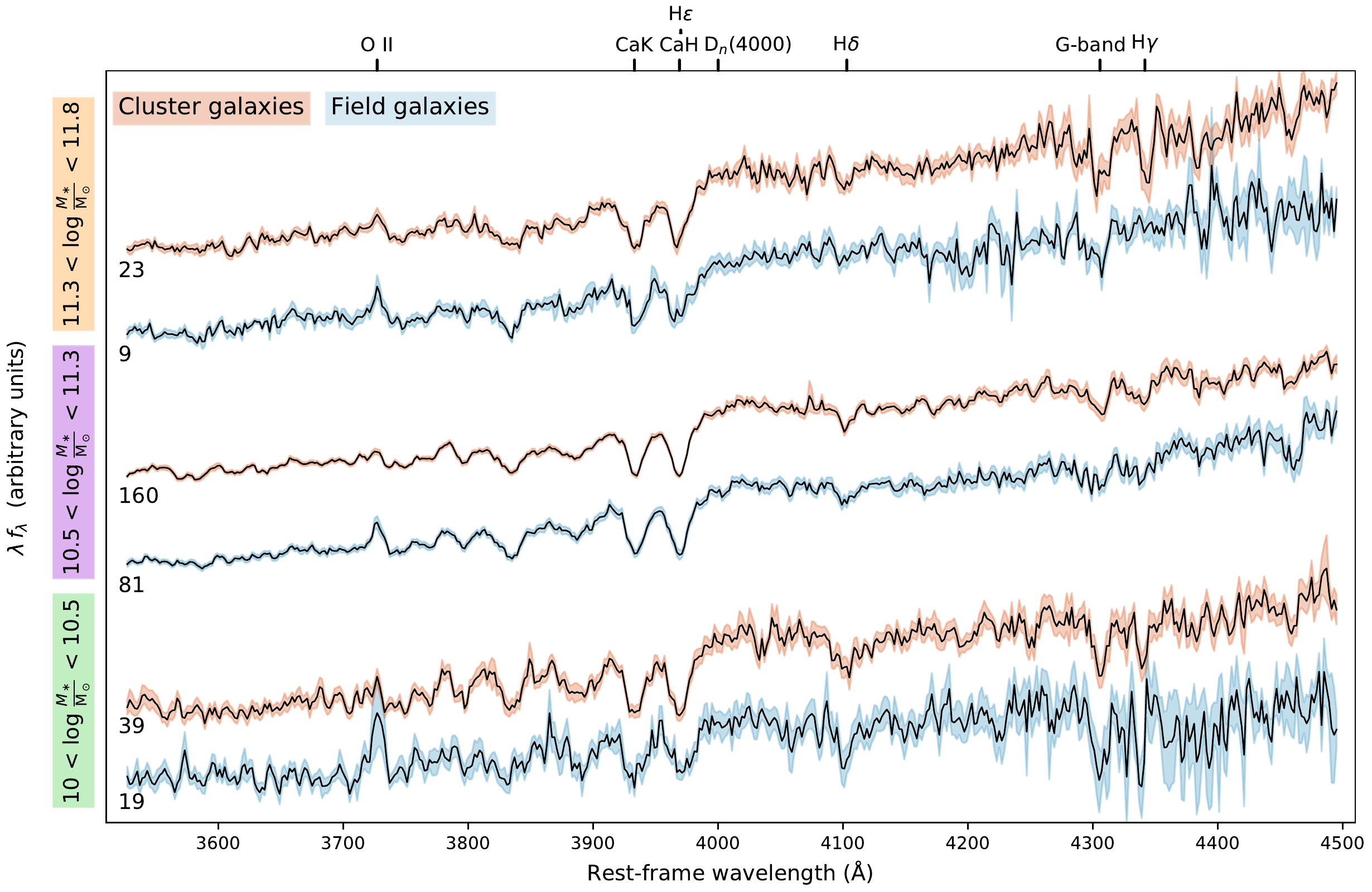}
      \end{center}
      \caption{Combined spectra of quiescent galaxies within mass and environment selections, shown within the wavelength region included in the SFH fitting procedure. The spectra in each subsample were de-redshifted, re-binned to a common wavelength sampling, flux normalised about 4120\,\AA, and then averaged. The uncertainty in the co-added spectra was determined from bootstrapping. Prominent spectral features are labelled on the top axis, and number of galaxies in each co-add are indicated on the left. An alternative to fitting the spectroscopy for individual galaxies and combined the posteriors (as we did in this paper), a common alternative is to combine the \textit{data} to create an average spectrum/SED, and measure physical parameters from that. While these two approaches do not necessarily give the same result, we qualitatively confirm the similarities between spectra of galaxies of equivalent mass with some exceptions: field galaxies have stronger \oii emission, cluster galaxies have slightly stronger H$\delta$ absorption. } \label{fig:uvj_stacks}
    \end{figure*}
    
    In this paper we have measured galaxy properties on individual galaxies, and then considered the statistics of those measurements. A common alternative in the literature is to combine the \textit{data} to create an average spectrum/SED, and measure physical parameters from that. As the parameters are nonlinearly related to SED shape, these two approaches do not necessarily give the same result.

Figure\,\ref{fig:uvj_stacks} shows co-added spectra of cluster galaxies and field galaxies in our sample, each separated into three stellar mass subsamples. Before stacking, the spectra were 
redshift corrected, binned to a common wavelength sampling, and flux normalised at 4120\,\AA. Spectra within a given stellar mass and environment subsample were then averaged and bootstrap sampled to determine the uncertainty. Combined galaxies within clusters are shown in orange, and within the field in blue, where the number of contributing galaxies to each spectrum is labelled on the left.

The average cluster and field spectra appear very similar overall, with only a few apparent differences. The field population has more prominent \oii emission at lower masses, while the cluster galaxies have stronger \oii emission at higher masses (although much weaker than in the field). This is likely related to the fact that \oii is not strictly related to recent star formation (e.g. from AGN and/or LINER; \citealt{heckman1980}, \citealt{yan2006}, \citealt{singh2013}). On the other hand, absorption lines from H$\delta$ appear stronger for cluster galaxies (except at the lowest stellar masses) suggesting that the cluster galaxies experienced, on average, more recent star formation.

\begin{figure}
      \begin{center}
        \includegraphics[width=\linewidth]{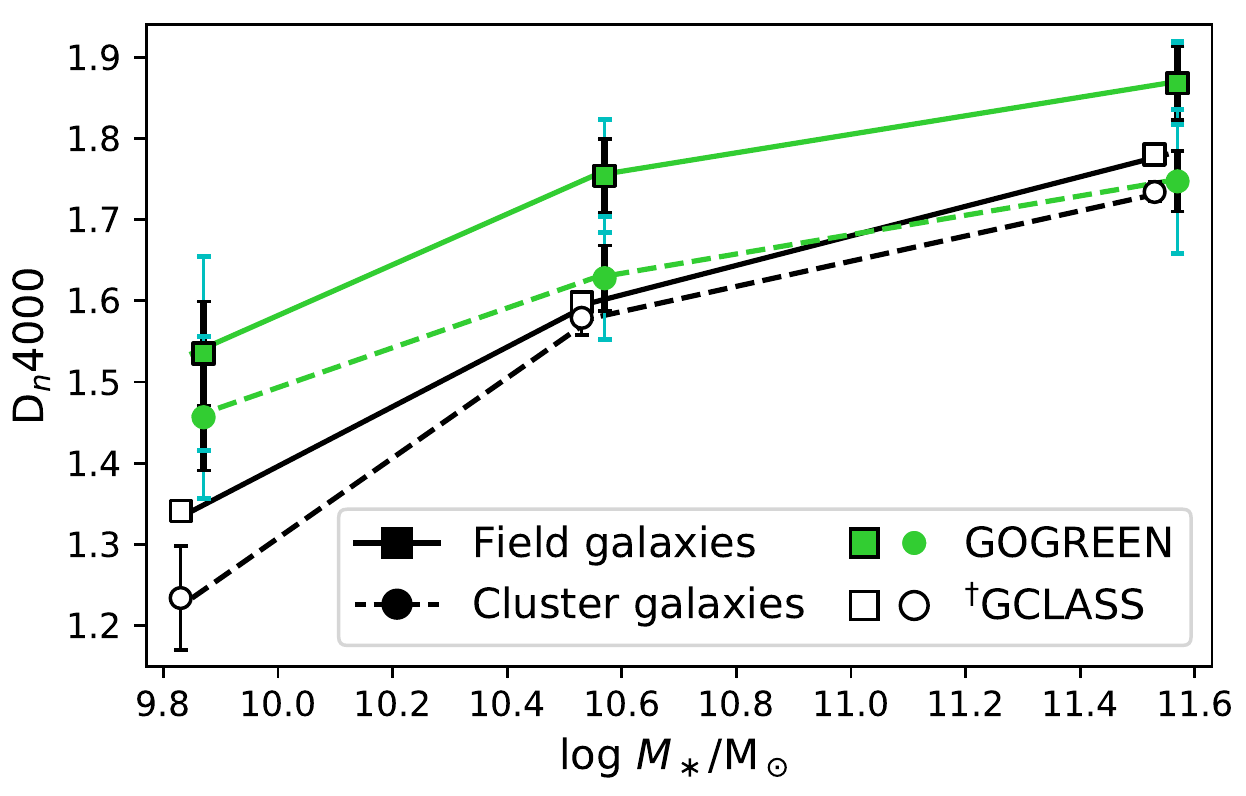}
      \end{center}
      \caption{\dbreak of \textit{averaged} spectra as a function of stellar mass, relative to equivalent results from GCLASS \citet{muzzin2012} -- \textit{not} the same mass binning as used throughout the paper or the co-added spectra shown in Figure\,\ref{fig:uvj_stacks}. Masses selected within bins of log\,$M_\ast$/M$_{\sun} {\in}$[9.45,10.15), [10.15,10.85), and [10.85,12.15) where a 0.2\,dex offset was applied to the selection of \citet{muzzin2012} based on the difference in mass measurement techniques (see Appendix\,\ref{sec:param_nonparam}). Points are shown slightly offset for clarity. Cluster galaxy values are marked with circles, field galaxy values with squares. Green colours mark measurements with GOGREEN, with black error bars corresponding the uncertainty in averaged \dbreak values, and cyan error bars showing the systematic error between methods of combining the values. Black outlined points show the measurements from \citet{muzzin2012} (taken from their table\,5, averaged over radial bins).
      While the GCLASS sample shows small differences in \dbreak between environments, on average we find larger differences in the \dbreak of the average spectrum of cluster galaxies than field galaxies for the GOGREEN sample -- consistent with the sense of the age difference we measure from fitting the SFHs of individual galaxies. } \label{fig:d4000_mass}
    \end{figure}

\dbreak is commonly used as an age indicator \citep[e.g.][]{balogh1999,kauffmann2003b,muzzin2012}, because it is insensitive to dust and, as a relatively wide feature, can be measured at high SNR relative to other indices. \dbreak is not sensitive to the SFH, however; a galaxy that quenched rapidly and one that quenched slowly can have the same \dbreak, depending on the relative timing of the quenching. Less apparent from the co-added spectra (and only statistically significant for the moderate mass galaxies) is that the field spectra have smaller \dbreak than cluster galaxies. A comparison is shown in Figure\,\ref{fig:d4000_mass} \textit{for different mass selections} than for the co-added spectra shown in Figure\,\ref{fig:uvj_stacks}, relative to values measured for galaxies in the GCLASS survey \citep[][averaged over radial bins -- see their table\,5]{muzzin2012}. We note that the GCLASS sample between $1{<}z{<}1.5$ is included in our GOGREEN sample. We increase the reported GCLASS masses and mass selections by 0.2\,dex to account for differences in how the stellar masses were estimated; see the discussion in Appendix\,\ref{sec:param_nonparam}: log\,$M_\ast$/M$_{\sun}$$\in$[9.45, 10.15), [10.15,10.85), and [10.85, 12.15). Black error bars indicate the uncertainties of the \dbreak measurements from the combined spectra, while cyan error bars indicate the systematic uncertainty due to how the spectra are combined (i.e., inverse-weighted averaged, or median combined). 

While the GCLASS sample shows small differences in \dbreak between environments, on average we find larger differences in the \dbreak of the average spectrum of cluster galaxies than field galaxies for the GOGREEN sample. This is consistent with the sense of the age difference we measure from fitting the SFHs of individual galaxies. The GCLASS sample is dominated by galaxies at $z{\sim}0.8$, particularly at low stellar masses. That we find larger age differences than in GCLASS could hint that the age difference evolves between $z{\sim}0.8$ and $z{\sim}1.2$.

\section{Age as a function of \textit{UVJ} colour} \label{sec:mwa_uvj}

\begin{figure}
      \begin{center}
        \includegraphics[width=\linewidth]{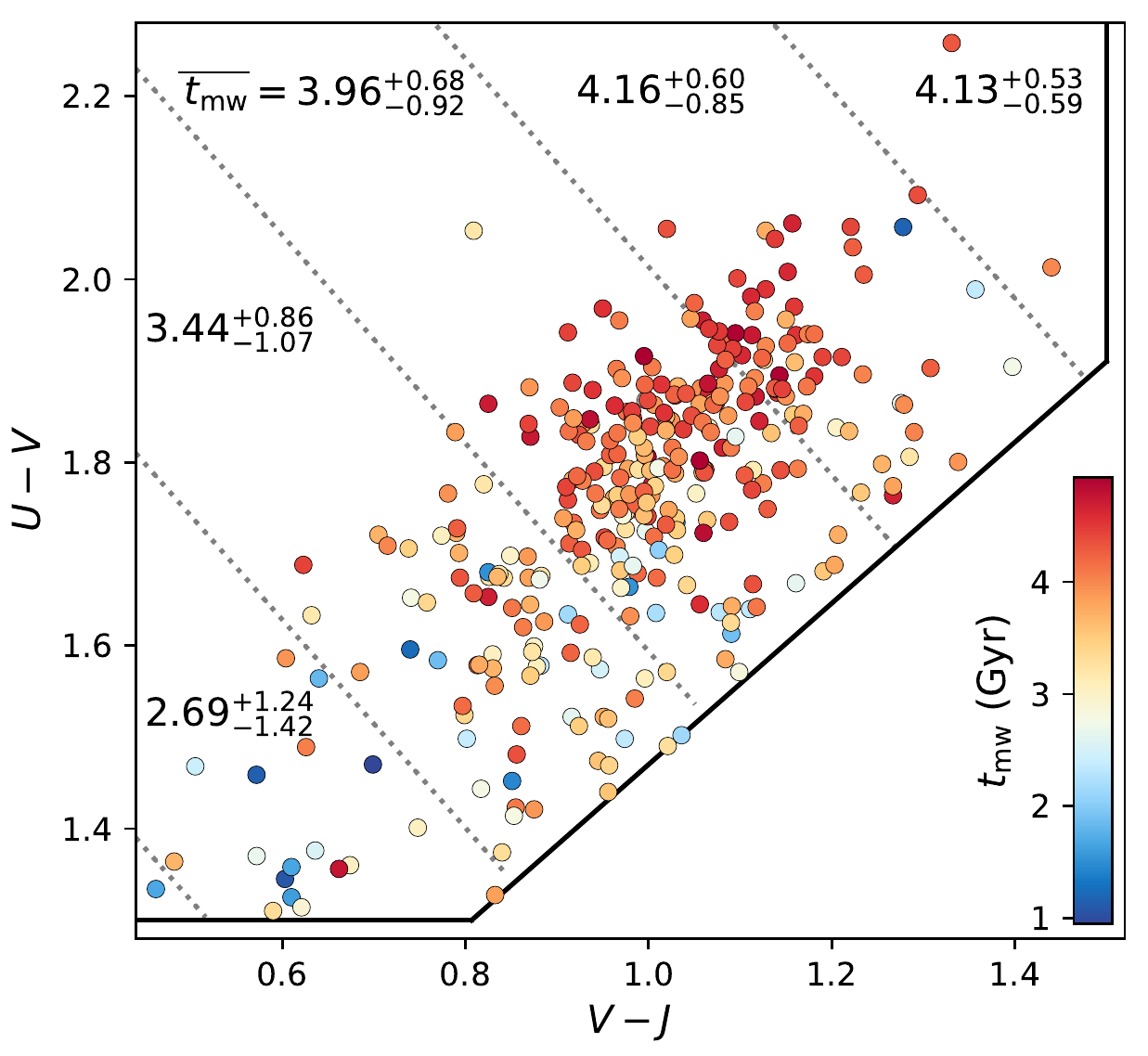}
      \end{center}
      \caption{Mass-weighted ages in rest-frame \textit{UVJ} colour space. The sample is 
      divided into five regions, where the median \mwa and 68 per cent credible regions for the galaxies in each bin are labelled. As expected, there is a positive trend between \mwa and rest-frame $U{-}V$ and $V{-}J$ colours, where the oldest galaxies are clustered towards the upper right of the quiescent region. The majority of galaxies in the `red clump' are the oldest galaxies in our sample, but otherwise there is not a smooth distribution of \mwa relative to \textit{UVJ} colours. } \label{fig:uvj_mwa}
    \end{figure}

Mass-weighted ages, \mwa, are shown in \textit{UVJ} colour space in Figure\,\ref{fig:uvj_mwa}. The sample is divided into five regions in \textit{UVJ}-colour space, delineated by dotted lines, and the median age (and 68 per cent credible regions) are labelled for each. As expected, there is a positive trend between \mwa and rest-frame $U{-}V$ and $V{-}J$ colours, where the oldest galaxies are clustered towards the upper right of the quiescent region. We find good consistency between our \textit{UVJ}-ages trend and trends in the literature \citep[e.g.][]{belli2019, estrada-carpenter2019, ferreras2019}, despite systematic or procedural differences between studies, for example: SFR parameterization, SED-fitting procedures, how the ages were measured (luminosity weighted, mass-weighted, median, etc.), and the mass or redshift range of the samples. The overall age gradient in \textit{UVJ}-colour space is flatter than predicted by \citet{belli2019}, which could be attributed to the aforementioned systematics. However, \citet{carnall2019b} report their sample of $1{<}z{<}1.3$ quiescent galaxies to have \mwa in good agreement with the \citet{belli2019} relationship despite having similar methodological differences. Although the systematics related to our fitting procedure are important when comparing to the literature, they are less important for the purposes of this study -- the differential comparison of cluster and field populations. Our age estimates are discussed further in Section\,\ref{sec:mwas}.

\section{Luminosity weighted ages} \label{sec:lwas}

The luminosity-weighted age is more sensitive to recent star formation, as younger stars dominate the integrated luminosity. For passively evolving galaxies, which formed all their stars a long time ago, the mass-weighted age and luminosity-weighted ages should be equivalent. We calculate the luminosity-weighted age from the SFH posteriors,
\begin{equation}
t_\mathrm{lw} = \frac{ \int_{t_\mathrm{obs}}^{0} t \, \mathrm{SFR}(t) \, L(t) \, \mathrm{d}t }{ \int_{t_\mathrm{obs}}^{0} \mathrm{SFR}(t) \, L(t) \, \mathrm{d}t }
\end{equation}
\noindent where $L$ is the g-band luminosity. 

\begin{figure*}
      \begin{center}
        \includegraphics[width=\linewidth]{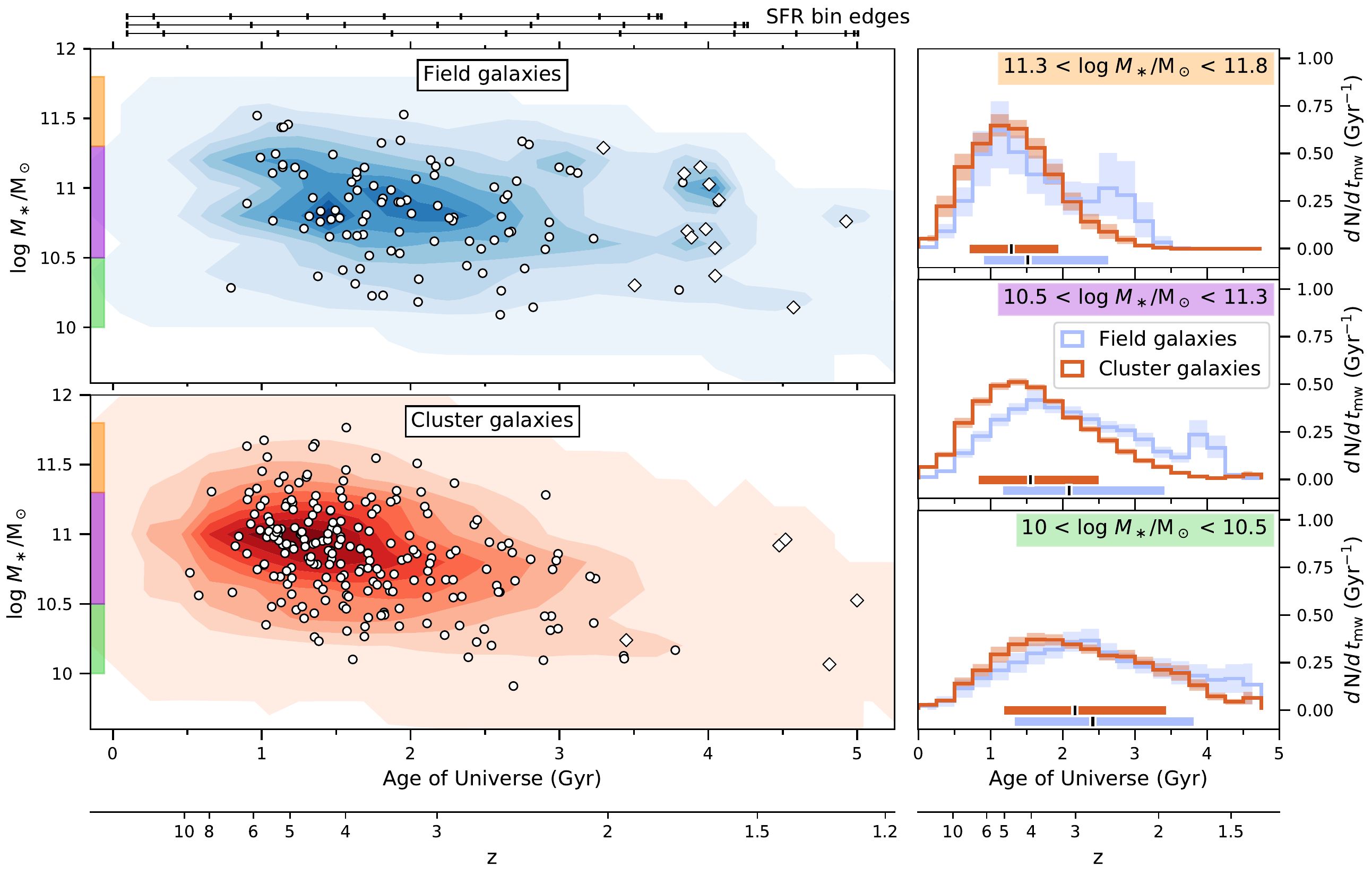}  
      \end{center}
      \caption{ 
      Comparison of stellar masses and luminosity-weighted ages between field (blue) and cluster (red) galaxies.
        Left: Combined posteriors of stellar masses and \lwa (in units of cosmic time), shown as contours. The medians of the individual posteriors are marked with white circles/diamonds. Diamonds indicate \frejuv${>}$0.1 galaxies (formed more than 10\,per cent of their stellar mass within the last 1\,Gyr). Horizontal bars at the top of the figure indicate the edges of the age bins for $z{=}1.5$ (top), $z{=}1.25$ (middle), and $z{=}1$ (bottom). The bins were defined in units of lookback time, and therefore do not match up for galaxies observed at different redshifts.
      Right: Combined \lwa posteriors for field and cluster galaxies, shown in three mass bins. The medians (black mark) and 68 per cent credible regions (coloured bar) of each distribution is marked at the bottom of each subplot. The shaded regions show the bootstrapped uncertainty of each histogram. 
      Although there are field galaxies that formed as early as the oldest cluster galaxies, and cluster galaxies that formed as late as the youngest field galaxies, \textit{on average} field galaxies formed at later times.  } \label{fig:diff_lwa}
    \end{figure*}

Figure\,\ref{fig:diff_lwa} shows the distribution of the stellar mass and luminosity-weighted ages, in units of cosmic time (similar to Fig\,\ref{fig:diff_mwa} for mass-weighted ages). Contours show the combined posteriors of the field (blue) and cluster (red) galaxies, where white points indicate the medians of the individual posteriors. Diamonds mark galaxies which have formed more than 10\,per cent of their stellar mass within the last 1\,Gyr, \frejuv${>}$0.1, discussed in Sec\,\ref{sec:rejuv}. Compared to the mass-weighted ages, the luminosity-weighted ages are younger on average, but not uniformly younger. As a result the age distributions are broadened. 

Following the same mass-matched cumulative age comparison as for \mwa, we find that cluster galaxies are on average ${0.39}_{^{-0.40}}^{_{+0.58}}$\,Gyr older than field galaxies, a ${\sim}$0.1\,Gyr larger difference. Figure\,\ref{fig:diff_mwa_cum_all} compares this age comparison to that with mass-weighted ages.

\bsp	
\label{lastpage}
\end{document}